\documentclass[12pt]{article}

\usepackage{epsfig}

\begin{document}

\title{UHECR propagation in the Galaxy: clustering versus isotropy}

\author{Jaime Alvarez-Mu{\~n}iz, Ralph Engel, and Todor Stanev \\
       { \it Bartol Research Institute, University of Delaware} \\
       { \it Newark, DE 19716, USA} }

%\date{ {\Large \bf DRAFT:} December 8, 2001}

\date{\today}

\maketitle

\begin{abstract}
Recently the AGASA Collaboration presented data suggesting a 
significant clustering of
ultra-high energy cosmic rays coming from the outer Galaxy region.
In this paper we calculate expected cosmic ray arrival distributions 
for several simple, limiting source location
scenarios and investigate the possibility of 
clustering and correlation effects.
The role of the Galactic magnetic field is discussed in detail.
\end{abstract}

\section{Introduction}

 The observations of cosmic rays of energy above 10$^{19}$ eV 
 reveal at least two features that we have not yet understood 
 and that appear to contradict each other. The cosmic ray energy
 spectrum does not seem to be cut off because of photoproduction
 interactions on the microwave background \cite{Greisen:1966,Zatsepin:1966}
 and extends above 10$^{20}$ eV \cite{Takeda:1998ps}.
 The existence of these particles suggests that the sources of
 ultra-high energy cosmic rays (UHECRs)
 are cosmologically nearby - within 20 Mpc or so \cite{Stanev:2000fb}.
 On the other hand, the UHECR arrival direction distribution, as far as
 we know it, seems isotropic on a large scale with a statistically
 improbable small scale clustering \cite{Uchihori:1999gu}  
(for a review on the subject see \cite{Nagano:2000ve}).

This small-scale clustering was first observed in the data of the AGASA
group \cite{Takeda:1999ap}. 
The current AGASA data set of 59 events of energy above
$4\times10^{19}$ eV  contains five doublets and one triplet
for a maximum separation angle of 
$2.5^\circ$~\cite{Takeda:2001icrc}. 
 The centroids of the triplet and one of the doublets are 
 less than 1 degree off the supergalactic plane (SGP). 
 When combined with data from  
 the Haverah Park, Yakutsk and Volcano Ranch experiments 
 the number of doublets (within $3^\circ$) 
 increases to eight and the number of triplets to two \cite{Uchihori:1999gu}. 
 The chance probability of observing these multiplets
 from a uniform distribution of sources is less than 
 $1.5~\%$ and less than $1~\%$ when a restricted region
 within $\pm10^{\circ}$ off the SGP is considered \cite{Uchihori:1999gu}.

Recently the AGASA group presented a self-correlation analysis of their
data with $E>10^{19}$ eV \cite{Takeda:2001icrc}.
Remarkable correlations are found in Galactic coordinates, supporting
further the previously found indications of clustering of UHECRs.
In a straight-forward interpretation one would link the existence of such
clusters directly to UHECR point sources 
\cite{Tinyakov:2001ic,Anchordoqui:2001qk} but
non-trivial effects such as possible clustering of sources or 
large-scale magnetic fields could also contribute to such a 
correlation \cite{Stanev:1996qj,MedinaTanco:1998aj,%
MedinaTanco:1998ap,MedinaTanco:1998yq,MedinaTanco:1999aj,%
Harari:2000he, Harari:2000az}.

The indications found in data
raise the interesting question of whether both an isotropic distribution
and a clustering of the arrival directions can be
explained consistently, assuming they are not a statistical 
fluctuation \cite{Goldberg:2000zq,Razzaque:2001tp}.

Potential models of UHECR origin, assuming they are charged
particles, 
are constrained by their ability to reproduce 
the measured energy spectrum and the approximately 
isotropic arrival distribution observed in the data 
\cite{Olinto:2000sa,Bhattacharjee:1998qc}. At the same time,   
their predictions should also be consistent with the small scale
clustering. 
In general, several key ingredients enter the models: 
(i) the distribution (locations) of the sources, (ii) their nature, i.e. 
whether they are sources emitting cosmic rays continuously for 
a long period of time or they are bursting sources, (iii)
the energy spectrum and total flux at injection, and (iv) the propagation 
of the cosmic rays including energy loss processes
and deflection due to magnetic
fields.
Clustering on small scales and overall isotropy seems to favour models 
based on several nearby sources (for example, in the Galactic halo).
Models of this type include
the acceleration of cosmic rays at magnetars \cite{Blasi:1999xm}, 
UHE neutrino interactions on the relic neutrino background
(Z-bursts)~\cite{Weiler:1997sh,Fargion:1997ft}, and the decay and 
annihilation of
superheavy Relics accumulated in the halo of the Galaxy 
\cite{Berezinsky:1997hy,Berezinsky:1998ed,Birkel:1998nx,%
Blasi:2001hr,Evans:2001rv,Sarkar:2001se}. 

In this paper we study general aspects related to cosmic ray 
propagation and the source distribution. The aim 
is to explore what a future confirmation of the correlations 
could reveal about the nature of 
the sources of UHECRs, their distribution and the strength
of the magnetic field in our Galaxy. 
To keep our results as general as possible we  
consider three simplified, limiting 
source location scenarios:\\
(i) Uniform distribution of sources, which is relevant to UHECR 
of extragalactic origin with sources of isotropic and homogeneous
distribution.\\
(ii) sources distributed uniformly within $\pm20^\circ$ off the
supergalactic plane (SGP), and\\
(iii) single and multiple point sources.\\ 
For each of these source scenarios we calculate arrival
distributions, one and two-dimensional correlation functions and, where
applicable, arrival time delays.

It should be emphasized that we do not attempt to 
reproduce AGASA data
with these source models. The interest of this work is merely the
investigation of characteristic features which are fairly
model-independent. For a more realistic analysis one needs 
a better knowledge of
the Galactic magnetic field, as well as full access to the
experimental data, including acceptance corrections.

The outline of the paper is as follows. In Sec.~\ref{sec:mag-field} we
discuss the Galactic magnetic field. The simulation method for cosmic
ray arrival distributions is explained in Sec.~\ref{sec:method} and the
different source location scenarios are presented in
Secs.~\ref{sec:uniform} through \ref{sec:multiple-point}.
Sec.~\ref{sec:discussion} applies the results of the previous sections to
AGASA multiplets and Sec.~\ref{sec:end} concludes the paper with 
a summary of our main results.

%%%%%%%%%%%%%%%%%%%%%%%%%%%%%%%%%%%%%%%%%%%%%%%%%%%%%%%%%%%%%%%%%

\section{Galactic magnetic field\label{sec:mag-field}}

Unfortunately the Galactic magnetic field (GMF) is not very well known.
The position of the Solar system makes it difficult to measure its
global structure. The most extensive and reliable information on the
GMF comes from measurements of polarized
synchrotron radiation and Faraday rotation of the
radiation emitted from pulsars and extragalactic sources. The
structure of the GMF as well as its local value at the Solar
system are still uncertain, mainly due to the
limited number of rotation measures available and the intrinsic
difficulties in distinguishing local small-scale features
from a large-scale one.
Faraday rotation measurements indicate that the GMF in the
disk of the Galaxy has a spiral structure with field reversals
at the optical Galactic arms (for a review see \cite{Beck:2001}). 
Recent work favours a bisymmetric spiral field (BSS) structure but 
an axisymmetric (ASS) field is not
excluded \cite{Han:1999mn,Han:2001i}.

In the following we use a BSS field model for the
regular magnetic field in the disk of the Galaxy.
The Solar System is at a distance $r_{\vert\vert}=R_\oplus=8.5$ kpc 
from the center of the Galaxy in the Galactic plane.
The local regular magnetic field in the vicinity of the Solar System
is assumed to be 
$\sim 1.5~\mu{\rm G}$ in the direction $l=90^{\rm o}+p$ where the
pitch angle is $p=-10^{\rm o}$ \cite{Han:1994aa}. New measurements
discuss larger total field strengths of up to 6 $\mu$G \cite{Beck:2001}.
Therefore our assumptions should be considered as rather conservative. 
To illustrate the effect of a stronger magnetic field we will also
perform a simulation with a local regular field strength of 
$3~\mu{\rm G}$.

\begin{figure}[htb!]
\centerline{
\epsfig{figure=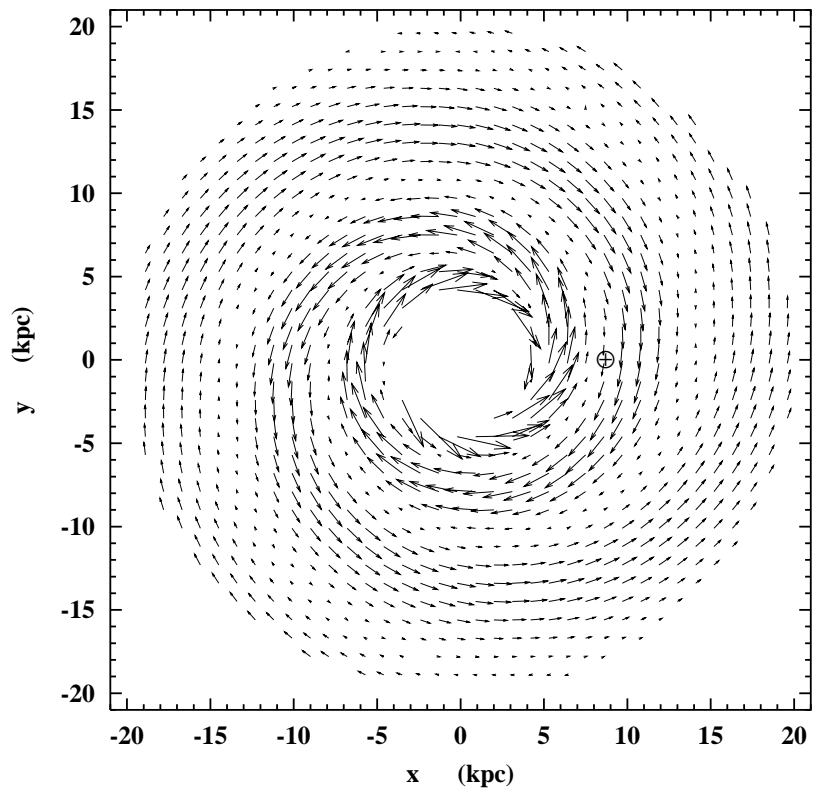,width=11cm}
}
\caption{
Magnetic field configuration in the Galactic plane. The vectors
indicate the field direction and their length is proportional to its
magnitude.
\label{fig:bss}
}
\end{figure}

The strength of the spiral field at a
point in the Galactic plane, described by the  polar 
coordinates $(r_{\vert\vert},\phi)$,
is given by:
\begin{equation}
B(r_{\vert\vert},\phi)=
B_0~\left({R_\oplus \over r_{\vert\vert}}\right)~
\cos\left(\phi - \beta \ln {r_{\vert\vert}\over r_0} \right)
\end{equation}  
where $B_0=4.4~\mu$G, $r_0= 10.55$ kpc 
%is the Galactocentric distance 
%of the location of the first maximum field strength in the plane 
and $\beta=1/\tan p=-5.67$.
The field decreases with Galactocentric distance as $1/r_{\vert\vert}$ 
and it is zero for $r_{\vert\vert}>20$ kpc. 
In the region around the Galactic center ($r_{\vert\vert} < 4$ kpc) 
the field is highly uncertain. For simplicity
we assume it is constant and equal to its
value at $r_{\vert\vert}=4$ kpc.
Fig.~\ref{fig:bss} shows the field in the Galactic plane. For clarity we
have not plotted the region $r_{\vert\vert}<3$ kpc.
%%%%%%%%%%%%%%%%%%%%%%%%%%%
%%%%%%%%%%%%%%%%%%%%%%%%%%%

Following \cite{Stanev:1996qj}
the spiral field strengths above and below the Galactic plane are taken to
decrease exponentially with two scale heights: 
\begin{equation}
\vert B(r_{\vert\vert},\phi,z)\vert = 
\vert B(r_{\vert\vert},\phi)\vert 
\left\{
\begin{array}{lcl}
\exp(-z) & \hspace*{5mm}: \hspace*{5mm} & \vert z \vert \leq 0.5~ {\rm kpc} \\ 
\exp(-3/8)~\exp(-z/4) & \hspace*{5mm}: \hspace*{5mm} & \vert z \vert > 0.5~ {\rm kpc} 
\end{array}
\right.
\label{eq:b-height}
\end{equation} 
where the factor $\exp(-3/8)$ makes the field continuous in $z$.
The BSS spiral field we use is of even parity, i.e.
the field direction is preserved at disk crossing. 

Observations show that 
the field in the Galactic halo is much weaker than that in the disk.
The currently discussed options favour either a quadrupole field or a
simple dipole field \cite{Beck:2001,Han:1997aa,Han:2001ii}.
In this work we assume that the regular field corresponds to a 
A0 dipole field as suggested in \cite{Han:2001ii}. 
The dipole field is toroidal and its strength
decreases with Galactocentric distance as $1/r^3$.
In spherical coordinates $(r,\theta,\varphi)$ 
the $(x,y,z)$ components of the halo field are given by:
\begin{eqnarray}
B_x=-3~\mu_{\rm G}~{\sin\theta \cos\theta \cos\varphi}/
r^3 \nonumber \\ 
B_y=-3~\mu_{\rm G}~{\sin\theta \cos\theta \sin\varphi}/
r^3 \\
B_z=\mu_{\rm G}~{(1-3\sin^2\theta)}/r^3 ~~~~~~~ \nonumber
\label{eq:bdipole}
\end{eqnarray}
where $\mu_{\rm G}\sim 184.2~{\rm \mu G~kpc^3}$ is the magnetic moment
of the Galactic dipole. The dipole field is very strong in the central region 
of the Galaxy, but is only 0.3 $\mu$G in the vicinity
of the Solar system, directed toward the North Galactic Pole.

There is a significant turbulent component, $B_{\rm ran}$,
of the Galactic magnetic field.
Its field strength is difficult to measure and results found in
literature are in the range of $B_{\rm ran} = 0.5 \dots 2 B_{\rm reg}$
\cite{Beck:2001}.
To simulate it we add to both regular components 
a random field with a strength of 50\% of the local regular field
strength and a coherence length of 100 pc.
The possible time dependence of the turbulent magnetic field component is
neglected. This could be an over-simplification if this component
changes significantly over time scales of $10^3$ years.

%%%%%%%%%%%%%%%%%%%%%%%%%%%%%%%%%%%%%%%%%%%%%%%%%%%%%%%%%%%%%%%%%

\section{Simulation and analysis methods\label{sec:method}}

\subsection{Simulation}

We simulate the propagation of protons of energy above $10^{19}$ eV in
the Galactic magnetic field (GMF) 
by numerically integrating the equations of motion in a magnetic
field. The results can be easily re-scaled in rigidity for heavier
nuclei. 

Since we are not interested in extragalactic propagation it is
sufficient to consider a number of typical Galactocentric 
source distances, namely $r_{\rm
src}=15, 20$ and 40 kpc. The 40 kpc distance corresponds to a scenario
of distant halo sources or extragalactic sources if the extragalactic 
magnetic field is weak, and the two others are
chosen to study the distance dependence within the Galaxy. The energy
loss of protons can be neglected for these distances. Therefore we use
a differential injection spectrum which is similar to
the observed spectrum, $dN/dE \propto E^{-2.7}$.

Protons are injected at the corresponding source locations
and each trajectory is followed until:\\
(i) it reaches a Galactocentric distance larger than 100 kpc,\\
(ii) the total path traveled by the proton is larger than
$3~r_{\rm src}$, or\\
(iii) it intersects a spherical detector of radius $r_{\rm det}$
centered at the position of the Solar System.\\
As will become clear later, the total length of trajectories reaching Earth
is always much smaller than the limit imposed by (ii).

The radius of the detector is $r_{\rm
det}=100~(r_{\rm src}/20~{\rm kpc})$ in pc. This guarantees the same geometric
efficiency for different source distances and optimizes the efficiency of
the propagation, keeping the unavoidable smearing in arrival
angle smaller than $0.5^\circ$. 

In all calculations we consider isotropically emitting sources.
However, to increase the efficiency of the simulations,
we only inject protons within a cone of half opening angle of
$30^\circ$, pointing toward the detector.
We have checked that the number of protons that reach the detector and
are injected at an angle larger than $30^\circ$ from the line-of-sight
is negligible for all energies and distances considered here.

%%%%%%%%%%%%%%%%%%%%%%%%%%%%%%%%
\begin{figure}[htb!]
\centerline{
\epsfig{figure=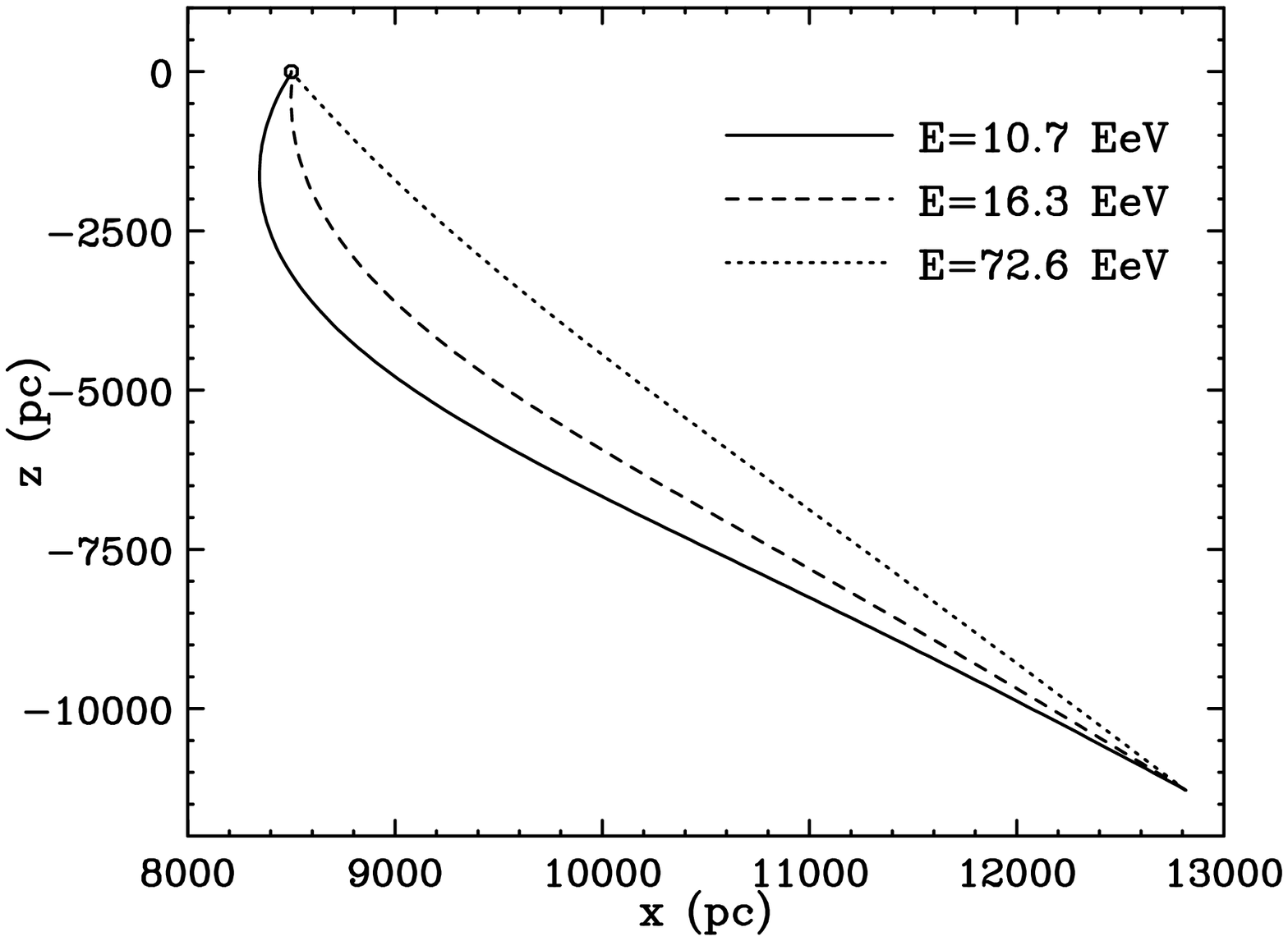,width=7.9cm}
\epsfig{figure=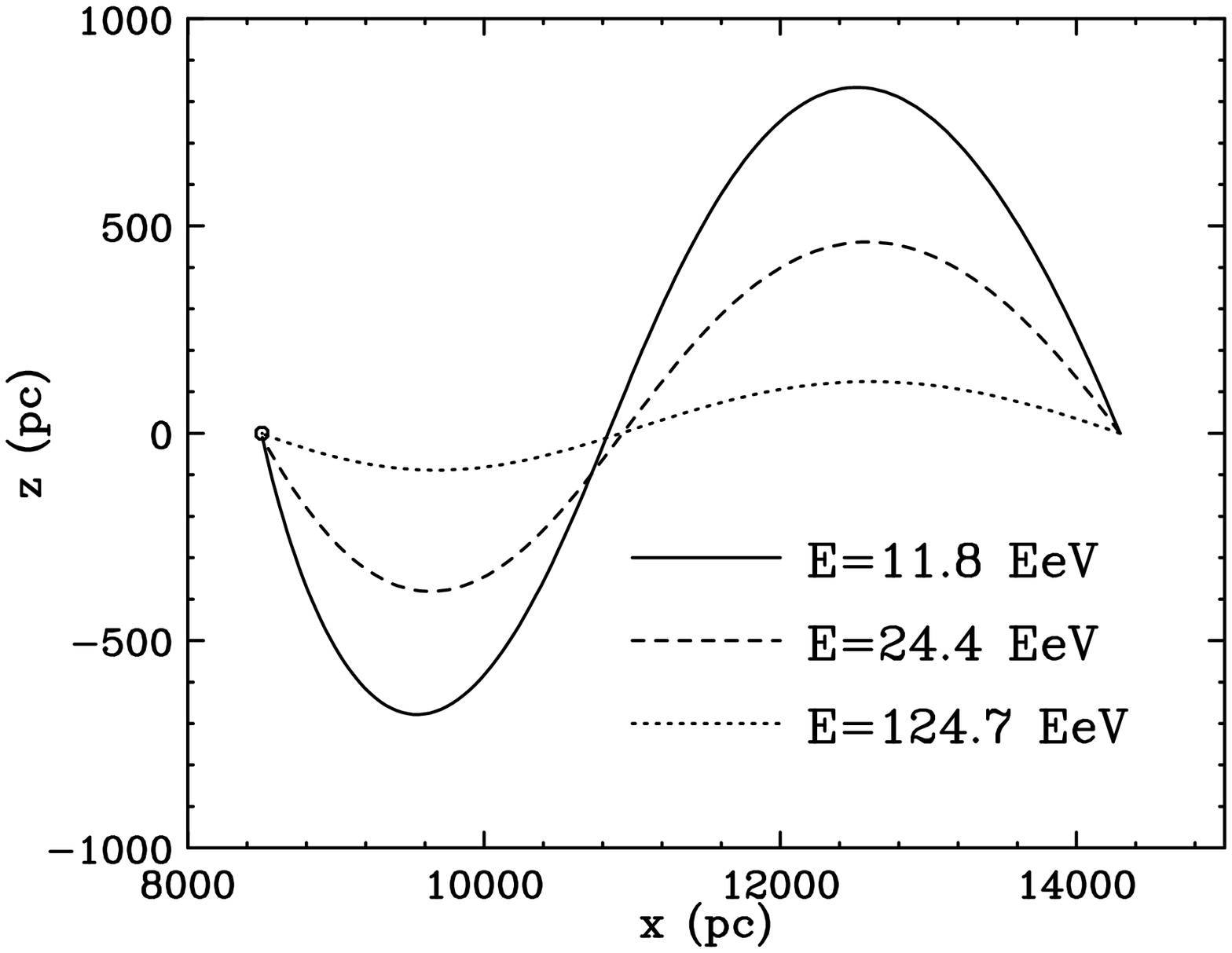,width=7.9cm}
}
\caption{
Examples of proton trajectories projected onto the $xz$
plane, being orthogonal to the Galactic plane. 
The left (right) panel corresponds to a source located at
$b=-45^\circ,~l=112.5^\circ$ ($b=0^\circ,~l=112.5^\circ$)
in Galactic coordinates at a distance of $r_{\rm src}=20$ kpc.
The tracks are 
labeled with their corresponding energies. The dot marks the 
position of the detector. 
Note the different scales of the horizontal and vertical axes.
Only the
regular GMF has been used in this simulation.
\label{fig:tracks-1}
}
\end{figure}
%%%%%%%%%%%%%%%%%%%%%%%%%%%%%%%% 
To illustrate our simulation procedure, Fig.~\ref{fig:tracks-1}
shows the projection of several representative proton trajectories
onto the $xz$ plane. Only the
regular GMF has been used in this simulation. The coordinate system is
defined by $+z$ pointing toward the North Galactic pole and the
location of the Solar system being at $\vec{r}_\oplus=(8.5,0,0)$ in units
of kpc.
The trajectories shown in
the right panel clearly reflect the structure of the BSS field with its
field reversals.

\subsection{Analysis}

One of the significant deviations from isotropy observed by the
AGASA detector (and supported by the other air shower data sets)
is the non uniformity in the distribution of space angle between
the UHECR events. This is caused by the large fraction of the
experimental statistics above $10^{19.4}$ eV that is in the form 
of doublets - 13 out of 59 individual showers. 
% if the triplet is reduced to three doubles.

%Following the recent AGASA approach \cite{Takeda:2001icrc}
%we select events coming from the Outer Galaxy, i.e.
%$90^\circ < l < 180^\circ$ and $-60^\circ < b < 60^\circ$.  

The number of multiplets for a given opening angle depends
on the total number of cosmic rays observed within the field of view of
the detector. With increasing number of detected cosmic rays the number
of ``artificial'' multiplets grows simply due to the decreasing mean
space angle between the cosmic rays. This is a  particular problem for
numerical simulations which are not restricted in statistics.
Therefore we will, following the similar approach of AGASA 
\cite{Takeda:2001icrc},
calculate the significance of one- and two-dimensional self-correlations.

For each of the scenarios we calculate
the self-correlation in separation angle 
of the events in the simulated sample.
The separation angle between pairs of events is obtained and
the distribution is divided by the solid angle of a concentric ring
centered in each angular bin. 
The statistical significance of any deviation from the expected
isotropic background is shown in the significance plots as
(signal-background)/sigma-of-background. 
The background is given by the mean self-correlation
function for many sets with the same number of cosmic rays but 
arrival directions sampled according to 
%phase space 
a locally isotropic flux.

The role of the Galactic magnetic field as well as possible bias due to 
cosmic ray selection criteria can be studied by 
two-dimensional self-correlations in Galactic coordinates $l,b$.
The difference in Galactic longitude ($\Delta l$) and 
Galactic latitude ($\Delta b$) 
between each pair of cosmic rays is calculated and plotted in a 2 dimensional
map.
The significance plots are meant to represent graphically the deviation
of a certain arrival direction distribution from a local isotropic
distribution. The function plotted is: 
\begin{equation}
\rho(\Delta l,\Delta b) = \frac{f_{\rm dat}(\Delta l,\Delta b)
-\langle f_{\rm bkg}(\Delta l,\Delta b) \rangle}{
\sqrt{\langle f^2_{\rm bkg}(\Delta l,\Delta b) \rangle - 
\langle f_{\rm bkg}(\Delta l,\Delta b) \rangle^2}}.
\label{eq:significance}
\end{equation}
The background (bkg) is calculated by sampling from an isotropic distribution
many sets of events containing the same number of events as the simulated data
set. A large number of these sets has to be generated since the background 
depends strongly on the number of events falling
into the $\Delta l \times \Delta b$ window of the correlation plot.
Typically we generate 10,000 such background configurations. 

%The correlation function $f$ is calculated without normalization because
%it cancels in (\ref{eq:significance}). 
The experimental resolution is important. 
We assume the error of a measured arrival direction is
Gaussian. The distribution in arrival directions is then given 
by
\begin{equation}
g(l,b) = \sum_i \frac{1}{\pi \sigma_i^2}
\exp\left\{ - \frac{(l - l_i)^2}{\sigma_i^2} -
\frac{(b-b_i)^2}{\sigma_i^2} \right\}, 
\end{equation}
where $\sigma_i$ is the error of the cosmic ray $i$ with arrival
direction ($l_i$, $b_i$). For simplicity we have assumed the same
uncertainty in Galactic longitude and latitude.

The correlation function reads
\begin{eqnarray}
f_{\rm dat}(\Delta l, \Delta b) = 
\int dl \int db\ g(l,b) g(l-\Delta l, b- \Delta b)\\
=\sum_{p} \frac{1}{2 \pi
\sigma_i\sigma_j}
\exp\left\{ - \frac{(\Delta l - \Delta l_p)^2}{\tilde\sigma_p^2} -
\frac{(\Delta b-\Delta b_p)^2}{\tilde\sigma_p^2} \right\} .
\end{eqnarray}
The sum runs over all pairs $p=(i,j)$ of cosmic rays with $i\ne j$
and $\Delta l_p = l_i - l_j$, $\Delta b_p = b_i - b_j$.
The new effective error is
\begin{equation}
\tilde\sigma_p = \sqrt{\sigma_i^2+\sigma_j^2},
\end{equation}
as expected from standard error propagation.
The plots are made with an energy-independent, constant error of
$\sigma_i = 1.47^\circ$.
Note that these plots are by construction symmetric, i.e.
every pair of particles enters the plots twice.

The significance plots can be interpreted as positive or negative
deviation (over- or underdensity) in units of the uncertainty 
given by the expected background fluctuations. Consequently, the
significance of an excess seen in a plot with only a few cosmic rays
falling in the $\Delta l \times \Delta b$ window 
is always small. The large background fluctuations reduce
the significance of any possible signal.

Since the particle gyroradius, and thus the degree of deflection
in the GMF depends on its energy, the physics of the
deflection of cosmic rays in the Galactic magnetic
field is better revealed by looking at energy-ordered pairs.
For example, knowing the sign of the charge of the cosmic rays, 
the energy-ordered
correlation function can be used to derive the mean magnetic field
direction along the path of weakly deflected cosmic rays coming 
from the same source.
In the following we apply the convention
\begin{equation}
\Delta l = \left\{
\begin{array}{lcl}
l_i - l_j & \hspace*{5mm}: \hspace*{5mm} & E_i > E_j \\
l_j - l_i & \hspace*{5mm}: \hspace*{5mm} & E_i \le  E_j
\end{array}
\right.
\label{eq:corr-ordered}
\end{equation}
and the corresponding definition for $\Delta b$. 
For cosmic rays of the same charge, the corresponding
correlation function is not symmetric in $\Delta l$ and $\Delta b$ if
there exists a magnetic field component orthogonal to the particle
trajectories (see below).

%%%%%%%%%%%%%%%%%%%%%%%%%%%%%%%%
\begin{figure}[htb!]
\centerline{
\epsfig{figure=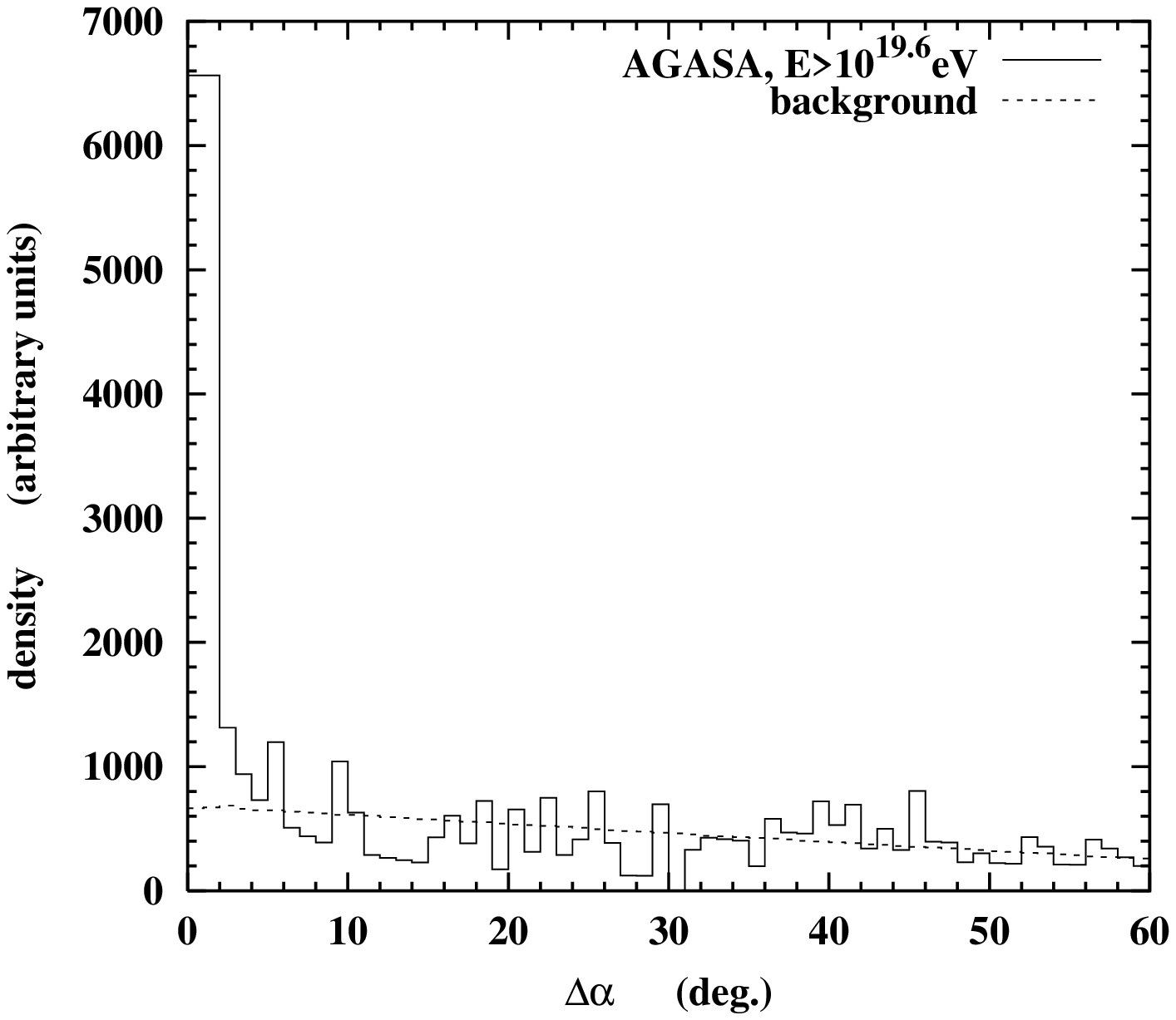,width=7.9cm}
\epsfig{figure=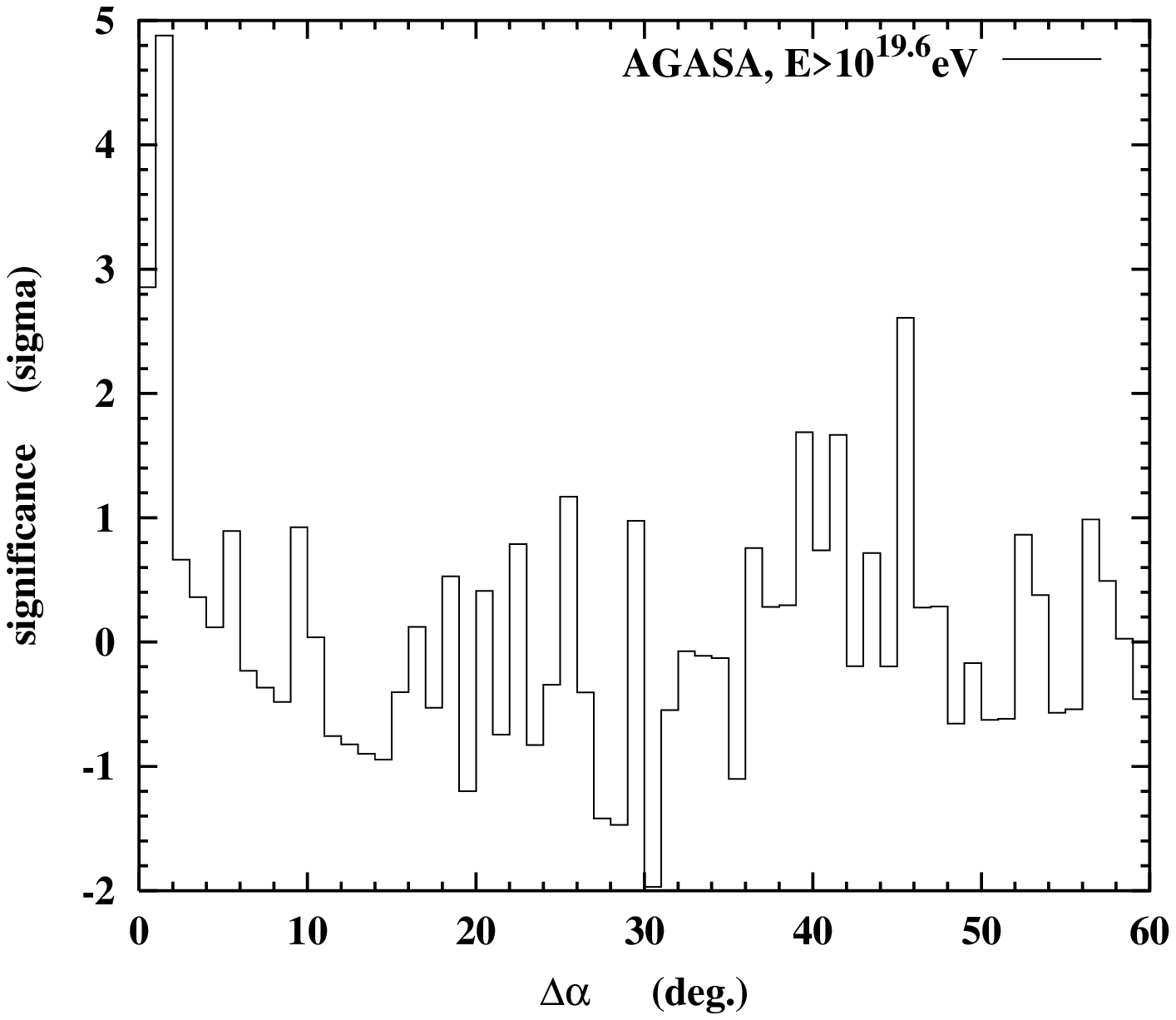,width=7.9cm}
}
\caption{
1D correlation function for published AGASA data with $E>10^{19.6}$ eV.
The left panel shows the correlation function normalized to the phase
space. The dotted line represents the expected distribution for an
isotropic flux.
The right panel is the corresponding statistical significance
of the deviation from the background.
\label{fig:AGASA-1dc-19p6}
}
\end{figure}
%%%%%%%%%%%%%%%%%%%%%%%%%%%%%%%%
To check our analysis methods and to 
demonstrate the importance of the newly introduced quantities 
(Eqs.~(\ref{eq:significance})  and (\ref{eq:corr-ordered}))  
we analyse the published list of AGASA events
with energy above $10^{19.6}$ eV, which consists of 58 events
containing 1 triplet and 5 doublets \cite{Hayashida:2000zr}.

Following the AGASA analysis \cite{Takeda:2001icrc} we 
apply the cuts  $90^\circ < l < 180^\circ$ and
$-60^\circ < b < 60^\circ$.  This cut in arrival direction will be
referred to as the ``outer Galaxy'' cut from now on.
From the AGASA data 25 events with $E > 10^{19.6}$ eV pass the outer
Galaxy cut.  They contain two of the doublets and the triplet.

The left panel of Fig.~\ref{fig:AGASA-1dc-19p6} 
shows the 1D angular correlation
of the AGASA data above $10^{19.6}$ eV coming from the 
outer Galaxy. The significance of the deviation from
an uniform distribution is plotted in the right panel.
The significance approaches 5$\sigma$ for a space angle of
2$^\circ$ and is consistent with no deviation for
larger space angles.

%%%%%%%%%%%%%%%%%%%%%%%%%%%%%%%%
\begin{figure}[htb!]
\centerline{
\epsfig{figure=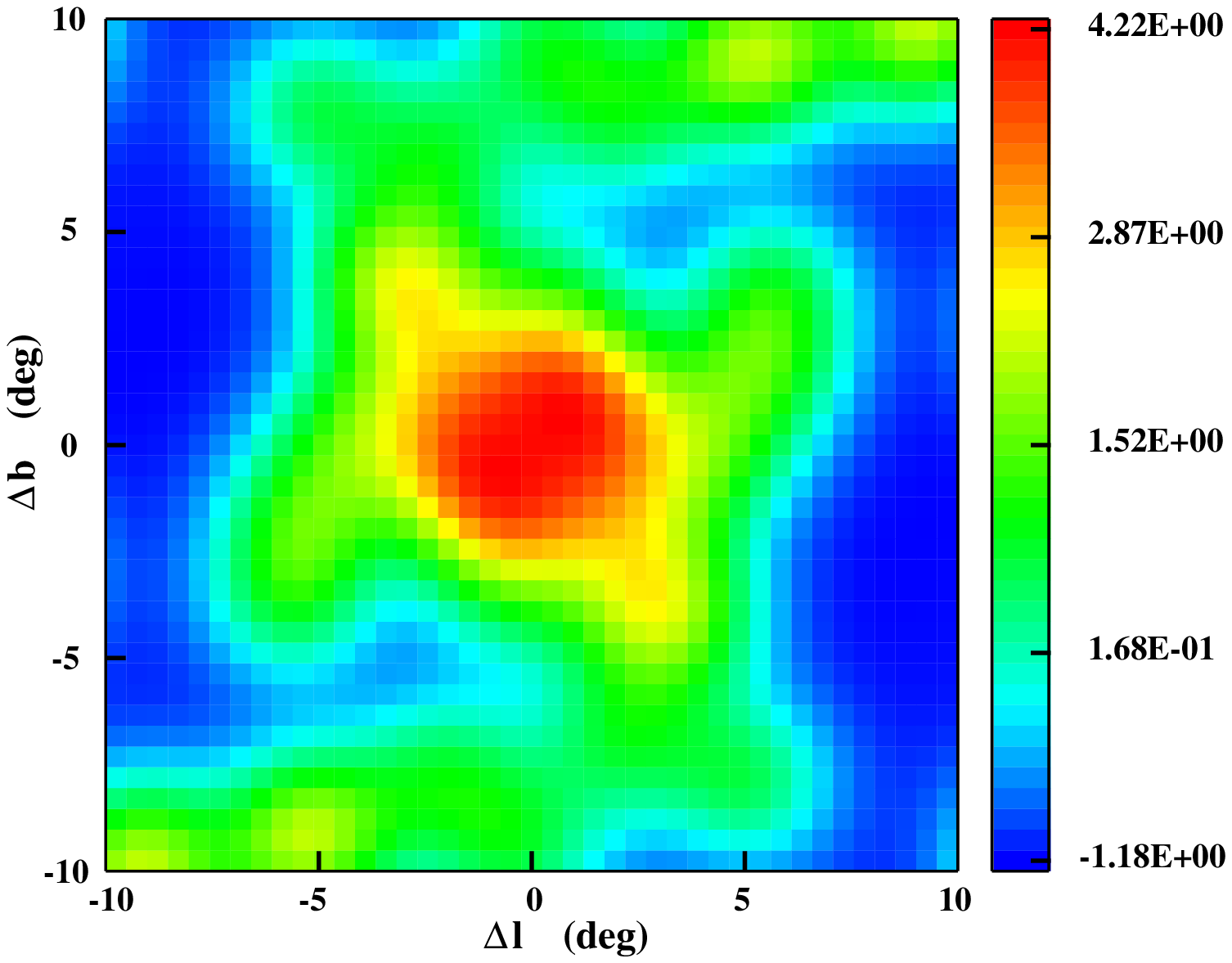,width=7.9cm}
\epsfig{figure=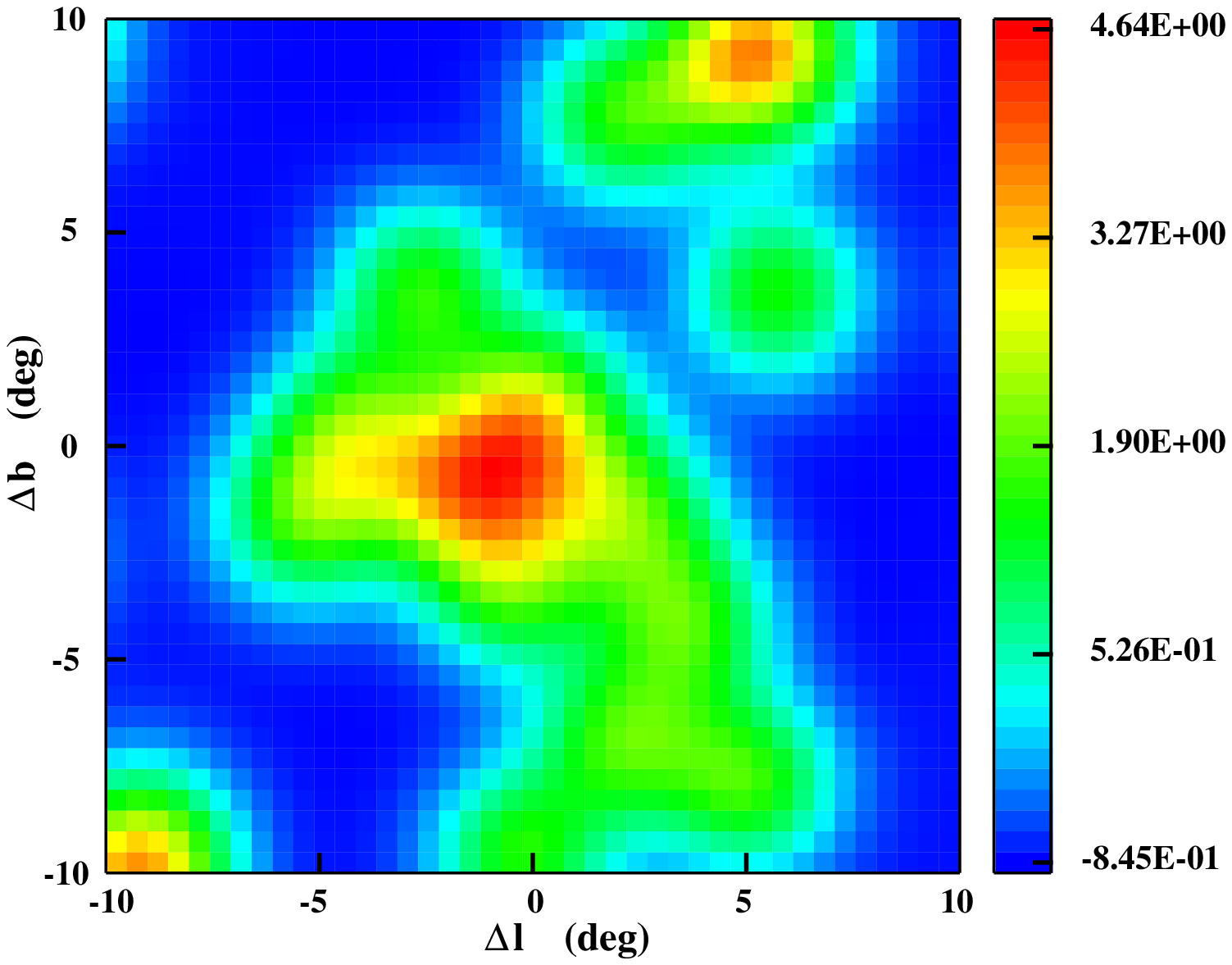,width=7.9cm}
}
\caption{
Significance of the self-correlation for published
AGASA data with $E > 10^{19.6}$ eV \protect\cite{Hayashida:2000zr}
relative to an isotropic flux at Earth.
The left panel is the significance of the symmetrized correlation and 
the right panel shows the energy-ordered
significance for the same events. 
Note that 1D and 2D self-correlation plots, similar
to those showed here, were first presented by the AGASA
collaboration at the ICRC 2001 \cite{Takeda:2001icrc}.
\label{fig:AGASA-2dc-19p6}
}
\end{figure}
%%%%%%%%%%%%%%%%%%%%%%%%%%%%%%%%
Both the symmetrized and the  energy-ordered significance 
of the self-correlation
$\rho(\Delta l,\Delta b)$ of the same event sample 
are given in Fig.~\ref{fig:AGASA-2dc-19p6}. Note that the
significance of these 2D self-correlations is respectively 
4.2$\sigma$ and 4.6$\sigma$. The energy-ordered plot also
shows the direction of deflection as a function of the energy.
The significance peaks at $\Delta l\approx \Delta b \approx -1^\circ$.

%%%%%%%%%%%%%%%%%%%%%%%%%%%%%%%
\begin{figure}[htb!]
\centerline{\epsfig{figure=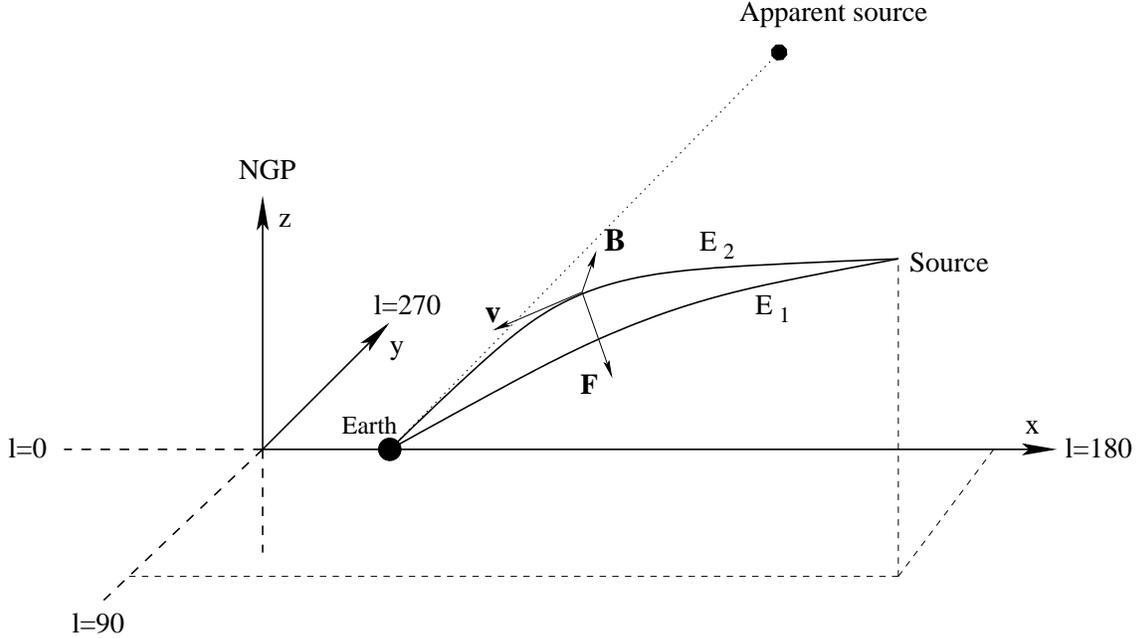,width=15cm}}
\caption{
Sketch of two proton trajectories and the corresponding 
magnetic field direction. For the depicted configuration of $E_1 > E_2$
one gets $b_1<b_2$ independent of the latitude of the source.
\label{fig:sketch}
}
\end{figure}
%%%%%%%%%%%%%%%%%%%%%%%%%%%%%%%
Assuming that this correlation is due to point sources, i.e. several 
of the cosmic rays originate from the same source, one can interpret the
energy-ordered plot as follows. The highest energy cosmic rays of the sample
propagate with the least deflection. For negative $\Delta b$ 
this means the lower energy cosmic rays from the same source 
appear to come from a higher Galactic latitude (see
Eq.~(\ref{eq:corr-ordered})). This is only possible if they get
deflected towards the Galactic south relative to the path of the higher
energy cosmic rays. Depending on the sign of the charge
this deflection determines the effective direction of the GMF transverse to 
the cosmic ray trajectory. However, only the effective 
direction integrated along the trajectory can be derived.
For example, the GMF has to have a component 
parallel to the $y$-axis for a positively charged particle from the
source direction $l\approx 180^\circ$, as shown in
Fig.~\ref{fig:sketch}.
The absolute value of the displacement of the correlation 
maximum from the center ($\Delta l= \Delta b = 0$) 
is given by the strength of the magnetic field and the
energy difference between the cosmic rays comprising pairs with small
$\Delta l$ and $\Delta b$. Of course, if the number of cosmic rays with
positive and negative charge are approximately equal, one expects a
symmetric significance pattern even for the energy-ordered correlation
plot.

We will discuss the AGASA data, including three of the observed 
multiplets, in more detail 
in Sec.~\ref{sec:discussion}.

%%%%%%%%%%%%%%%%%%%%%%%%%%%%%%%%%%%%%%%%%%%%%%%%%%%%%%%%%%%%%%%%%%%%%%%
\clearpage

\section{Uniformly distributed sources\label{sec:uniform}}

The first scenario we explore is that of uniformly distributed sources.
We inject protons of energy above  $10^{19}$ eV from sources distributed
uniformly on a sphere around the Galactic center and of radius
$r_{\rm src}$. In the simulation each cosmic ray is injected from 
a new source position. 
Astrophysically this would, depending on the value of $r_{\rm src}$,
correspond to either sources 
%gravitationally trapped 
in the Galactic halo, or a uniform and homogeneous distribution
of extragalactic UHECR sources in the absence of large-scale 
extragalactic magnetic fields. Here it is assumed that each source
injects a very small flux, contributing
not more than one particle to the cosmic ray spectrum
seen at Earth.

Even in the absence of the Galactic magnetic field one does not expect 
a completely isotropic flux at Earth. For isotropically emitting sources
with locations symmetric about the Galactic center one does find an
enhanced flux from the directions $l=90^\circ$ and $270^\circ$.
The enhancement is caused by the offset of the position of the
Earth from the center of the sphere where the sources are distributed
(i.e. from the Galactic center). The enhancement occurs because
for any longitude but $l=0^{\circ}$ and $l=180^{\circ}$, the $1/r^2$
behavior of the flux does not fully compensate the increase in the area
(as seen from Earth) where the sources are located (see, for example, the
discussion in \cite{Lipari:2000wu}). The enhancement factor
is given by
\begin{equation}
F = \left[1 - \left({R_\oplus\over r_{\rm src}}\right)^2~
[1-\cos^2(l-180^\circ)]  \right]^{-1/2}
\label{eq:PaoloL}
\end{equation}   
Here $R_\oplus=8.5$ kpc is the Galactocentric distance of the position
of the Solar System. For Galactocentric source distances of
$r_{\rm src}=15, 20$ and 40 kpc the maximum enhancement 
factor is $\sim 1.214 ,1.105$ and 1.023 respectively. 

\begin{figure}[htb!]
\centerline{
\epsfig{figure=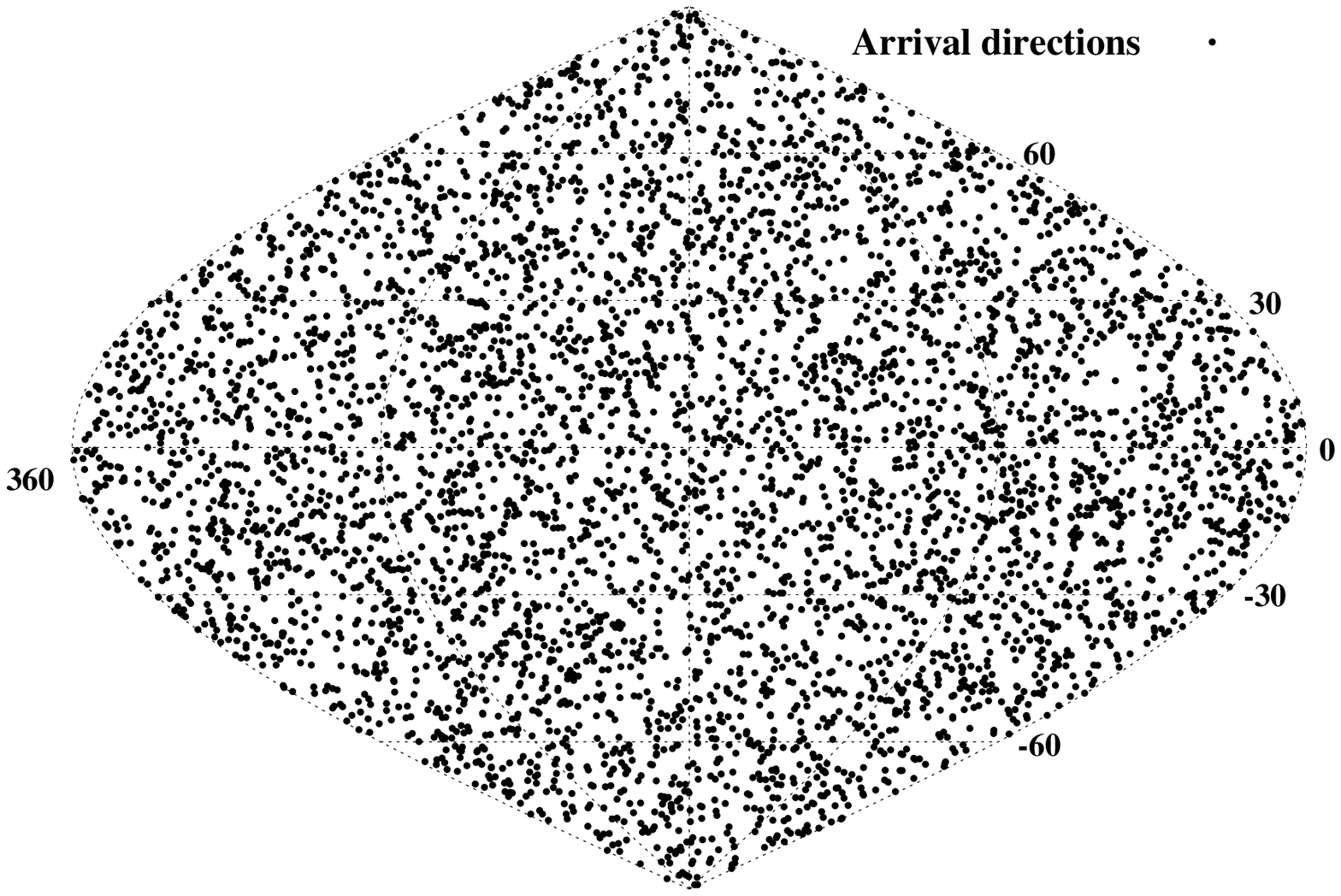,width=8cm}
\epsfig{figure=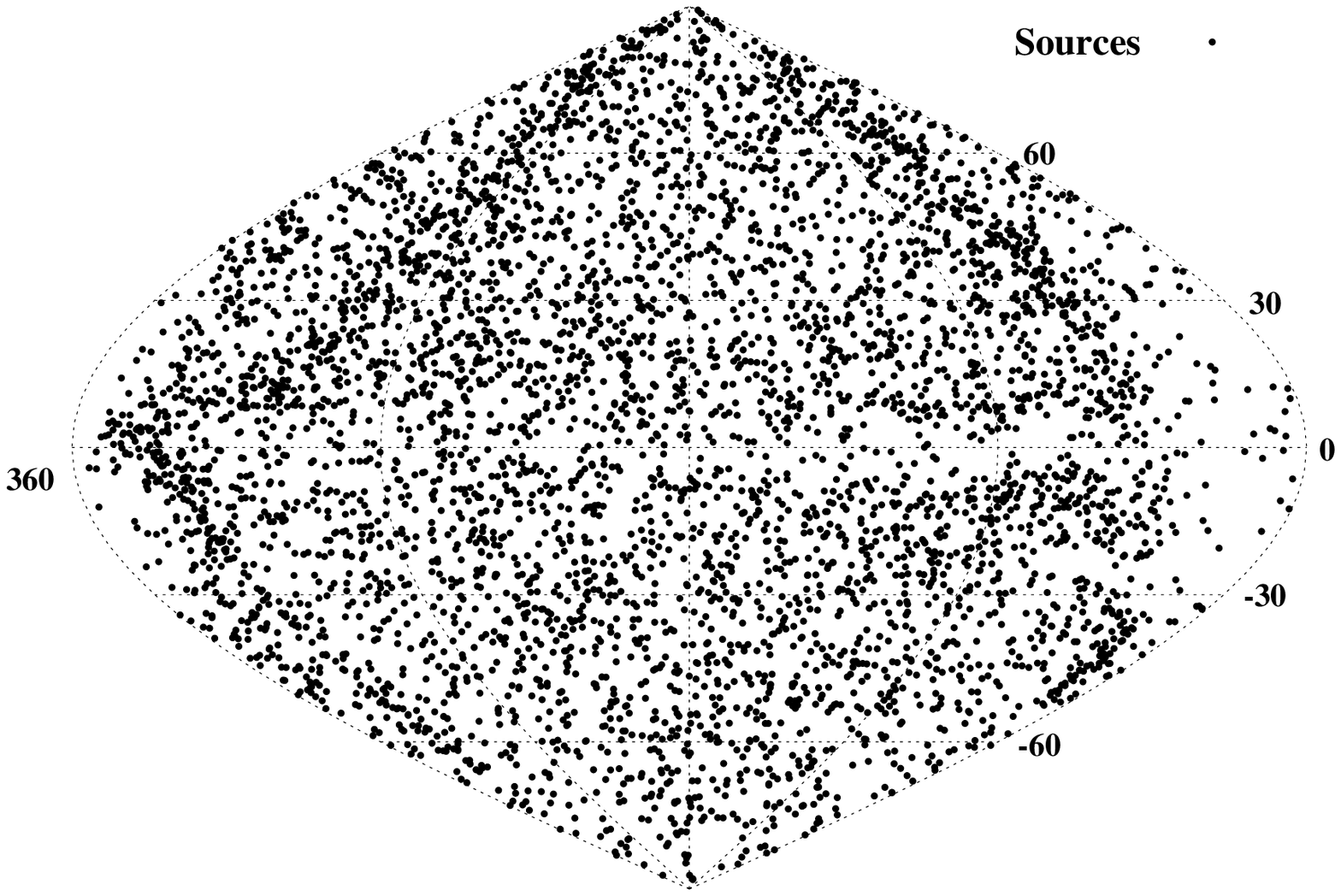,width=8cm}
}
\caption{
Left panel: equal area sky map of the
arrival directions (in Galactic coordinates) 
of the cosmic rays with $E>10^{19.4}$ eV.
The sources from which protons are injected
are uniformly distributed in a sphere centered around the
Galactic center and radius $r_{\rm src}=40$ kpc.
Right panel: sky map of 
of the position of the sources actually contributing to cosmic rays
that arrive at Earth.
\label{fig:uniform-sky}
}
\end{figure}

Neglecting for the moment this small flux enhancement, one expects an
isotropic flux at Earth even in the presence of the Galactic magnetic
field.  This is shown in the left--hand panel of Fig.~\ref{fig:uniform-sky}
and it is an effect of Liouville's theorem. 
The magnetic field does not change the magnitude of the cosmic ray momentum.
Hence along each particle trajectory
the phase space is preserved and the flux constant. If
there exists a trajectory for any direction at Earth which can be
followed back to the outer sphere of sources, the flux at Earth has to
be the same as it is just outside the Galactic magnetosphere. If the
flux is isotropic outside the GMF then it is also isotropic at Earth.

%%%%%%%%%%%%%%%%%%%%%%%%%%%%%%%%
%%%%%%%%%%%%%%%%%%%%%%%%%%%%%%%%
Nonetheless it is interesting to see that the mapping of the sources to
the arrival directions at Earth is realized
in a non-trivial way. The sources that actually give rise to the 
flux at Earth are
clustered in certain directions as shown in the right--hand panel of 
Fig.~\ref{fig:uniform-sky}, despite the fact source positions 
were drawn from an
uniform distribution.
The mapping reveals that sources in the Galactic plane are less
efficient in producing CRs arriving at Earth. The reason for this can
be understood with the aid of the right panel of Fig.~\ref{fig:tracks-1},
which shows
the projection onto the $xz$ plane of three CR trajectories arriving at
the detector from the same source in the Galactic plane. CRs have to
travel along different magnetic arms in the Galactic disk in order
to arrive at the detector. Each time an arm is crossed the field
reverses sign deflecting the CR in opposite directions.
Earth can only be reached by tracks that cross the Galactic plane at
roughly the same location in between the source and the detector.
This reduces the number of potentially detectable tracks compared
to other source locations.

Fig.~\ref{fig:uniform-sky} also shows a deficit of sources in the
direction of the Galactic center. The explanation of this effect
is the large strength of the dipole field
near the center of the Galaxy which strongly deflects
the particles. 
% Sources located behind the Galactic center
% would still have small efficiency, independently from the shape
% of the GMF, if they are indeed that strong around the Galactic
% center. 

These are two important features of the
GMF model described above which will be reflected in all the scenarios
we explore in this paper. In a model with different Galactic halo field,
or different ratio between the strengths of the BSS and dipole fields,
these effects would change.

%%%%%%%%%%%%%%%%%%%%%%%%%%%%%%%%
\begin{figure}[htb!]
\centerline{
\epsfig{figure=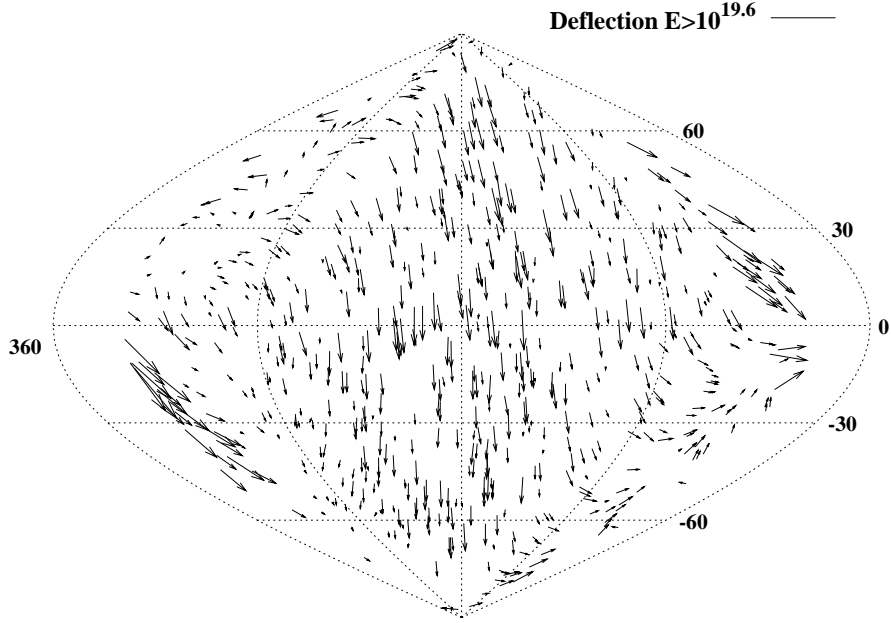,width=12cm}
}
\caption{Sky map of the deflections of the cosmic rays
that arrive at the detector. The arrows
join the position of the source and the
arrival direction of the
cosmic rays coming from it. Only cosmic rays with 
$E>10^{19.6}$ eV are shown. 
The sources are located at a sphere of 20 kpc radius.
\label{fig:uniform-sky2}
}
\end{figure}
%%%%%%%%%%%%%%%%%%%%%%%%%%%%%%%%
Fig.~\ref{fig:uniform-sky2} shows another interesting feature of the 
GMF which is a tendency to deflect the CRs in the direction from the
northern Galactic hemisphere towards the Galactic South.
This effect is caused by deflection in the BSS field, which is 
considerably stronger than the large-scale dipole field of the halo.
The North to South deflection is caused by the direction
of the local BSS field. This deflection is
similar, although of much smaller
magnitude, to the deflection in the Parker type halo field 
considered by Biermann
{\it et al.} \cite{Biermann:2000fd}.

%%%%%%%%%%%%%%%%%%%%%%%%%%%%%%%%
\begin{figure}[htb!]
\centerline{
\epsfig{figure=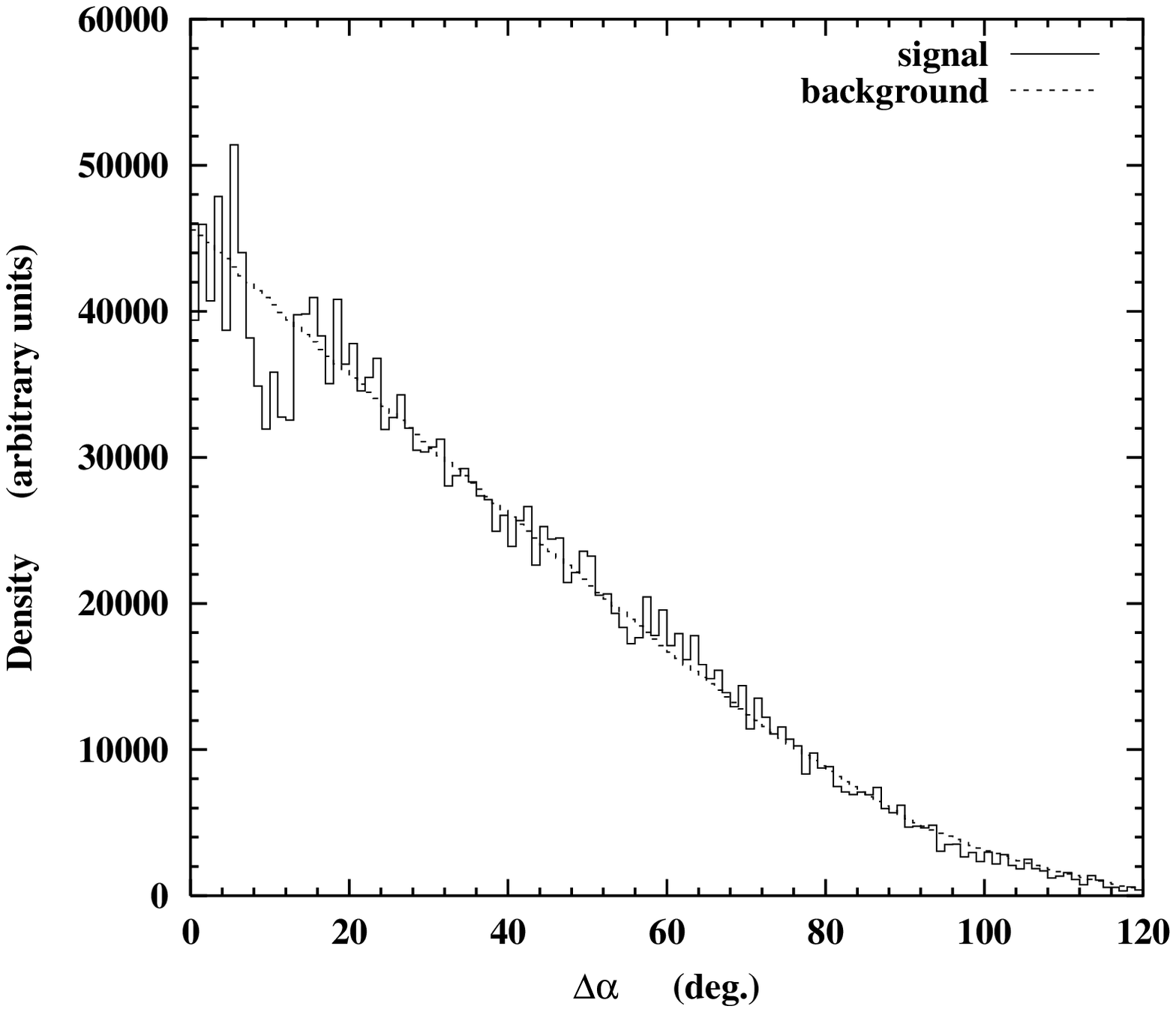,width=7.9cm}
\epsfig{figure=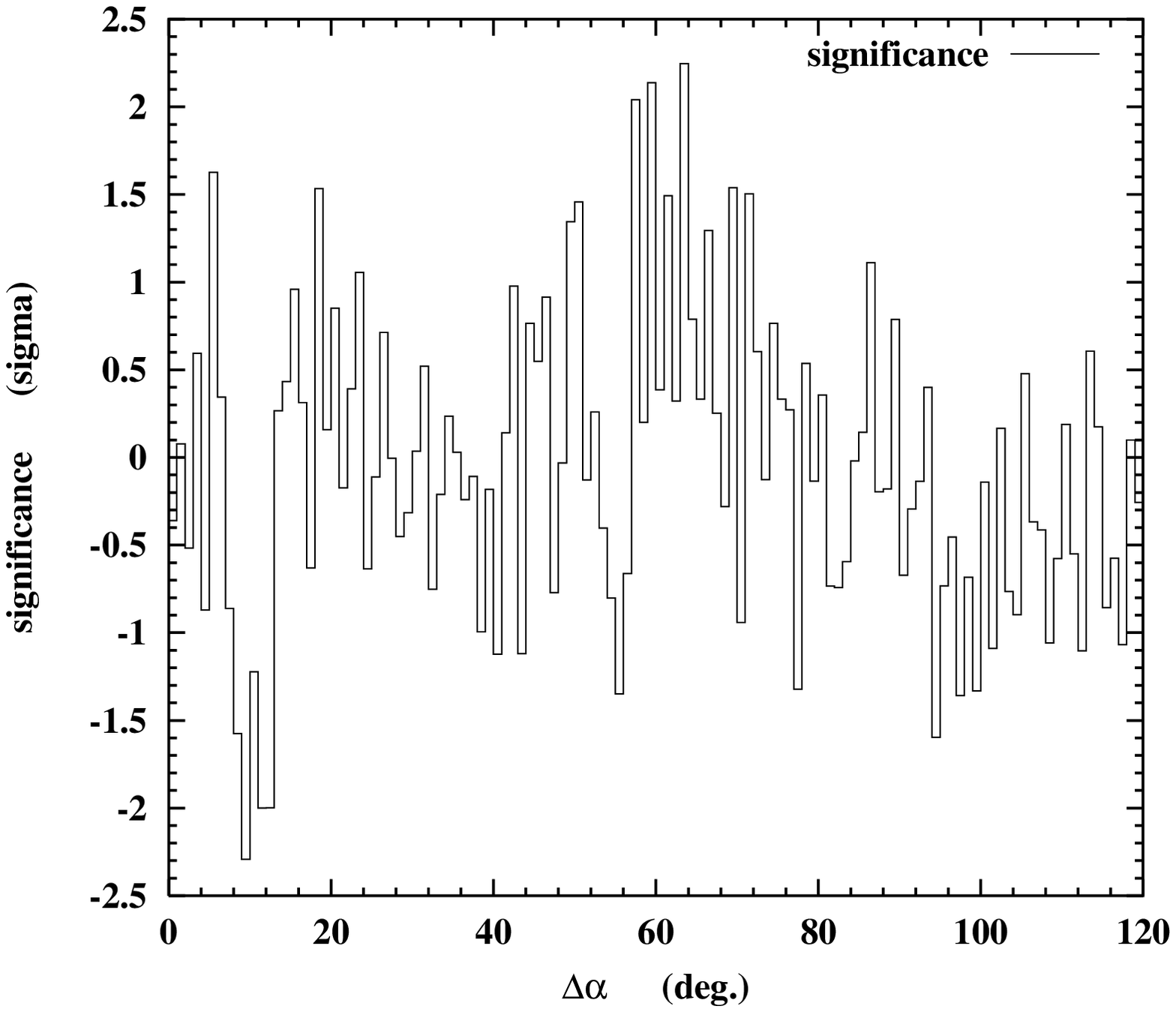,width=7.9cm}
}
\caption{Self correlation of arrival directions
of MC generated events (uniform
distribution of sources $r_{\rm src}=40$ kpc) with $E > 10^{19.6}$ eV.
Left panel: correlation function normalized by the phase space.
The self correlation for an isotropic background is also shown.
Right panel: significance of the correlation (i.e.
[signal-background]/width).
\label{fig:uniform-1dc}
}
\end{figure}
%%%%%%%%%%%%%%%%%%%%%%%%%%%%%%%%
As is expected from Fig.~\ref{fig:uniform-sky}, the 
one-dimensional correlation function shown in
Fig.~\ref{fig:uniform-1dc} is fully consistent with the
expectations
from an isotropic flux. Assuming the self-correlation
observed by AGASA is not a statistical fluctuation, a uniform and
isotropic distribution of low-flux sources in the
Galactic halo is not supported by current observations.

% It can be seen in Fig.~\ref{fig:uniform-1dc} that the expected 
% 1D angular correlation of events arriving at the detector does not  
% match the correlation seen in AGASA data. Clearly
% a uniform and isotropic distribution of sources in the 
% Galactic halo is not supported by current observations.
% Fig.~\ref{fig:uniform-1dc} shows the expected correlation 
% only for sources at $r_{\rm src}=20$ kpc, but
% the same conclusion holds for $r_{\rm src}=15$ and 40 kpc.
% Even an isotropic distribution of sources at different distances
% wouldn't match the AGASA data because the GMF strength is expected
% to be low at distances larger than 20 kpc. Under this assumption,
% distributing the sources on the surface of the 20 kpc sphere 
% would be equivalent to having a distribution of sources emitting
% isotropically at different distances from the center of the
% Galaxy, with the exception of the geometric enhancement of 
% Eq.~\ref{PaoloL}. The 2D correlations (not shown) support this view.
% Its significance is very small $\sim 2.5 \sigma$.  

\clearpage

%%%%%%%%%%%%%%%%%%%%%%%%%%%%%%%%%%%%%%%%%%%%%%%%%%%%%%%%%%%%%%%%%%%%%%%%%

\section{Supergalactic plane\label{sec:sgp}}

It is a long-standing question whether there is a correlation of the
highest energy cosmic rays with the local matter enhancement, the so
called supergalactic plane (SGP).
An indication for a large scale correlation was first reported in
\cite{Stanev:1995my}. The significance of such a correlation
decreased with the increasing world statistics \cite{Stanev:1999icrc} 
and was replaced by
evidence for small scale clustering.
A recent analysis of the world data on UHECRs concludes
that the chance probability of the observed clustering within
$\pm 10^\circ$ off the SGP is less than $1 \%$ \cite{Uchihori:1999gu}.
Furthermore the triplet and 
one of the two doublets in the AGASA data passing the outer Galaxy cut
are located at less
than $1^\circ$ off the SGP. 
 
%%%%%%%%%%%%%%%%%%%%%%%%%%%%%%%%
\begin{figure}[htb!]
\centerline{
\epsfig{figure=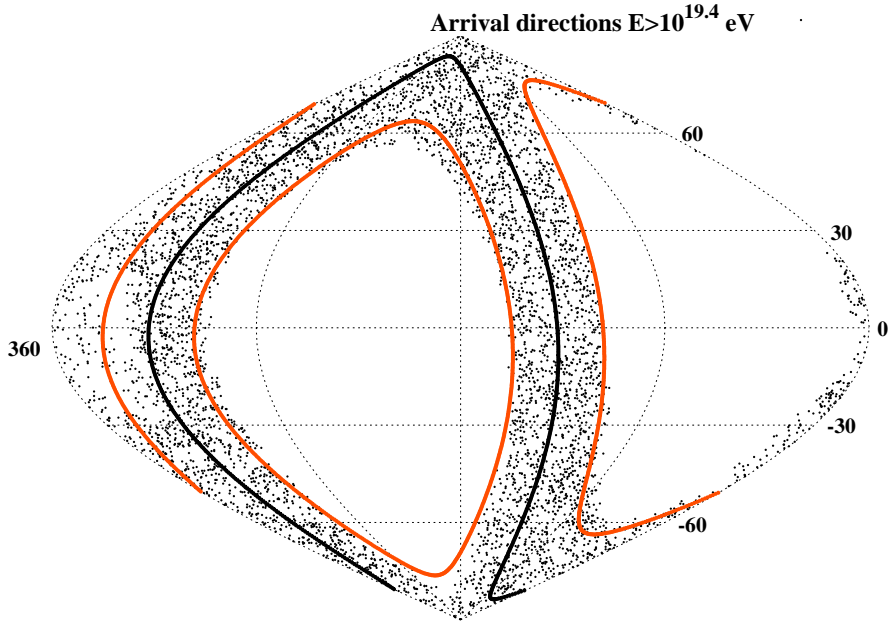,width=12cm}
}
\caption{Equal area sky map (in Galactic coordinates) of the
arrival directions of the cosmic rays with energy above 
$10^{19.4}$ eV for
$r_{\rm src}=20$ kpc. The sources are
concentrated in the supergalactic plane (SGP). The solid line
in the center marks the position of the SGP, whereas the
left and right line around it bound the location where the sources
are uniformly distributed ($\pm 20^\circ$ in supergalactic latitude).
\label{fig:sgp-sky}
}
\end{figure}
%%%%%%%%%%%%%%%%%%%%%%%%%%%%%%%%
To simulate sources in the SGP we sample source locations
uniformly distributed in the direction of the
SGP with a maximum offset of $20^\circ$ in supergalactic latitude.
Fig.~\ref{fig:sgp-sky} shows the corresponding arrival direction of 
protons emitted from sources at a Galactocentric distance of 20 kpc and
$E>10^{19.4}$ eV. 
Again we consider here the low-flux limit:
each source emits only one particle.
The total source efficiency is a convolution of the
area within 20$^\circ$ off the supergalactic plane and the
geometric effect of the right--hand panel of Fig.~\ref{fig:uniform-sky}.

%%%%%%%%%%%%%%%%%%%%%%%%%%%%%%%%
\begin{figure}[htb!]
\centerline{
\epsfig{figure=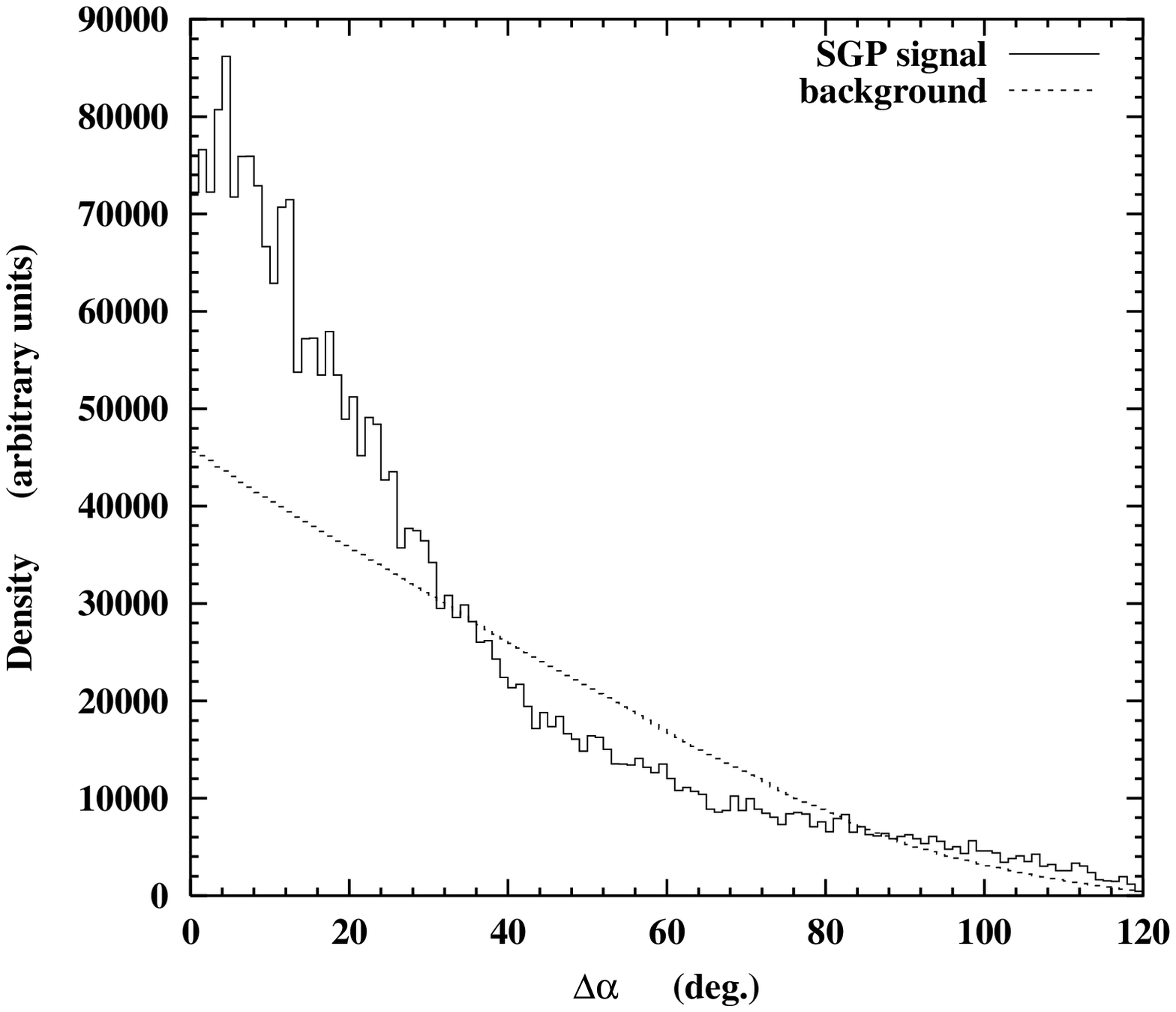,width=7.9cm}
\epsfig{figure=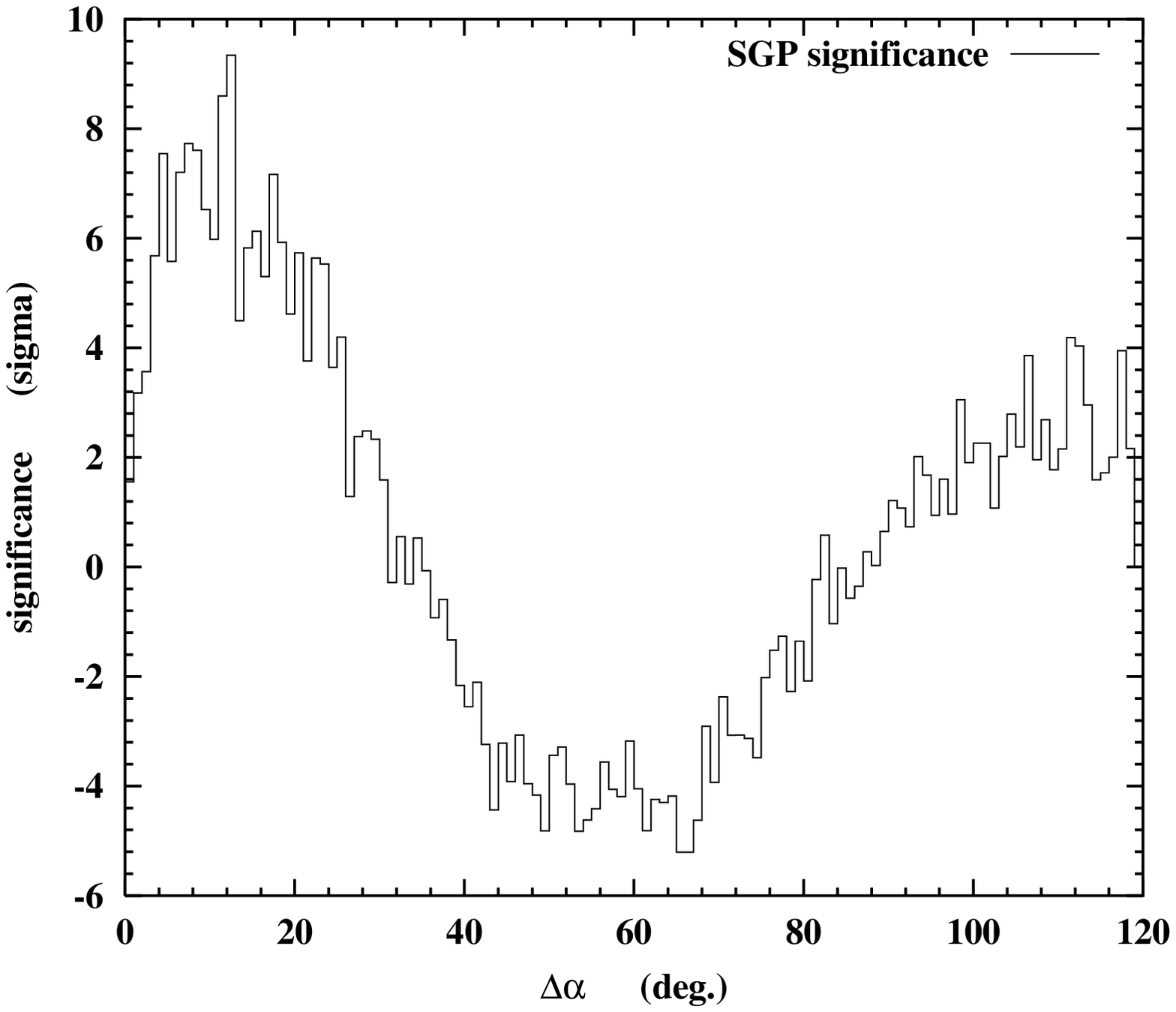,width=7.9cm}
}
\caption{
Self-correlation of arrival directions
of MC generated events with $E > 10^{19.6}$ eV. The
sources are distributed within $\pm 20^\circ$ off the SGP
at a Galactocentric distance $r_{\rm src}=20$ kpc. 
Left panel: phase-space normalized 
correlation function. The self correlation
for isotropic background events is also shown. Right panel: significance of
the correlation in units of $\sigma$.
\label{fig:sgp-1dc}
}
\end{figure}
%%%%%%%%%%%%%%%%%%%%%%%%%%%%%%%%
The mapping of the supergalactic plane to Galactic coordinates
leads to a strong correlation in Galactic longitude. However the
1D correlation in Fig.~\ref{fig:sgp-1dc} is too broad as
compared to data. The significance of the correlation is larger
than $4 \sigma$ in an angular region of width $\sim 30^\circ$
which approximately corresponds to the band around the SGP where
the sources are distributed. 
In addition, the features of the GMF already emphasized in the
uniform source scenario are also seen. Cosmic rays are preferentially
deflected to the south and certain source positions have a higher
efficiency in producing observable cosmic rays.
Both effects lead also to a very broad maximum in the two-dimensional
correlation with a significance of up to $8 \sigma$.

It is interesting to note that the difference in the real direction to
the source and the arrival direction is large in latitude but not in
longitude. Consequently the plane structure of the SGP in Galactic
coordinates is approximately preserved by the GMF and the flux observed
at Earth does not appear to be isotropic.

We conclude that the SGP is disfavoured 
as the source of the events seen by AGASA, assuming  
a weak extragalactic magnetic field. The expected 1D correlation
could come closer to the observed correlation if the distribution 
of sources was restricted to a  narrower band, say $\pm 10^\circ$
off the SGP. However, the resulting arrival direction distribution would
then be highly anisotropic, following the direction of the
supergalactic plane.  It can be seen in Fig.~\ref{fig:sgp-sky} 
that the region around Galactic latitude $60^\circ$, where AGASA
has seen several events, is largely unpopulated in this scenario
even for sources within 20$^\circ$ off the SGP.

%%%%%%%%%%%%%%%%%%%%%%%%%%%%%%%%%%%%%%%%%%%%%%%%%%%%%%%%%%%%%%%%%%%%%%
\clearpage

\section{Single point source\label{sec:single-point}}

The apparently isotropic arrival distribution of UHECRs strongly
disfavours any scenario of a nearby, dominant single point source,
unless a strong
coherent large-scale magnetic field is assumed \cite{Biermann:2000fd}.

The purpose of considering here single point sources 
is to study the correlation
expected from different source locations and source distances.
The results are directly applicable to the interpretation of
multiplets seen in data, under the assumption that they originate from
point sources.

%%%%%%%%%%%%%%%%%%%%%%%%%%%%%%%%
\begin{figure}[htb!]
Point source at $l=112.5^\circ$, $b=0^\circ$:\hfill\\
\centerline{
\epsfig{figure=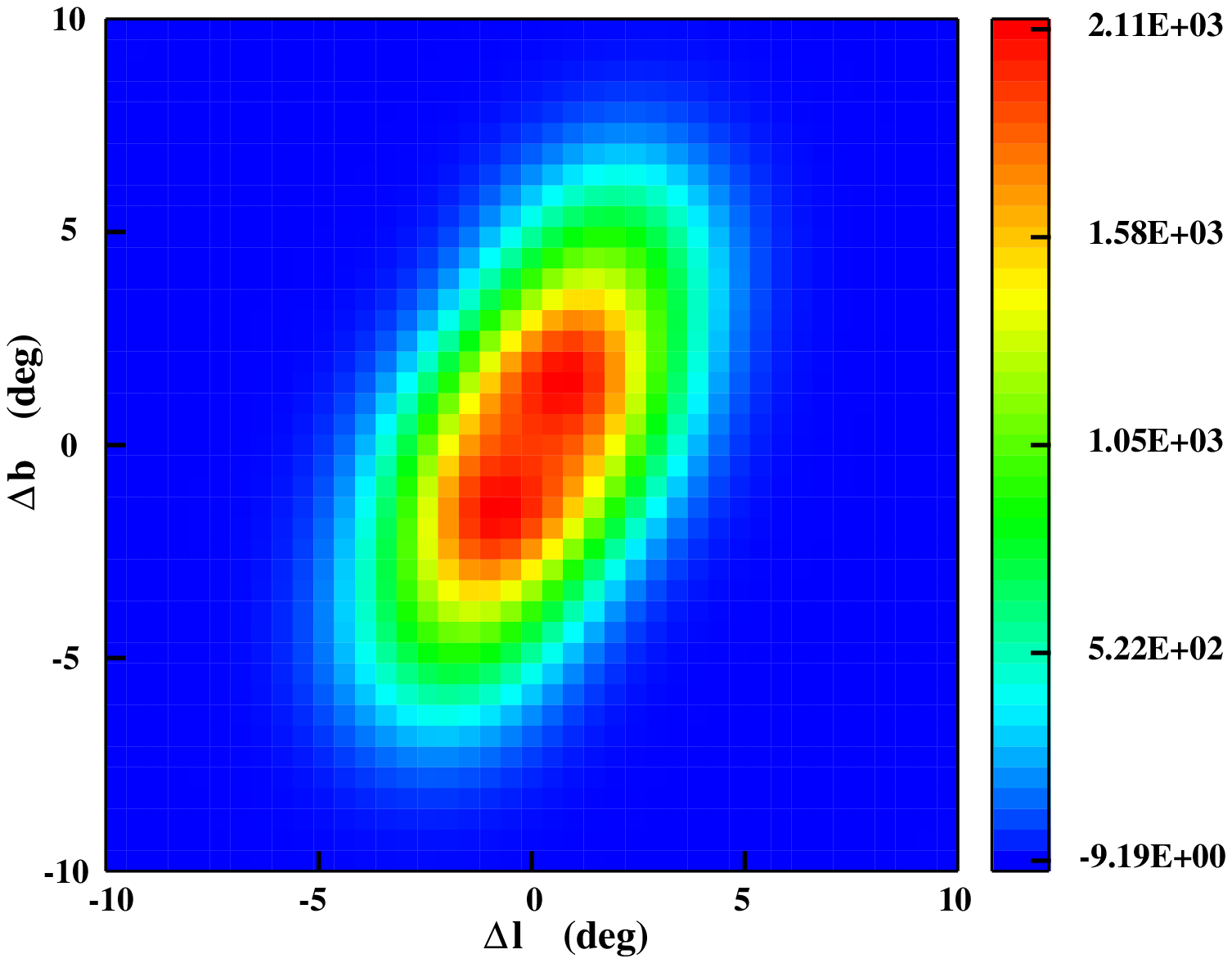,width=5.cm}
\epsfig{figure=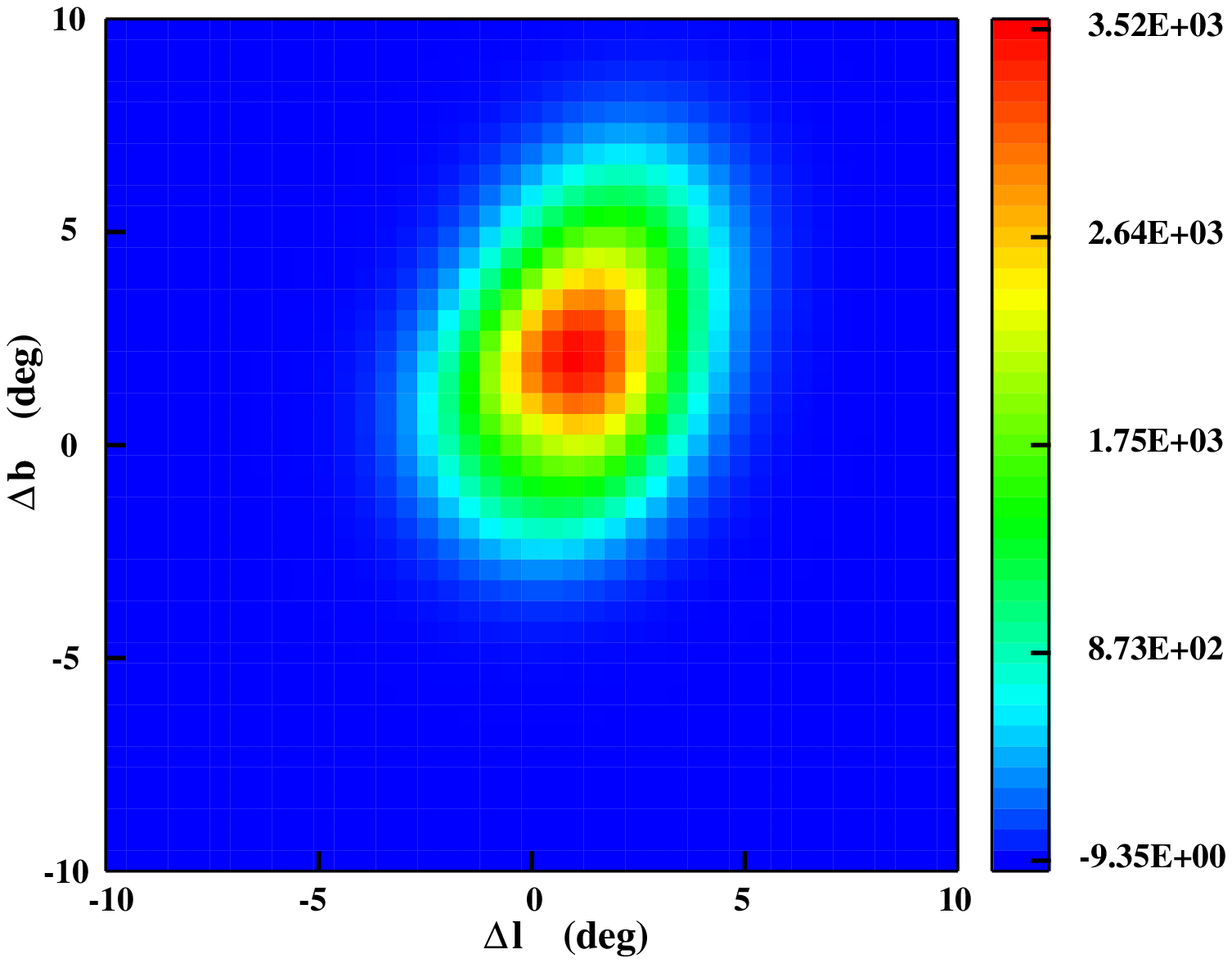,width=5.cm}
}
Point source at $l=112.5^\circ$, $b=45^\circ$:\hfill\\
\centerline{
\epsfig{figure=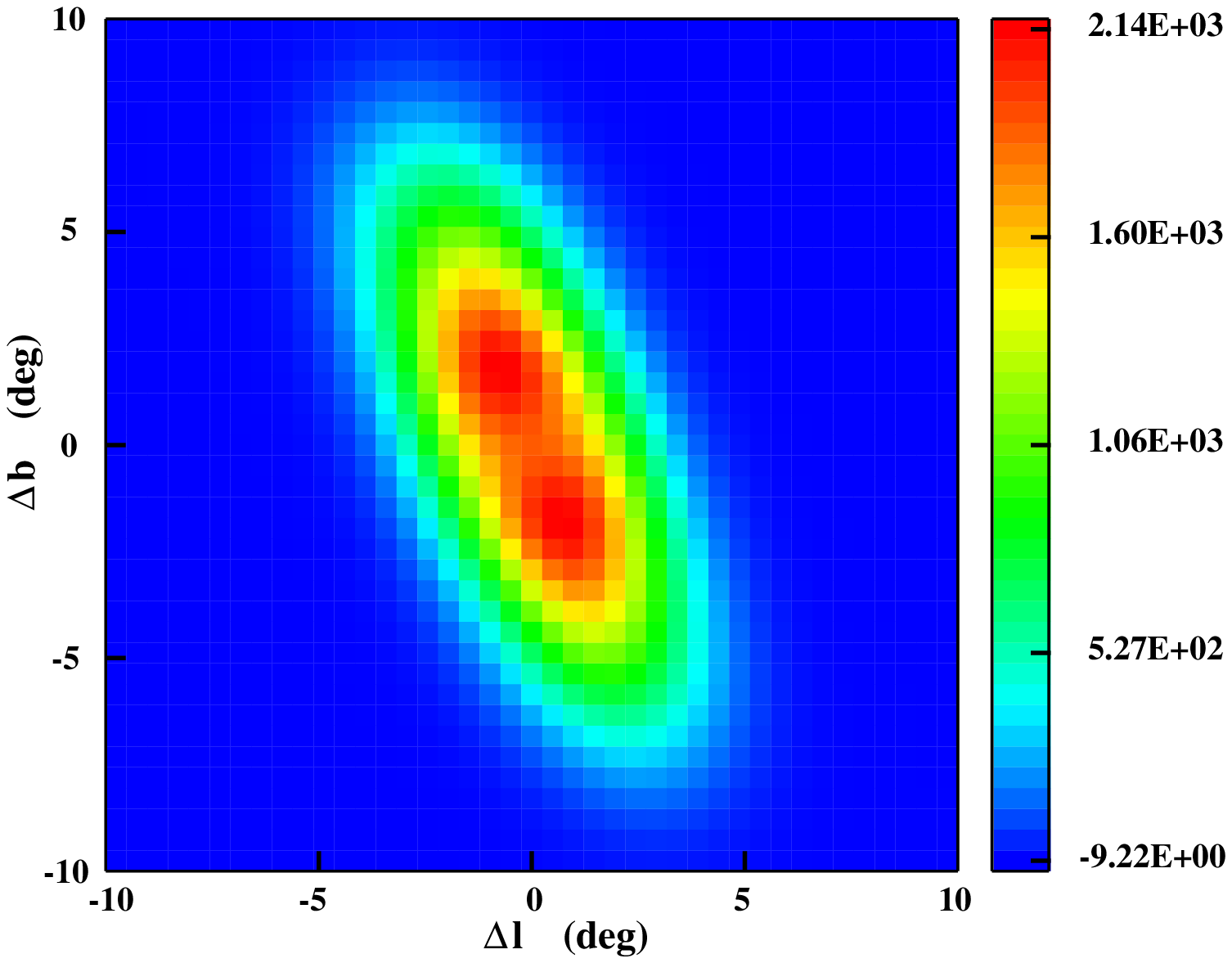,width=5.cm}
\epsfig{figure=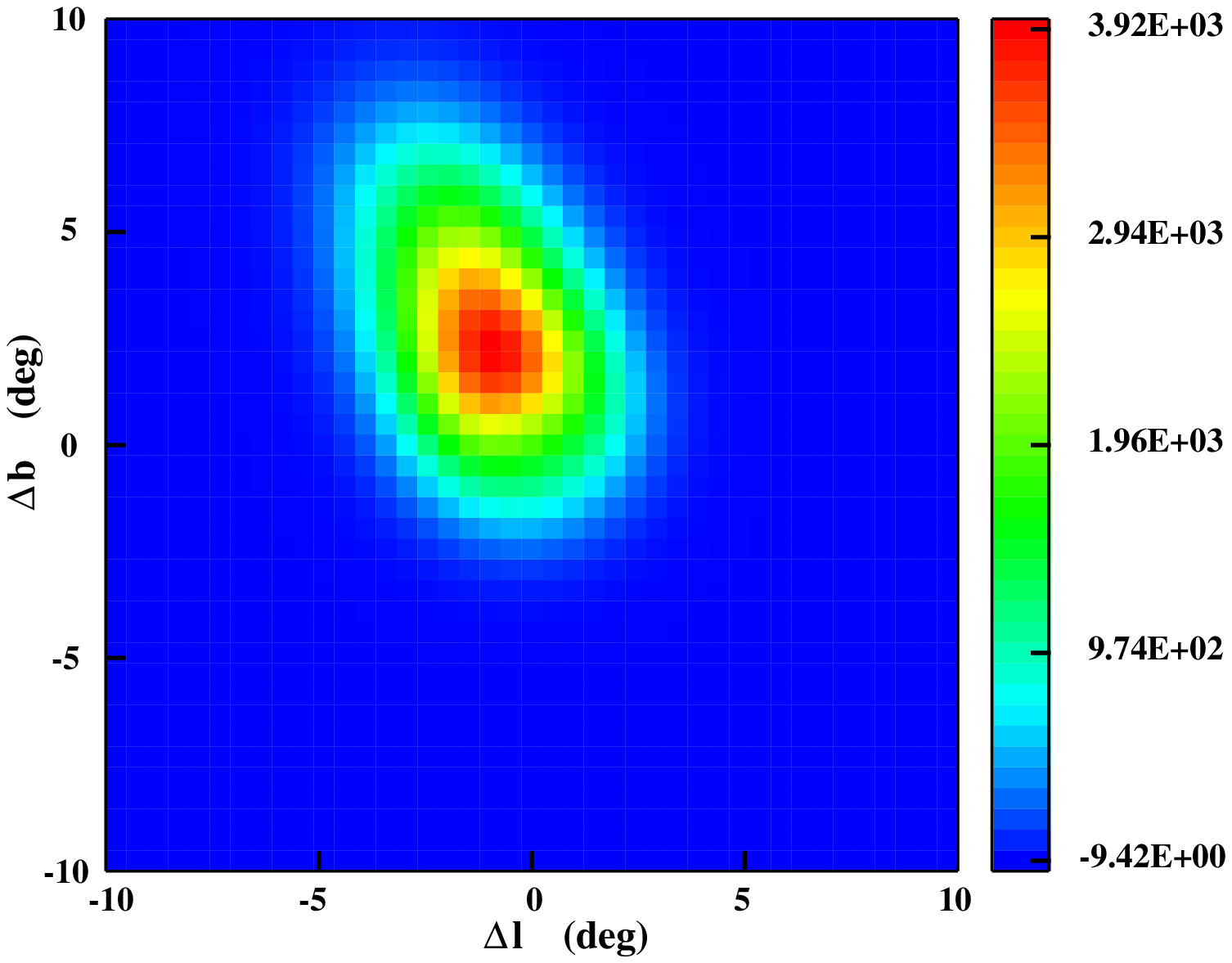,width=5.cm}
}
Point source at $l=112.5^\circ$, $b=-45^\circ$:\hfill\\
\centerline{
\epsfig{figure=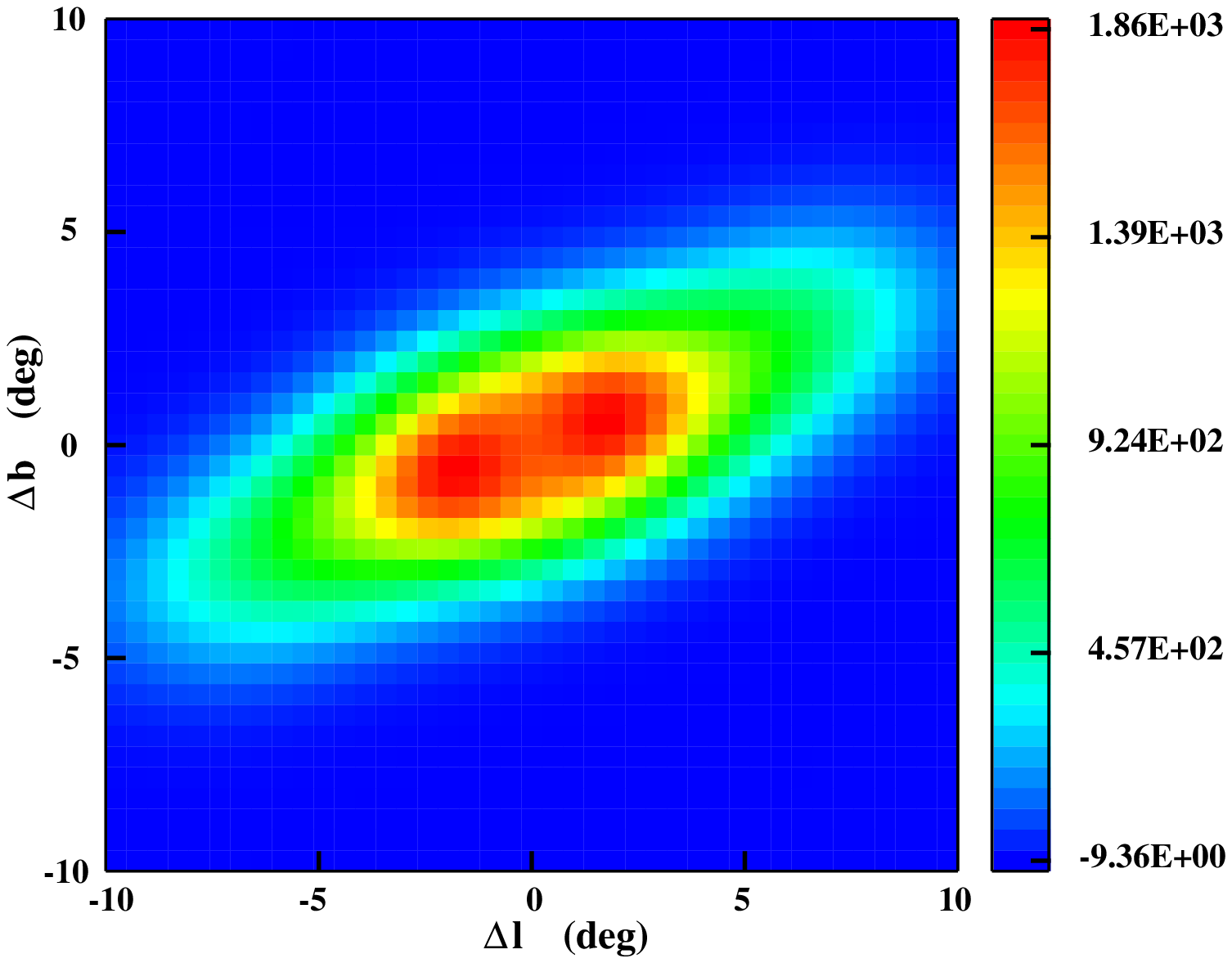,width=5.cm}
\epsfig{figure=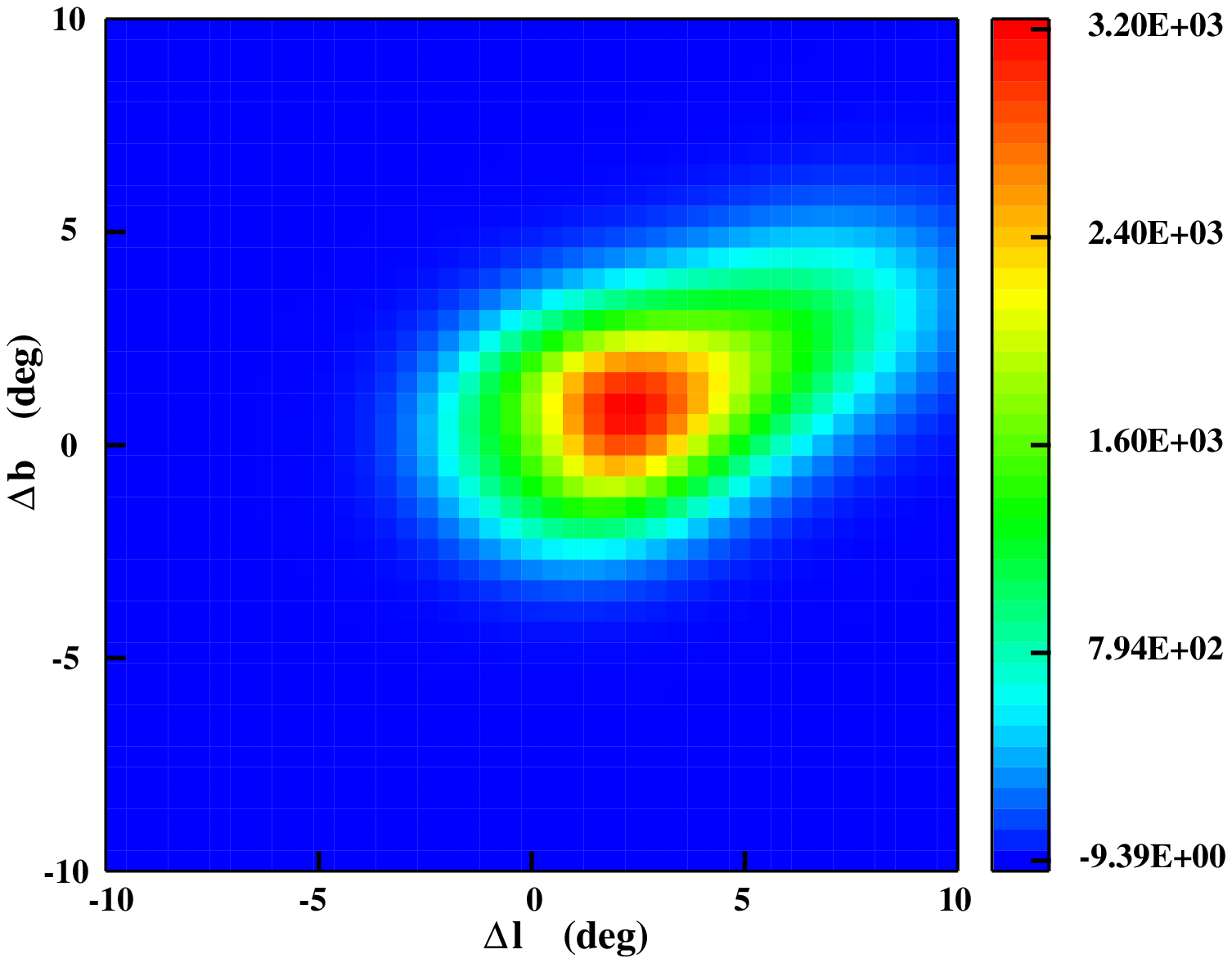,width=5.cm}
} 
Point source at $l=157.5^\circ$, $b=45^\circ$:\hfill\\
\centerline{
\epsfig{figure=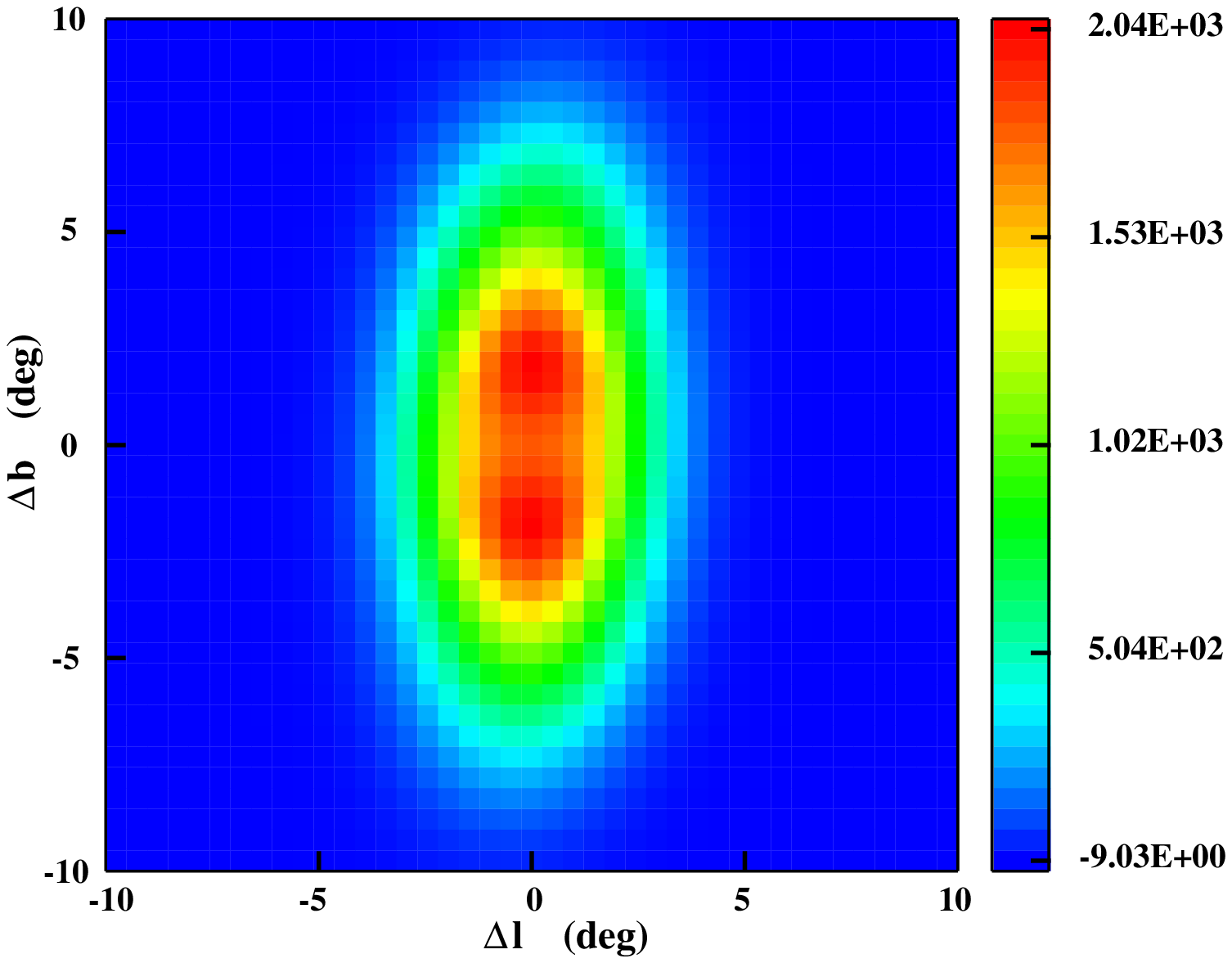,width=5.cm}
\epsfig{figure=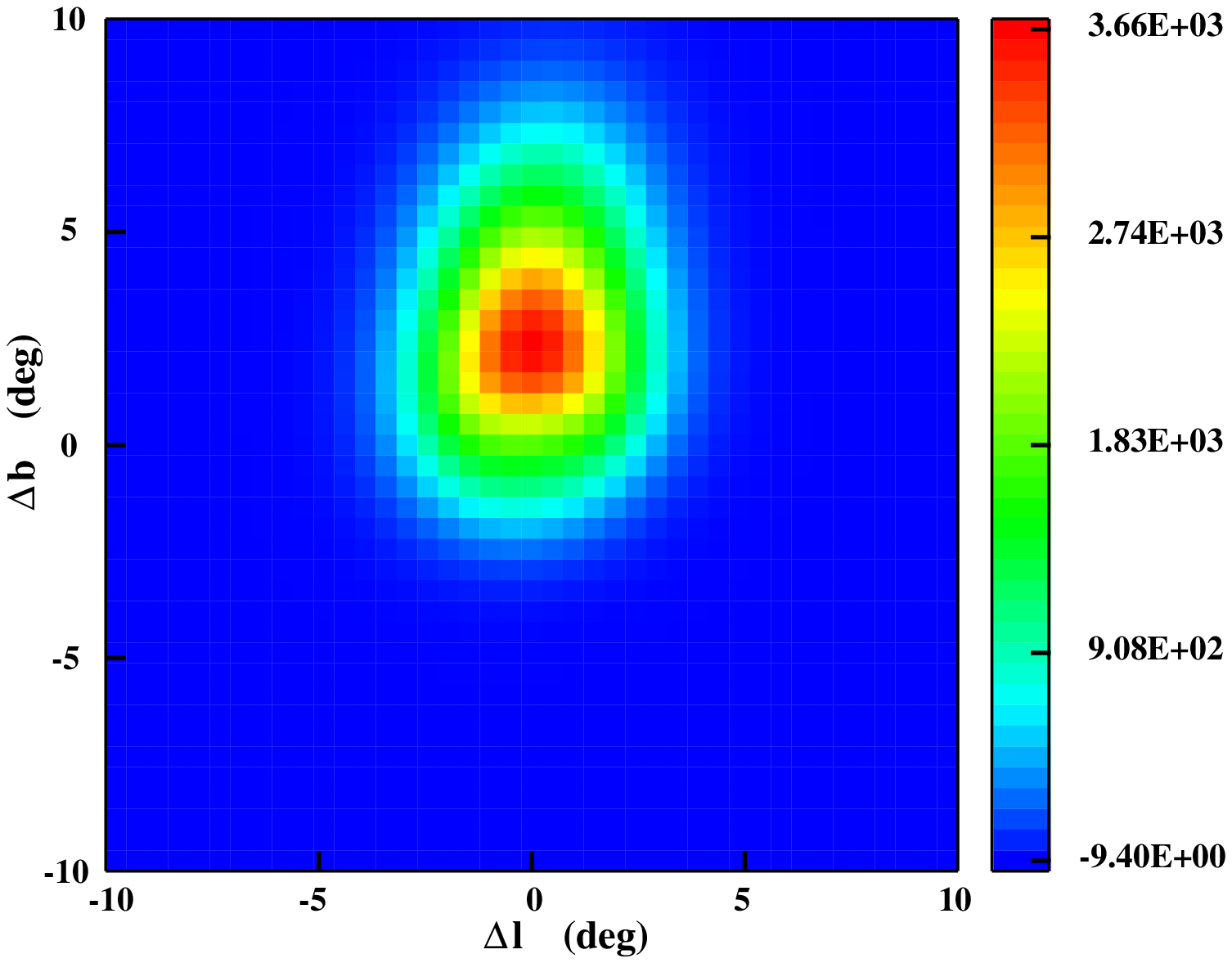,width=5.cm}
} 
\caption{
Symmetrized (left) and energy-ordered (right) significance of
self-correlation for $E>10^{19.6}$ eV.
The plots were made from 200 cosmic rays arriving at Earth from sources
with a Galactocentric distance of $r_{\rm src}=20$ kpc.
\label{fig:single-2dc}
}
\end{figure}
%%%%%%%%%%%%%%%%%%%%%%%%%%%%%%%%
In Fig.~\ref{fig:single-2dc} we show the significance plots
of the symmetrized and energy-ordered self-correlation
for single sources at four different directions. 
As discussed before, although the protons were 
only injected in a $30^\circ$ cone around the
line-of-sight to the Solar system the results apply to isotropically
emitting point sources.
Again the background is assumed to be an isotropic distribution, which
leads to a large significance of the self-correlation.
Only a number of representative examples of source locations in the AGASA
outer Galaxy window are shown. 

The shape of the 2D correlations is strongly dependent on the 
position of the source, consistent with the deflections shown in
Fig.~\ref{fig:uniform-sky2}. 
Due to the strong spiral component of the
magnetic field parallel to the Galactic plane, the 2D
correlation is more stretched in $\Delta b$ than
$\Delta l$. Cosmic rays coming from 
sources with longitudes close to $90^\circ$ get deflected toward larger
(smaller) longitudes in the northern (southern)
Galactic hemisphere. For sources in the vicinity of $l=180^\circ$ there
is almost no deflection in longitude.

The dipole field would always deflect positively charged cosmic rays to
larger Galactic longitudes. For the arrival direction this means that
the cosmic rays appear to come from smaller Galactic longitudes (see
Fig.~\ref{fig:sketch}).
However, the small field strength of the dipole component makes it
rather unimportant for the correlations considered here. Of course,
the situation is
different for cosmic ray trajectories coming close to the Galactic
center, which are excluded due to the ``outer Galaxy'' cut.
In particular for sources at high latitudes 
the non-trivial variation of the magnitude of the spiral field leads to a
drift-like deflection parallel to the Galactic plane.

The energy-ordered correlation plots show that, for positively charged
multiplets of cosmic rays, the particle with lower energy is 
deflected more to the Galactic north and hence appears to come from further
south (e.g. higher Galactic latitude in the southern hemisphere). 
The reason is the GMF component, parallel to the Galactic plane and
pointing from large to small longitudes (see Fig.~\ref{fig:bss}). 
Since there is good agreement between different measurements 
on the direction of
the GMF in the vicinity of the Solar system \cite{Beck:2001}, this
prediction can be used to derive the charge sign of UHECRs if point
sources are identified.

%%%%%%%%%%%%%%%%%%%%%%%%%%%%%%%%
\begin{figure}[htb!]
\centerline{
\epsfig{figure=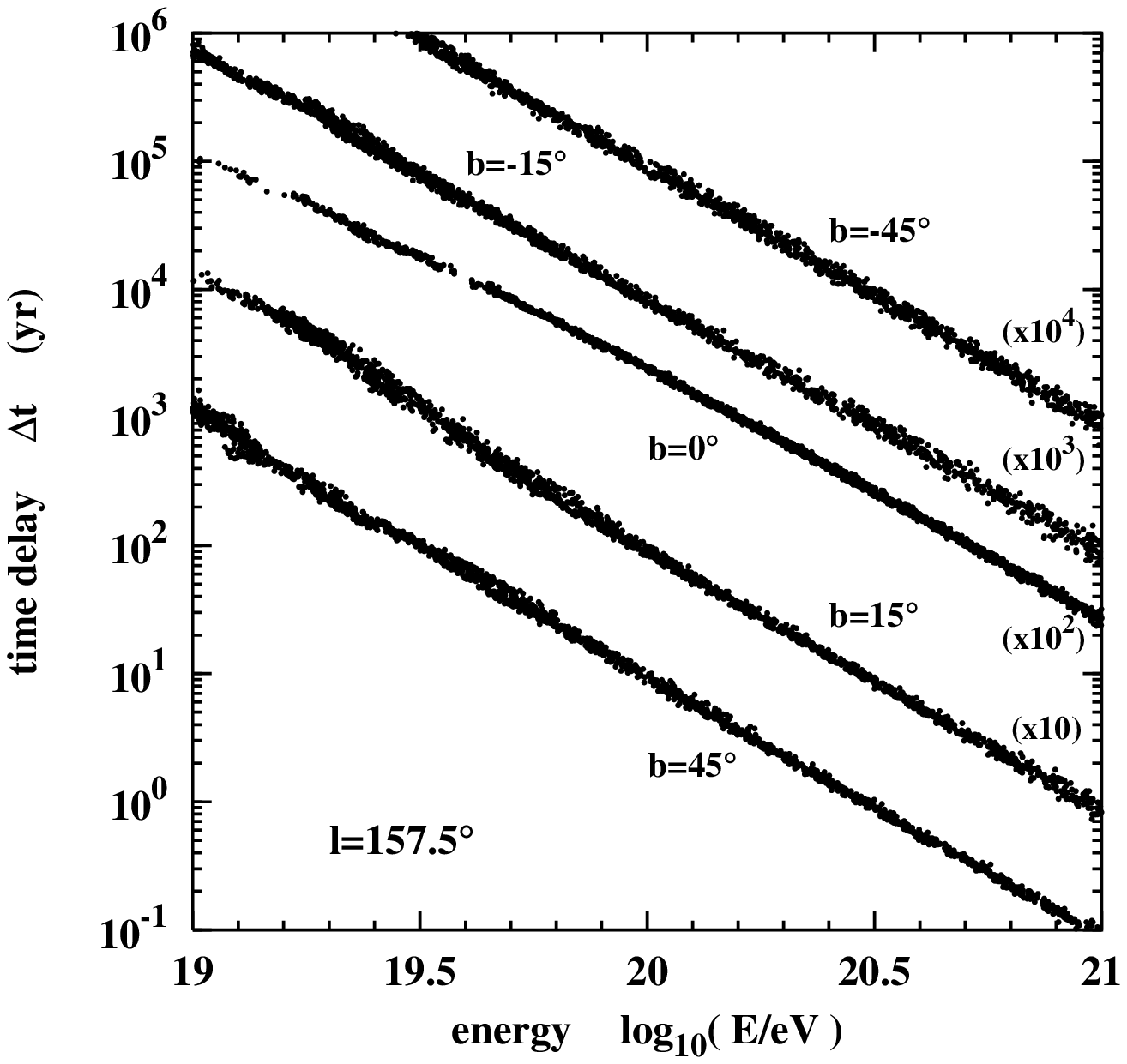,width=7.9cm}
\epsfig{figure=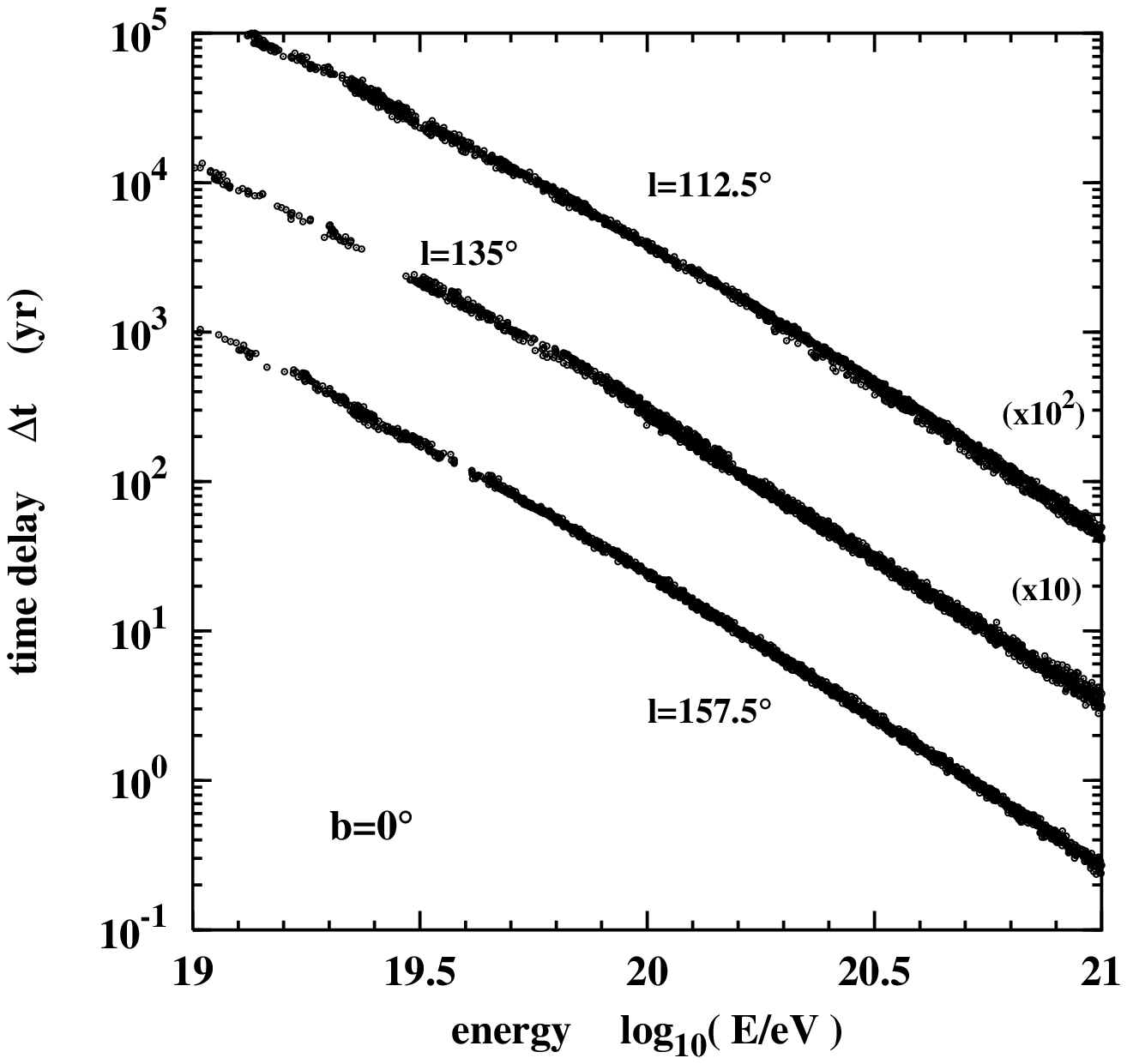,width=7.9cm}
}
\caption{
Arrival time delay of cosmic rays propagating from sources to Earth. 
The sources are again at a Galactocentric distance 
of $r_{\rm src} = 20$ kpc. Each simulated cosmic ray trajectory,
arriving at Earth from the ``outer Galaxy'' direction, is represented by
one point in this plot. 
Cosmic rays were injected with a $dN/dE \sim E^{-1.1}$ spectrum 
to obtain good statistics at high energy.
\label{fig:tdelay}
}
\end{figure}
%%%%%%%%%%%%%%%%%%%%%%%%%%%%%%%%
A number of UHECR models is based on 
point sources emitting cosmic rays only for a short period of time (e.g.
$Z$-bursts, decay of super-massive particles).
In Fig.~\ref{fig:tdelay} we show the arrival time delay of cosmic rays
coming from sources at a Galactocentric distance of 20 kpc. The arrival
time delay is defined as the time difference between the arrival of the
cosmic ray and a light signal emitted from the source at the same time.
For a given source, a strong correlation between 
energy and time delay is observed. Therefore one can estimate
the expected time difference of cosmic rays emitted simultaneously from
a point source simply by using
this energy-time delay relation. Of course, such a simple energy-time
delay relation is expected to hold only in the limit of small deflections. 
In this limit there exists essentially 
only one trajectory, up to small scale differences due to the local 
magnetic field fluctuations, which connects the source
with Earth for a given energy.

The width of the distributions shows how strong
the correlation is. The correlation will be wider if the 
fraction of the field strength coming from the turbulent component
is increased. The simulations confirm the expectation that high-energy
cosmic rays arrive earlier than low-energy ones. However 
exceptions are possible. The width of the correlation allows for the
inverted arrival time behaviour as long as the energies of the cosmic
rays are different by less than 30\%.

%%%%%%%%%%%%%%%%%%%%%%%%%%%%%%%%
\begin{figure}[htb!]
\centerline{
\epsfig{figure=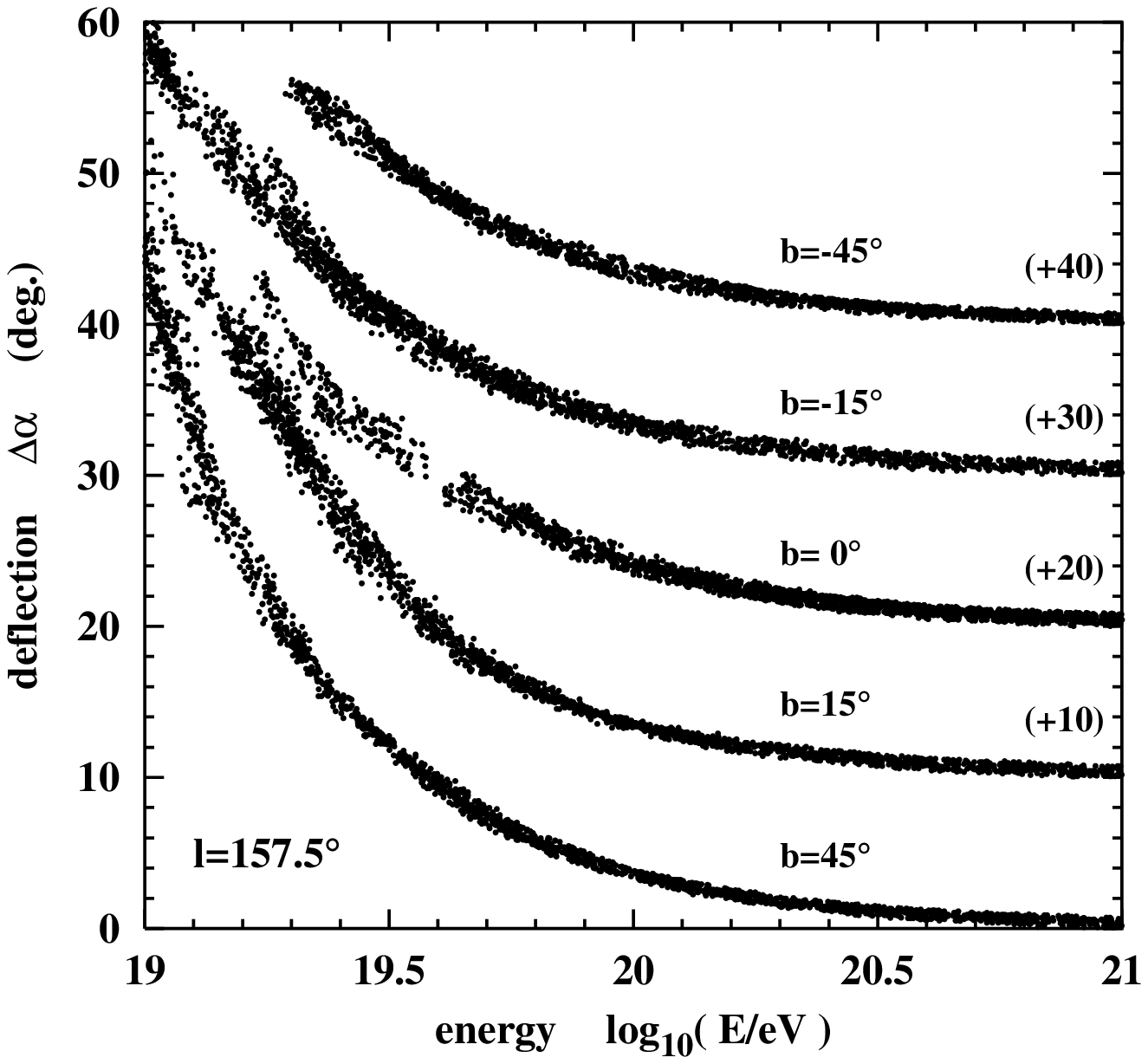,width=7.9cm}
\epsfig{figure=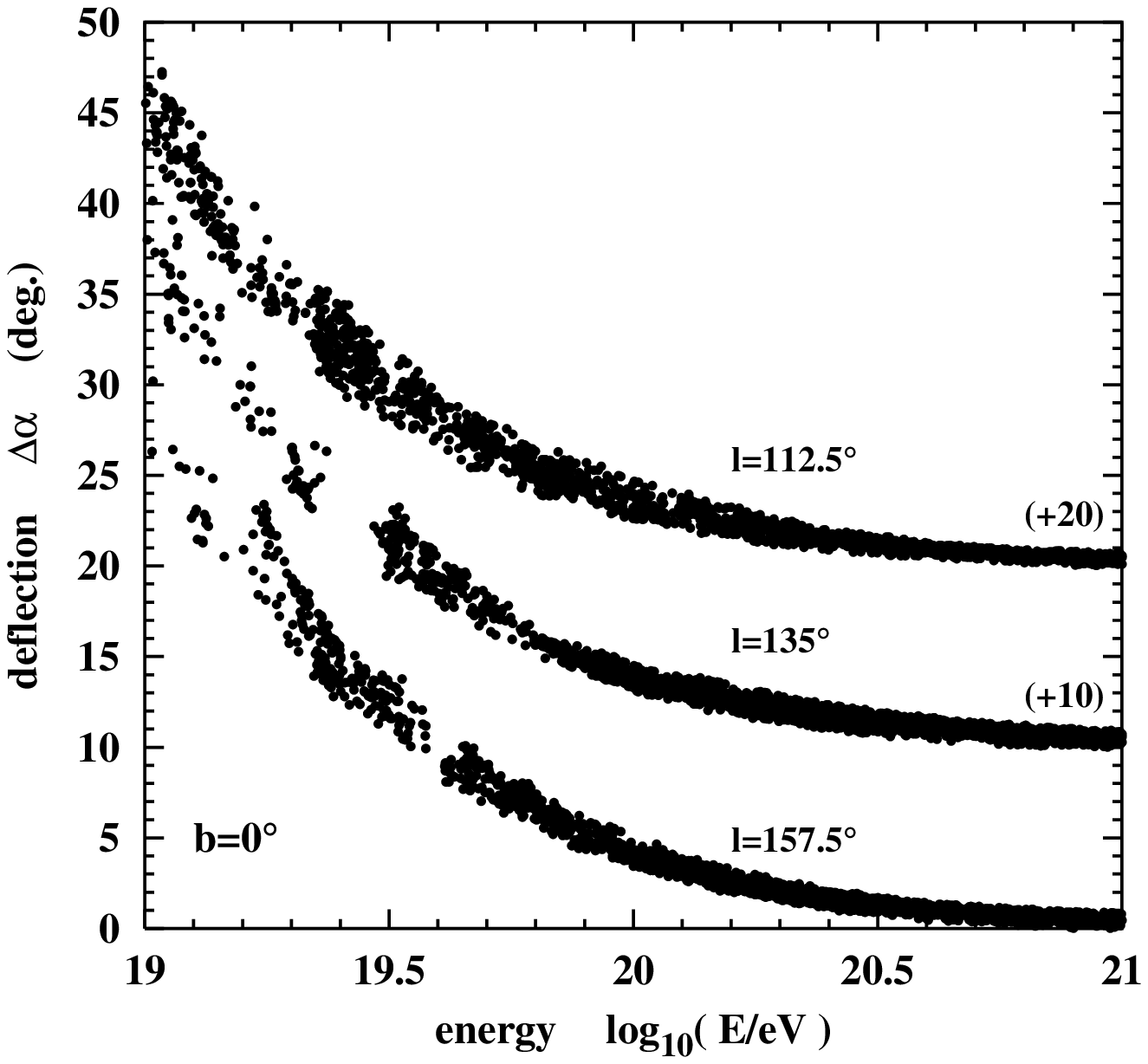,width=7.9cm}
}
\caption{
Angle between arrival and source direction $\Delta \alpha$ for
cosmic rays propagating from sources at a
distance of 20 kpc from the Galactic center.
\label{fig:alpha}
}
\end{figure}
%%%%%%%%%%%%%%%%%%%%%%%%%%%%%%%%            
Similarly, the angular difference between the arrival direction and the
line-of-sight to the source, $\Delta \alpha$,
is directly related to the energy. With the
2D correlation plots and the knowledge of this angle one can estimate
the source distance for a given multiplet and Galactic field configuration. 
Fig.~\ref{fig:alpha} shows the expected angle $\Delta \alpha$ for a
number of source directions. The correlations 
depend only weakly on the longitude for fixed-latitude sources.

%%%%%%%%%%%%%%%%%%%%%%%%%%%%%%%%
\begin{figure}[htb!]
\centerline{
\epsfig{figure=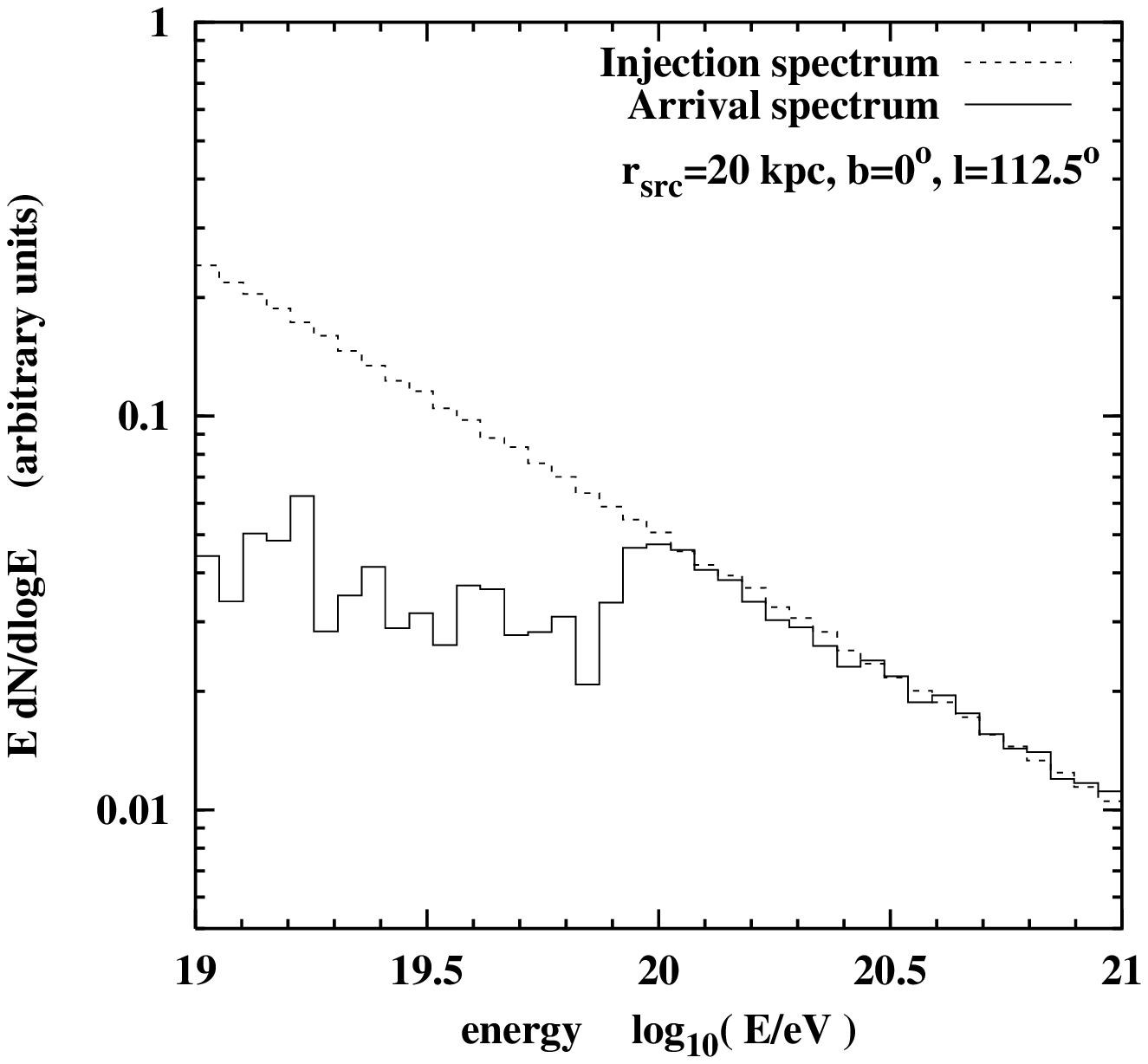,width=7.9cm}
\epsfig{figure=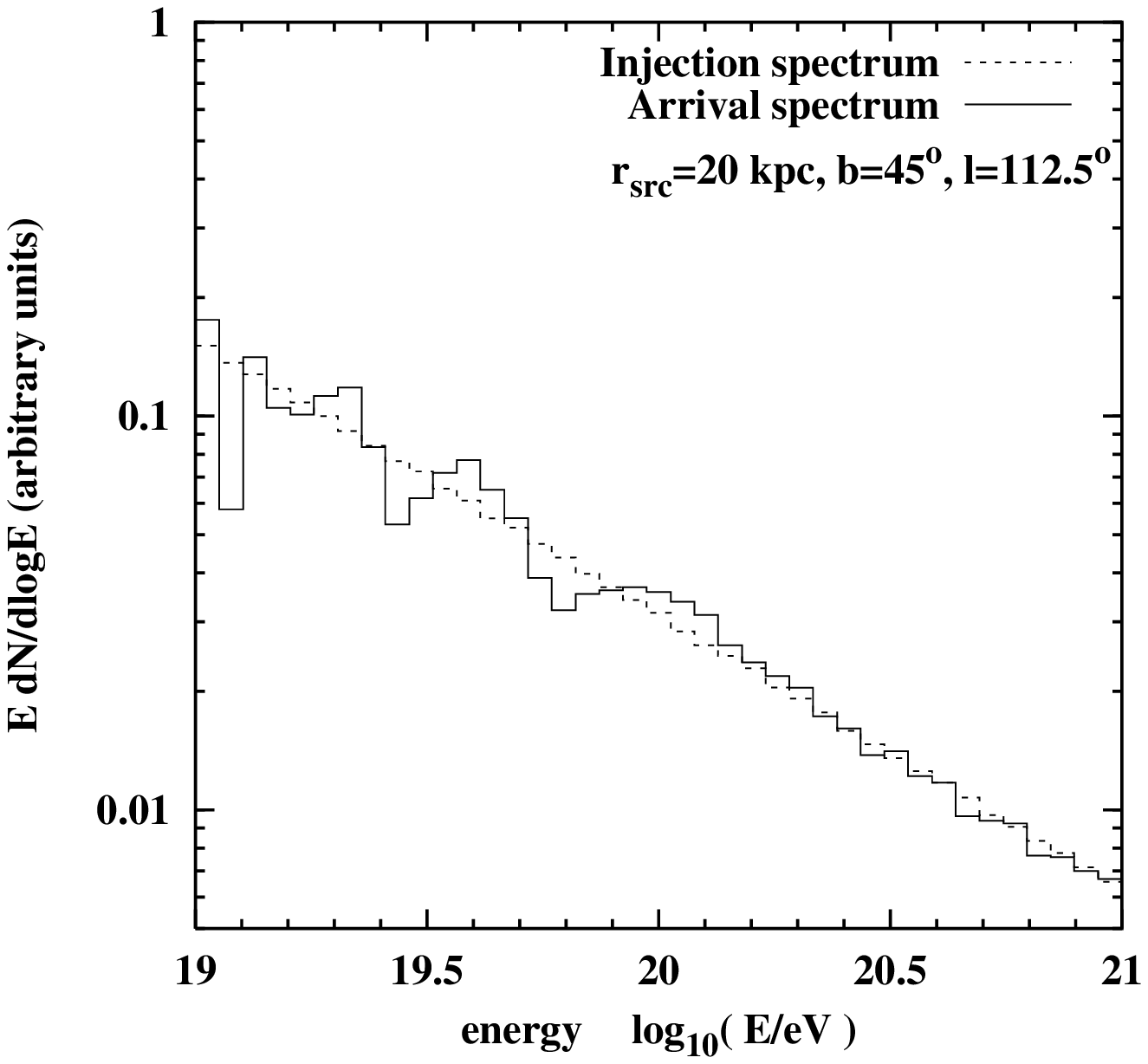,width=7.9cm}
}
\caption{
Injection and arrival spectra of cosmic rays for two source positions.
The injection spectrum is proportional to $E^{-2.7}$.
The normalization is arbitrary. 
}
\label{fig:spectra}
\end{figure}
%%%%%%%%%%%%%%%%%%%%%%%%%%%%%%%%
Also clearly seen is the selection effect of the magnetic field. At
certain energies almost no cosmic rays arrive at Earth form a given
source. This can be
studied in detail  by comparing the injection spectrum with that of
observed cosmic rays, see Fig.~\ref{fig:spectra}. The low-energy
flux from sources in the Galactic plane is strongly suppressed. On the
other hand, the flux from sources at high Galactic latitudes does not suffer 
energy-dependent modulation.

%%%%%%%%%%%%%%%%%%%%%%%%%%%%%%%%
\begin{figure}[htb!]
\centerline{
\epsfig{figure=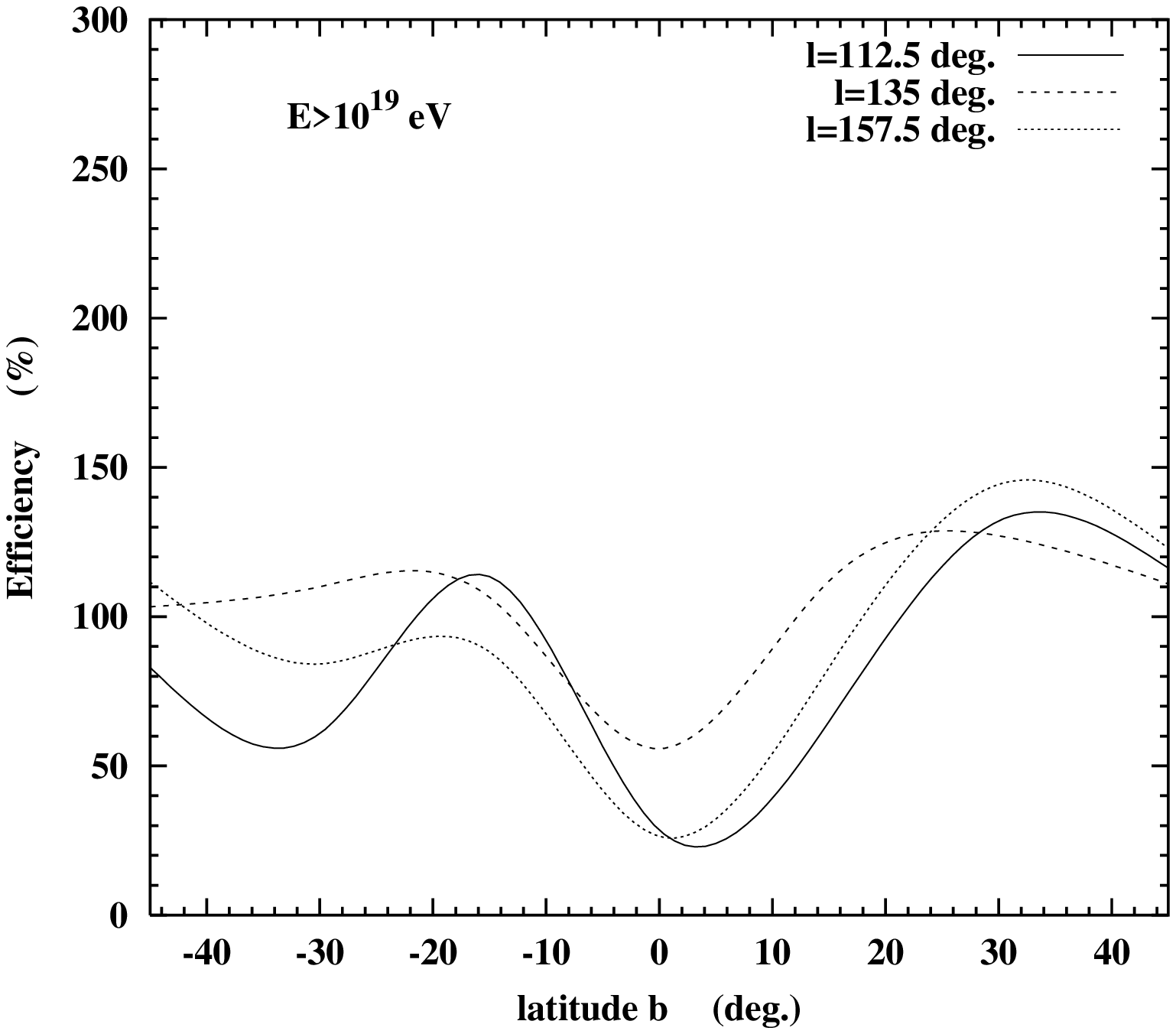,width=7.9cm}
\epsfig{figure=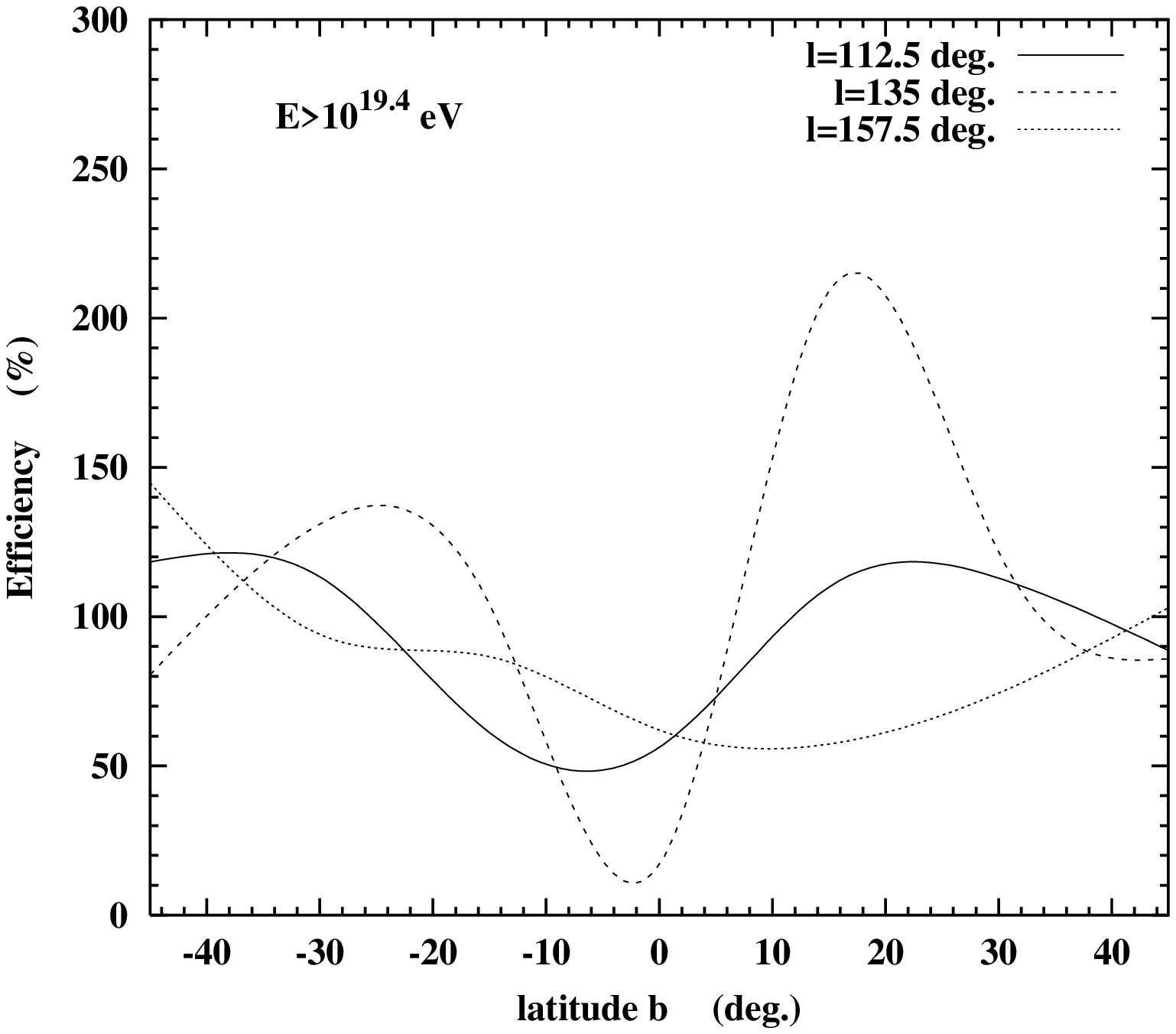,width=7.9cm}
}
\caption{
Efficiency (in \%) of a single source for different source locations
in Galactic coordinates and $r_{\rm src}=20$ kpc. 
In the left panel the efficiency is calculated for
cosmic rays with $E>10^{19.0}$ eV and in the right panel 
for cosmic rays with $E>10^{19.4}$. The differential injection spectrum 
is proportional to $E^{-2.7}$. 
\label{fig:efficiency}
}
\end{figure}
%%%%%%%%%%%%%%%%%%%%%%%%%%%%%%%% 
Another way of characterizing the selection effect due to the GMF is the
source efficiency, defined here as the ratio of the flux of detected CRs
to the expected flux for vanishing GMF.
Fig.~\ref{fig:efficiency} shows the efficiency of a point source
for different positions. 
The magnetic field has a focusing or defocusing effect on the 
flux of cosmic rays emitted from a source depending on its position.
Sources in the Galactic plane are 
less efficient than sources away from it as explained before. 
It is also interesting to note the slight asymmetry in 
source efficiency between the northern and southern Galactic
hemispheres. For a fixed source position, the efficiency 
is a non-monotonous function of the energy. 
At sufficiently high energy the deflection in the magnetic field
is small and the efficiency approaches $100 \%$. 

%%%%%%%%%%%%%%%%%%%%%%%%%%%%%%%%
\begin{figure}[htb!]
\centerline{
\epsfig{figure=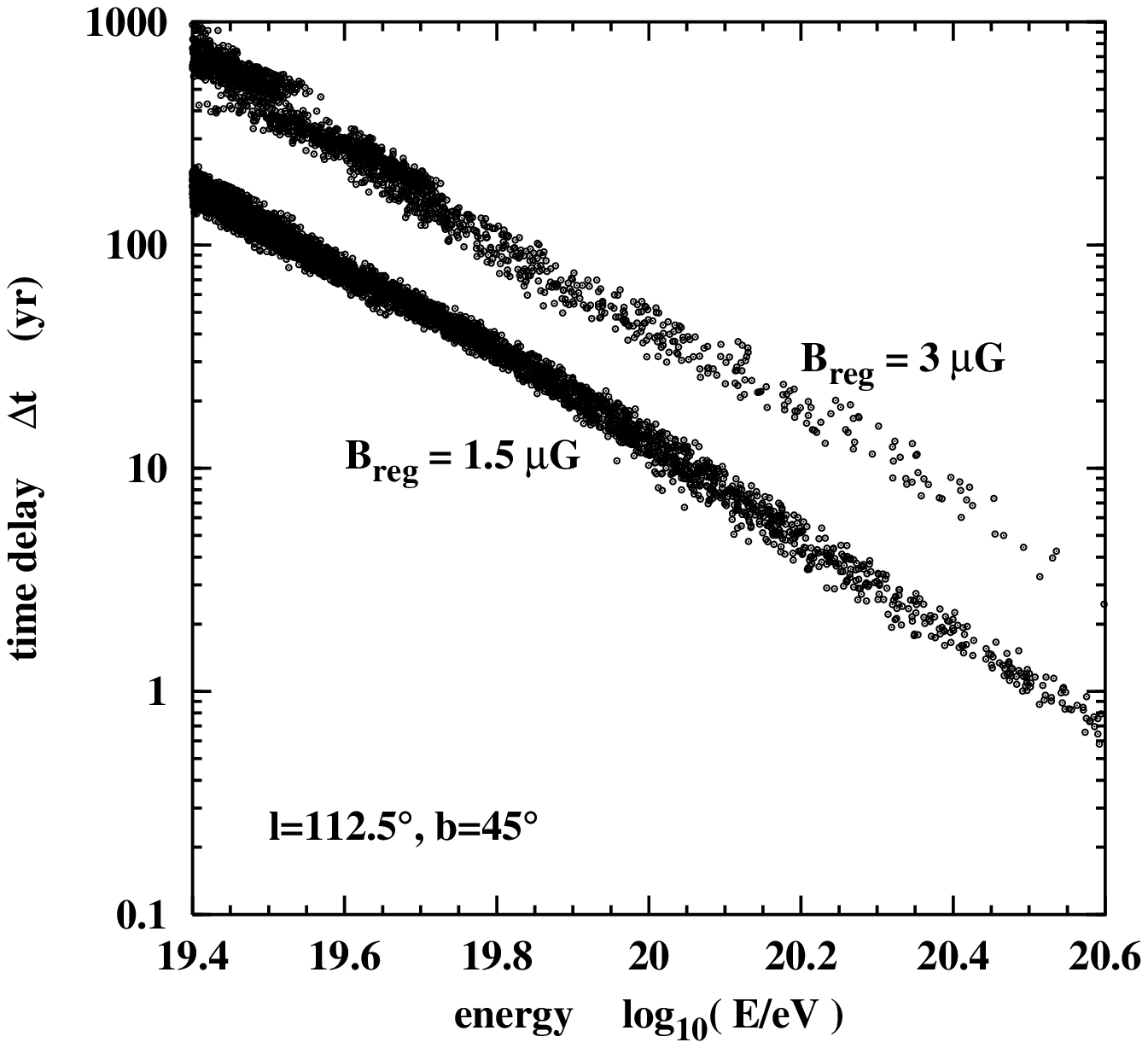,width=7.9cm}
\epsfig{figure=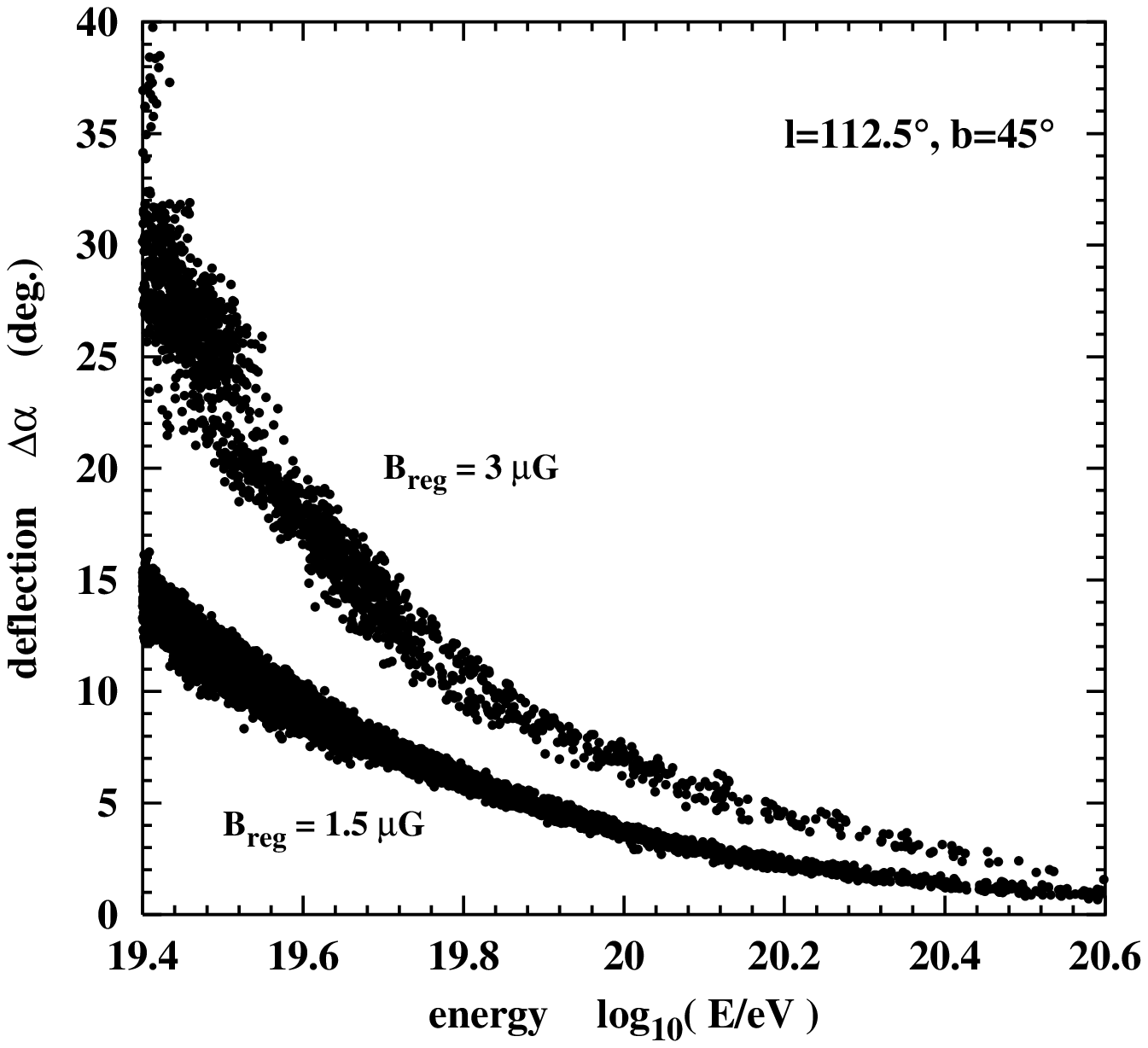,width=7.9cm}
}
\caption{
Arrival time delay and deflection angle for two different magnetic field
strengths at the position of the Solar system. 
\label{fig:strength}
}
\end{figure}
%%%%%%%%%%%%%%%%%%%%%%%%%%%%%%%%
So far we have considered only sources at a Galactocentric distance of
20 kpc and a Galactic magnetic field with a local strength of 1.5
$\mu$G. We have checked with simulations for $r_{\rm scr} =15, 20$ and
40 kpc that the arrival time delay as well as the
deflection angle depend only weakly on the distance to the source. 
However, as expected, the correlation of both quantities with the
arrival energy becomes weaker for increasing source distance.
In contrast to the distance-dependence the amplitude of the GMF 
is very important. Fig.~\ref{fig:strength} shows the
arrival time delay and deflection angle 
for a strength of 1.5 and 3 $\mu$G of the regular
component of the magnetic field in the vicinity of the Solar system.
Doubling the field strength leads to an approximately twice as large
deflection angle. The time delay increases by more than a factor of 2.

%\clearpage
%%%%%%%%%%%%%%%%%%%%%%%%%%%%%%%%%%%%%%%%%%%%%%%%%%%%%%%%%%%%%%%%%%%%%%%%%%

\section{Many point sources\label{sec:multiple-point}}

In the following we consider the case of many point sources. 
The flux of each of these point sources is large, i.e. each of
them might contribute up to several cosmic rays observed at Earth.
In astrophysical terms this scenario realizes some features expected in
models for UHECR production in the Galactic halo, such as $Z$-bursts, decay
or annihilation
of super-heavy particles, or acceleration by rapidly spinning magnetars.

In the simulation the number of cosmic rays arriving at Earth from a particular
source depends on the source location (i.e.\ its efficiency) and on 
the total number of cosmic rays injected per source.
The latter parameter is one of the major unknowns in the simulation.
In the following we assume that each source emits the
same number of cosmic rays isotropically with a differential 
energy spectrum $dN/dE \propto E^{-2.7}$, i.e. all sources have
the same cosmic ray luminosity.
Thus, by construction, one has fewer cosmic rays injected
from the same source at high energy than at low energy.
This leads to a maximum of the self-correlation
at some intermediate energy. At low energy ($\sim 10^{19}$ eV) the
Galactic magnetic field will destroy the correlation and at high
energy ($\sim 10^{20}$ eV) the number of cosmic rays from the same source is
small.

In the limit of many sources,
the multiple point source scenario
results in an almost perfectly
isotropic arrival distribution on large angular scales, as
already discussed in the uniform source distribution scenario. 
At the same time it can exhibit a significant self-correlation at small
scales.
Due to the large-scale isotropy 
we do not expect any significant differences in the
arrival flux for positively and negatively charged particles.
Although different sources contribute to the flux at Earth,
the observed flux will appear to be nearly isotropic. In
particular there will be no asymmetry between the northern
and southern Galactic hemispheres.

In practice we inject 20,000 cosmic rays per source within a
$30^\circ$-cone around the line-of-sight. The source positions are again
sampled from a uniform distribution on spheres around the Galactic center.
We do not restrict the position of the source,
although for the analysis we only select events that come from
the outer Galaxy as previously defined.
Simulations were done for the
energy cutoffs $E>10^{19}$ and $10^{19.4}$ eV. All cosmic rays arriving
at the detector were recorded. The significance of the self-correlation
in these data sets is then reduced by randomly diluting the samples of
cosmic rays accepted for the analysis. For example, a set is diluted by
a factor of 2 by accepting cosmic rays with a probability of 0.5.

In the following we consider two examples of typical correlations
obtained in the multiple point source setup. Since the number of
doublets depends on the total number of detected cosmic rays, we dilute our
simulated data sets until we obtain a two-dimensional self-correlation
of about 5-7 $\sigma$ at $\Delta l = \Delta b = 0$.

The first example shows the self-correlation plots for 200 
cosmic rays arriving at Earth with energy $E>10^{19.4}$ eV. 
The sources were located at a sphere with 20 kpc radius.
%%%%%%%%%%%%%%%%%%%%%%%%%%%%%%%%
\begin{figure}[htb!]
\centerline{
\epsfig{figure=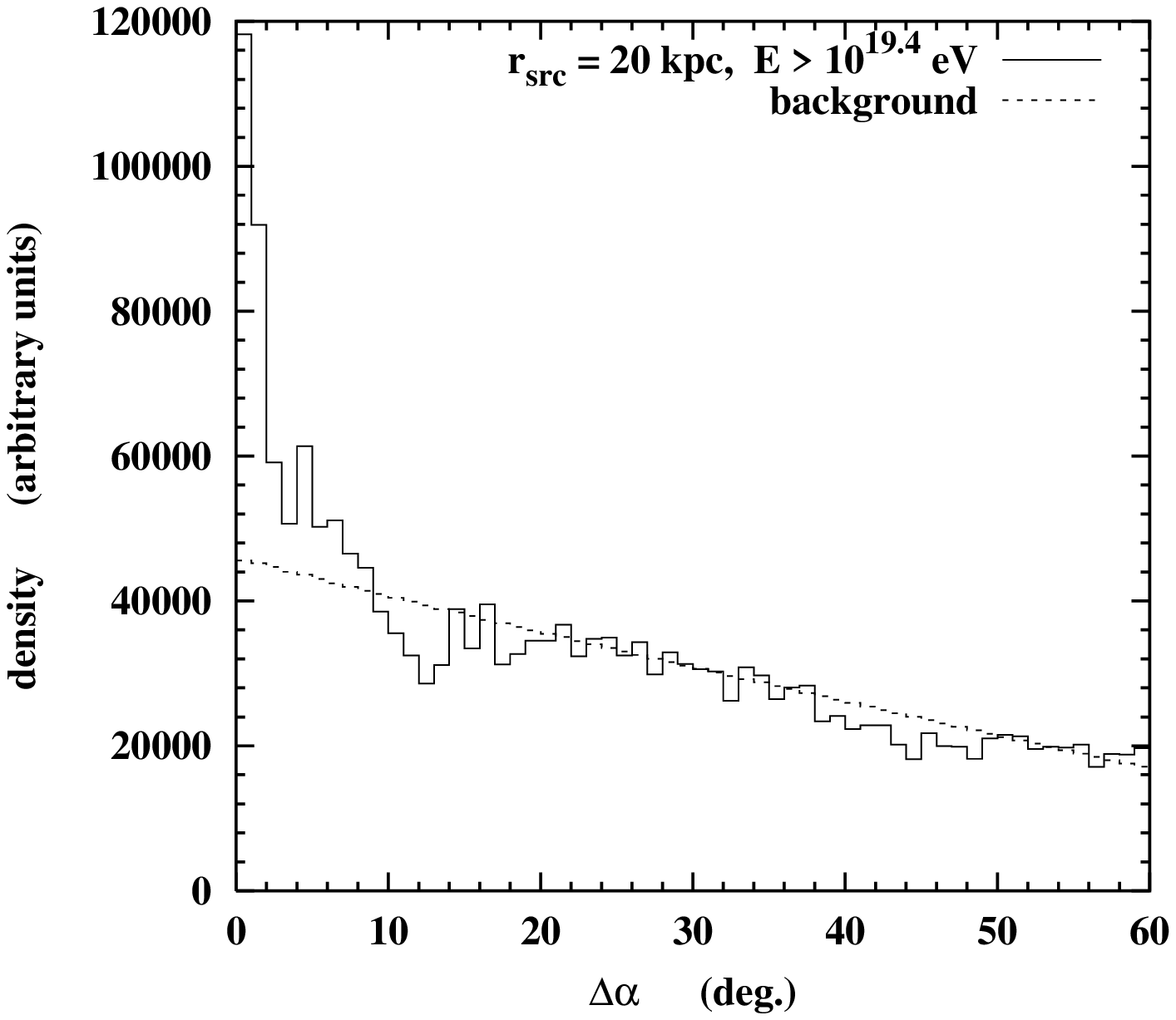,width=7.9cm}
}
\caption{
One-dimensional self-correlation 
for events with $E>10^{19.4}$ eV in a many point source scenario.
The correlation is calculated from a data set containing 200
cosmic rays observed at Earth from sources at 20 kpc 
from the Galactic center. 
}
\label{fig:dis-1dc-19p4}
\end{figure}
%%%%%%%%%%%%%%%%%%%%%%%%%%%%%%%%
The 1D angular self-correlation in Fig.~\ref{fig:dis-1dc-19p4} 
exhibits a structure which is qualitatively similar to that
seen in AGASA data: A large peak at small angular separation,
followed by a tail at large angles. The significance plot reveals
considerable statistical fluctuations due to the limited number of 200
cosmic rays used for this calculation. The observed cosmic rays
were produced by 139 different sources, i.e.\ about 1.4 particles 
per source reached the detector.

%%%%%%%%%%%%%%%%%%%%%%%%%%%%%%%%
\begin{figure}[htb!]
\centerline{
\epsfig{figure=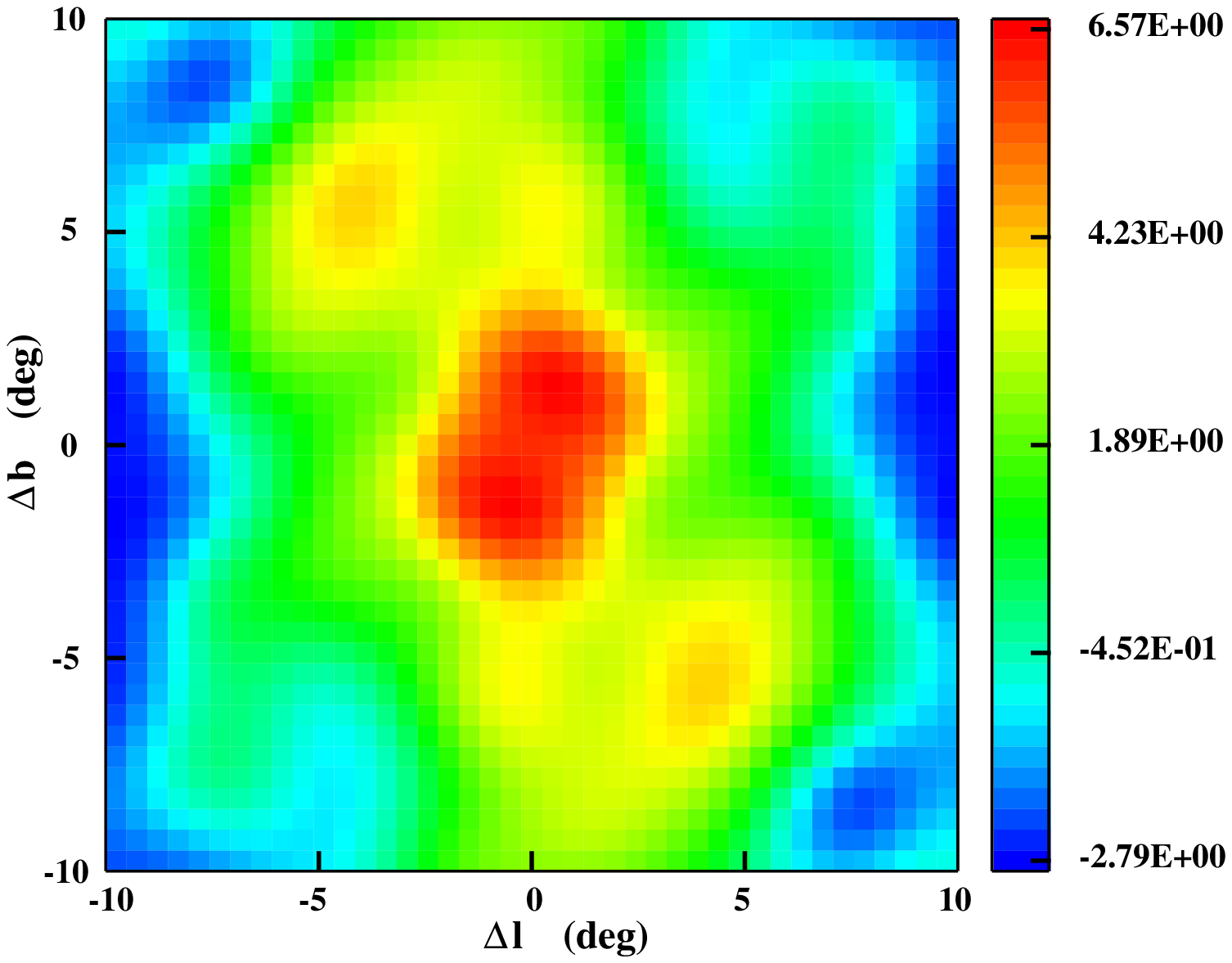,width=7.9cm}
\epsfig{figure=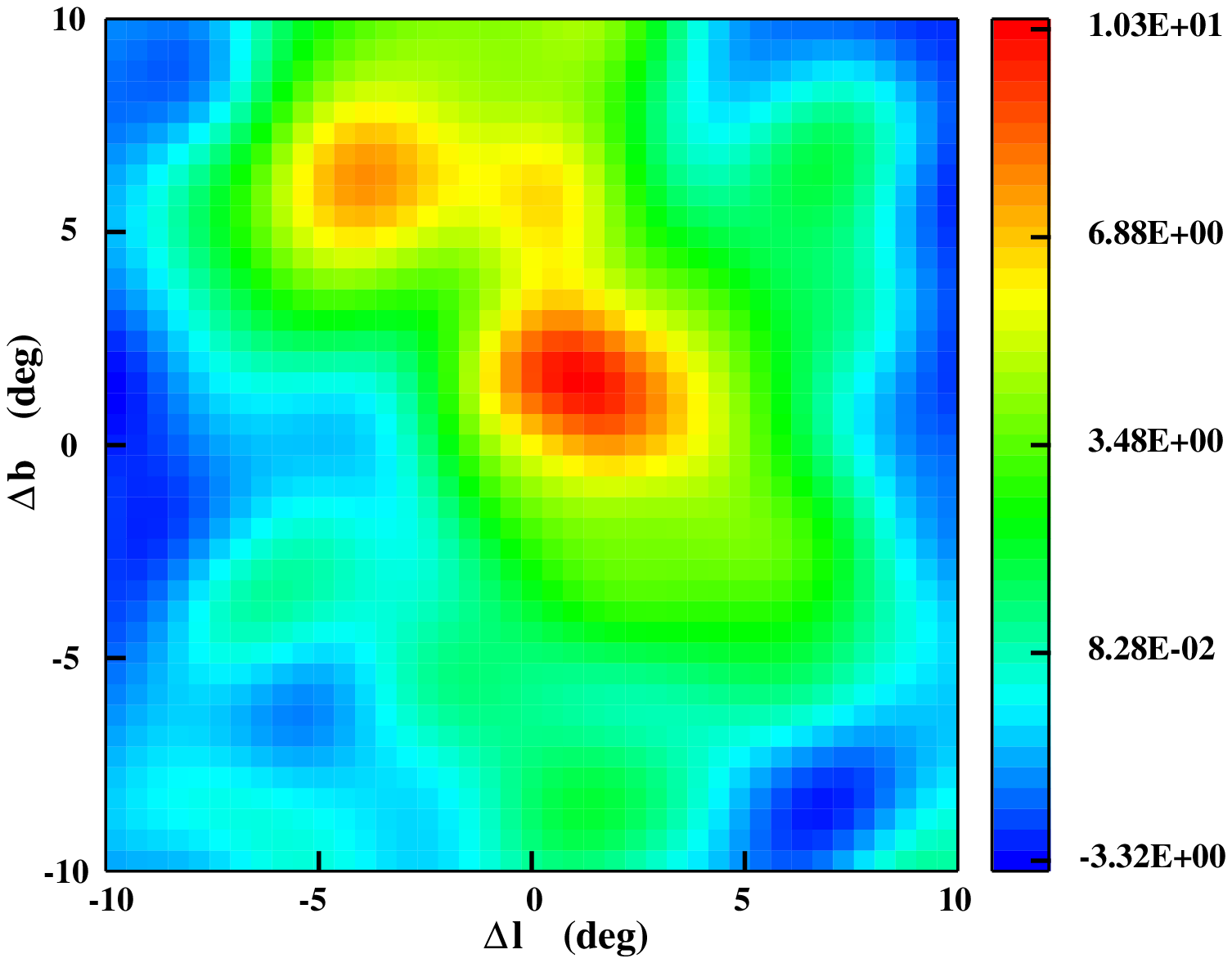,width=7.9cm}
}
\caption{
Symmetrized (left) and energy-ordered (right) significance of
self-correlation for $E>10^{19.4}$ eV.
The sources are located at a $r_{\rm src}=20$ kpc sphere around the Galactic
center. 
The 200 events used for the calculation
are the same as in Fig.~\ref{fig:dis-1dc-19p4}.
\label{fig:dis-2dc-19p4}
}
\end{figure}
%%%%%%%%%%%%%%%%%%%%%%%%%%%%%%%%
The 2D correlation calculated from the same events
is shown in Fig.~\ref{fig:dis-2dc-19p4}.
The structures of the correlation ellipse are qualitatively very similar
to the one observed in the AGASA data
(see Fig.~\ref{fig:AGASA-2dc-19p6}). This could be considered as a
strong support for our magnetic field model and point sources at 20 kpc
Galactocentric distance.

\begin{figure}[htb!]
\centerline{
\epsfig{figure=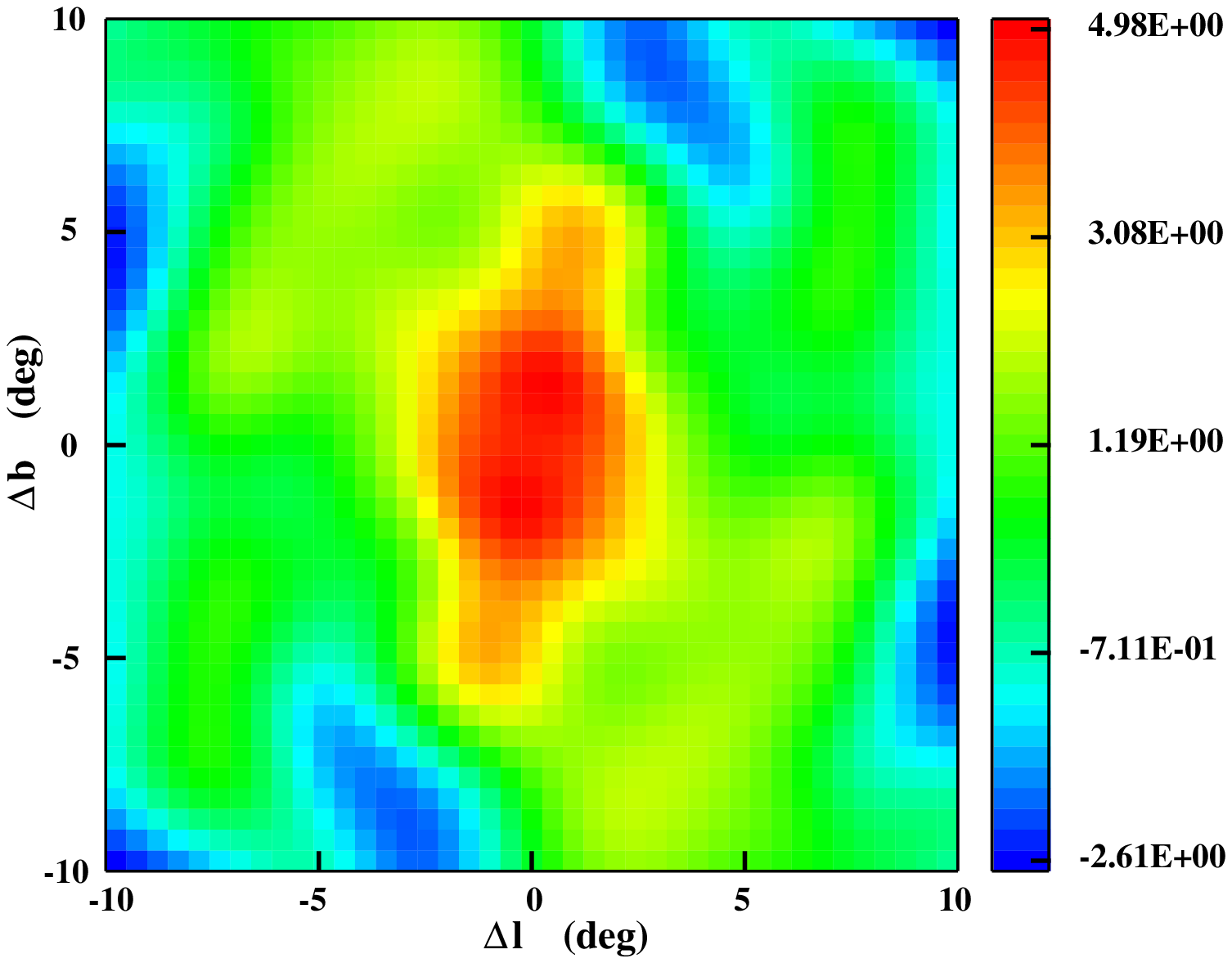,width=7.9cm}
\epsfig{figure=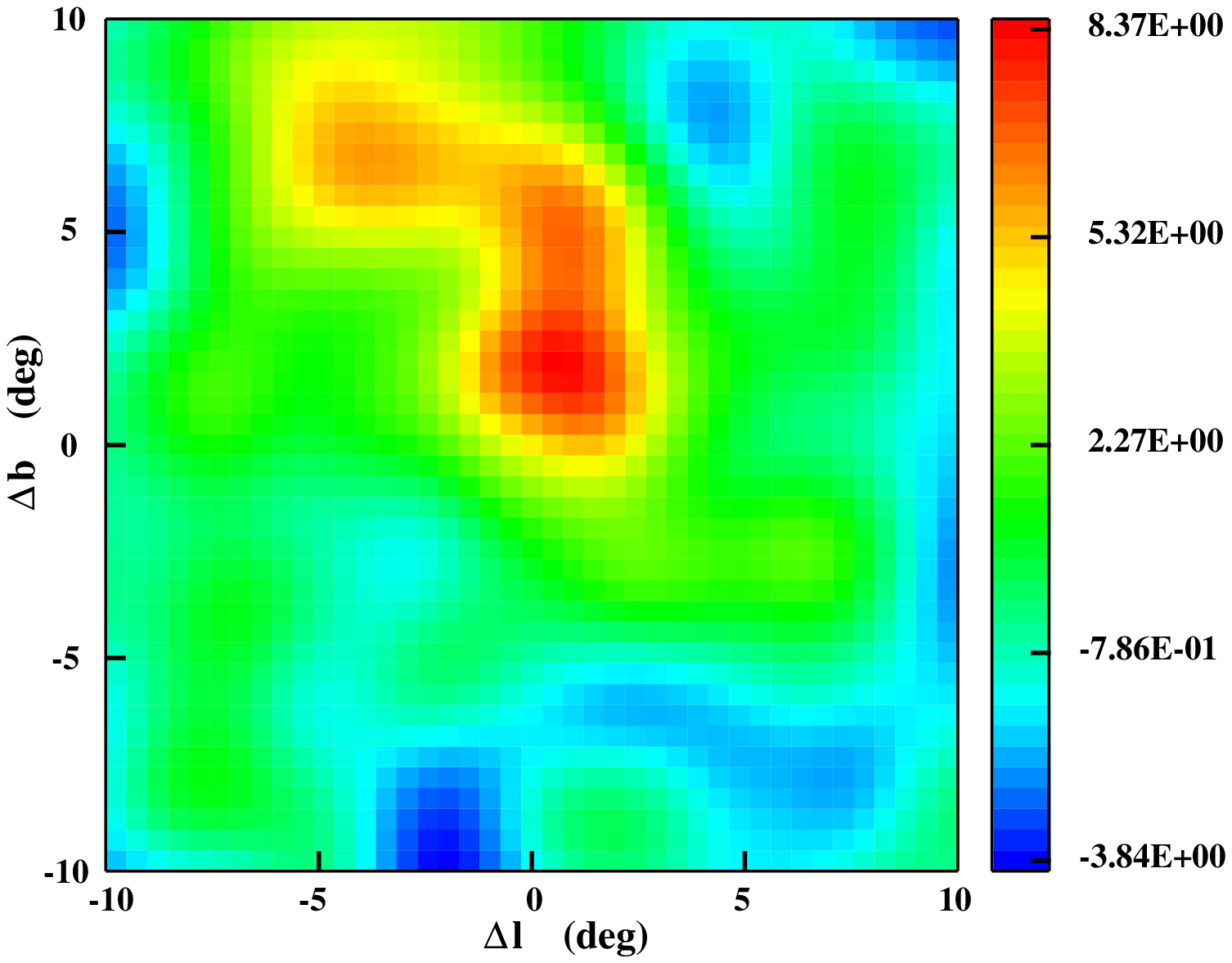,width=7.9cm}
}
\caption{
Symmetrized (left) and energy-ordered (right) significance of
self-correlation for $E>10^{19.4}$ eV.
The sources are located at a $r_{\rm src}=20$ kpc sphere around the
Galactic center. The correlation was calculated from 400 trajectories
reaching Earth.
\label{fig:dis-2dc-19p4-400}
}
\end{figure}

However, it should be emphasized that with only 200 cosmic rays the slope as
well as the lobe-structures of the correlation ellipse are not
statistical significant. 
To show this we calculate the self-correlation for 400 events, including
the 200 previously analysed events, in Fig.~\ref{fig:dis-2dc-19p4-400}.
The 400 cosmic rays at Earth were produced by 280 sources which
again corresponds to about 1.4 detected cosmic rays per source.
%%%%%%%%%%%%%%%%%%%%%%%%%%%%%%%%
%%%%%%%%%%%%%%%%%%%%%%%%%%%%%%%%
Although the self-correlation seems to be somewhat weaker and the slope
of the correlation ellipse has changed,
the significance of the excess at small $\Delta l$ and $\Delta b$ is
still obvious. 

Furthermore we want to emphasize that the energy-ordered
correlation is always asymmetric towards positive $\Delta b$. This means that
low-energy particles appear to arrive from directions with smaller $b$ than
high-energy ones. Such a result is expected from the field
direction of the magnetic arm next to our Solar system. However, there
is no such indication in the AGASA data. In case of a mixture of protons
and antiprotons as UHECRs, as expected from decays of super-massive
particles or neutrino annihilation, the energy-ordered correlation
function is nearly symmetric. Our simulations show that this expected
symmetry is only realized for samples with very high statistics.

\begin{figure}[htb!]
\centerline{
\epsfig{figure=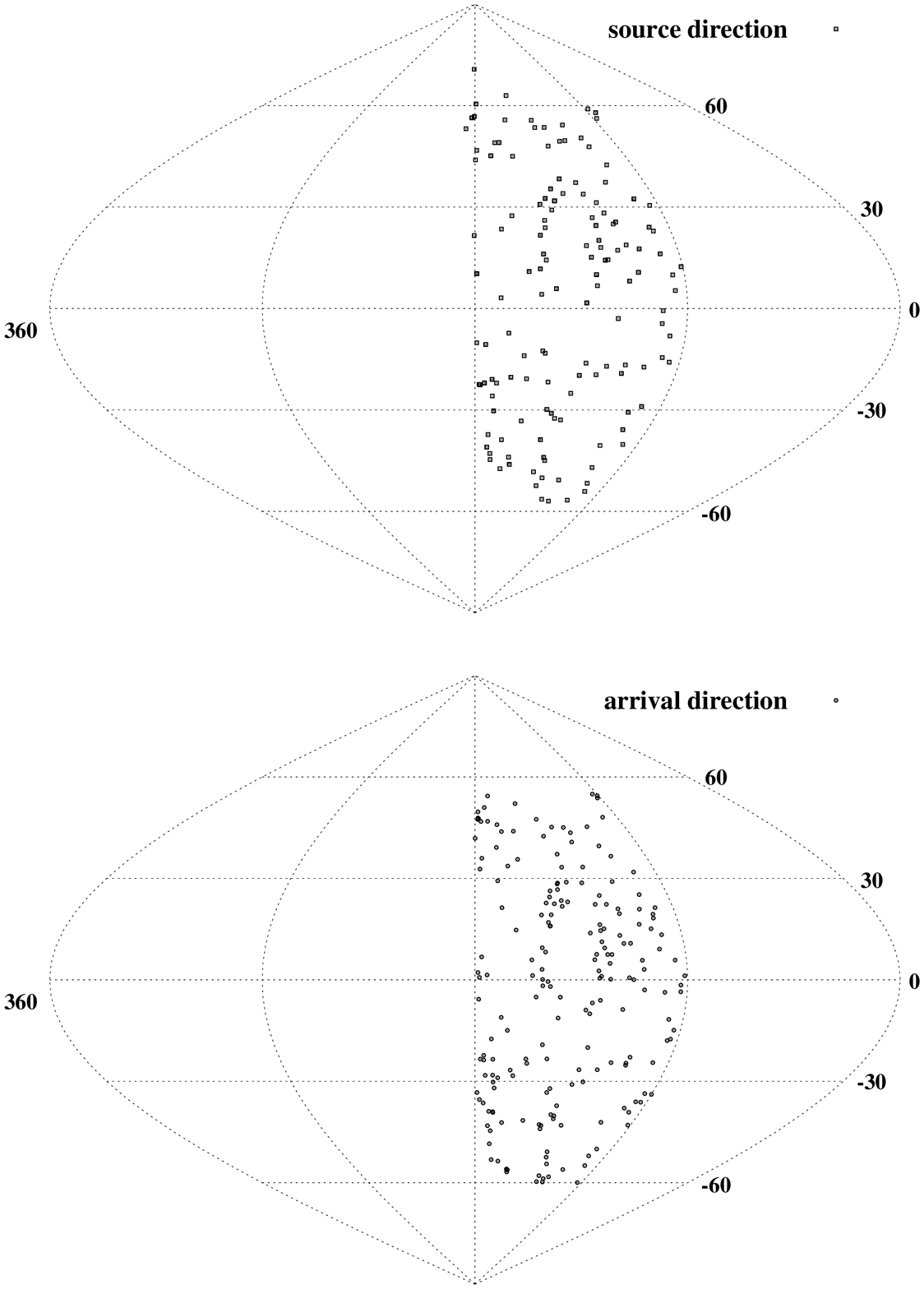,width=12cm}
}
\caption{
Source and arrival direction distribution for 200 simulated cosmic rays
with energy $E>10^{19.4}$ eV. The injection spectrum at the sources
was $dN/dE \propto E^{-2.7}$.
}
\label{fig:sky-map-20d-20k-19p4}
\end{figure}

Finally we show in Fig.~\ref{fig:sky-map-20d-20k-19p4}
the arrival distribution of the 200 cosmic rays used previously
in Figs.~\ref{fig:dis-1dc-19p4} and \ref{fig:dis-2dc-19p4}. 
The location of the sources is also shown as the limited number of
sources might contribute to some anisotropy.
A visual inspection of Fig.~\ref{fig:sky-map-20d-20k-19p4} 
reveals no significant large-scale anisotropy.
%%%%%%%%%%%%%%%%%%%%%%%%%%%%%%%%
%%%%%%%%%%%%%%%%%%%%%%%%%%%%%%%% 
We have also performed simulations restricted to 25 cosmic rays arriving
from the outer Galaxy, the same number as the AGASA data. The
arrival distributions appeared to be compatible with that expected for
an isotropic flux on large angular scales.

%%%%%%%%%%%%%%%%%%%%%%%%%%%%%%%%%%%%%%%%%%%%%%%%%%%%%%%%%%%%%%%%%
\clearpage

\section{Application to AGASA multiplets\label{sec:discussion}}

\begin{figure}[htb!]
\centerline{
\epsfig{figure=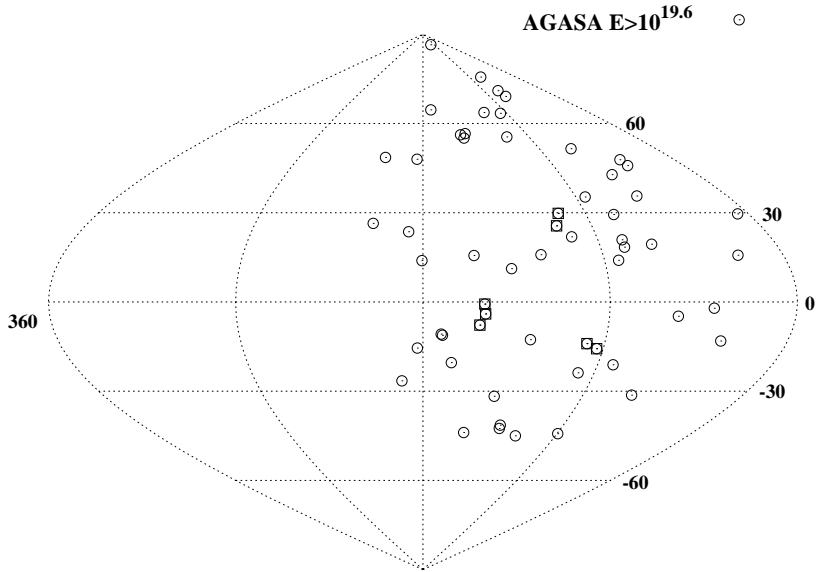,width=11cm}
}
\caption{
Arrival direction distribution of cosmic rays with $E>10^{19.4}$ eV as
published by AGASA \protect\cite{Hayashida:2000zr}. The three
multiplets listed in Table~\protect\ref{tab:agasaD} are marked 
by squares (see text).
\label{fig:AGASA-arr}
}
\end{figure}

As an application of our results we now discuss whether the clusters
found in the AGASA data above $10^{19.6}$ eV are compatible 
with the model of the GMF we have adopted  and the
hypothesis of single point sources for their origin.   
Fig.~\ref{fig:AGASA-arr} shows
the full data set with 2 doublets and 1 triplet from the ``outer
Galaxy'' \cite{Hayashida:2000zr}. 
%%%%%%%%%%%%%%%%%%%%%%%%%%%%%%%%%%%%%%%%%%%%%%%%%%%%%%%
%%%%%%%%%%%%%%%%%%%%%%%%%%%%%%%%%%%%%%%%%%%%%%%%%%%%%%%

For simplicity we use only protons in our simulations. 
The results do not change much
for antiprotons but would be considerably different for nuclei.
The approximate Galactic coordinates of the 
potential source position are estimated based on the energy of the
constituents of the multiplet and the expected angular deflection
of their tracks in the GMF. Then we inject protons from 
these potential sources and compute, for different distances
to the source, the time delay $\Delta t$ between the
arrival at the detector of a proton and a light ray emitted in 
the same direction,  
and the angular difference $\Delta \alpha$ between the line of sight 
from the detector to the source and the proton's arrival direction. 
Our results are summarized in Table
\ref{tab:agasaC} in which we also include the energy, arrival directions
and dates of detection of events belonging to clusters 
as measured by AGASA \cite{Hayashida:2000zr}. 
$\Delta t$ and $\Delta \alpha$ depend on the energy of the
events, the position of the source
and the injection angle from the source. 
Protons having the same energy encounter slightly different magnetic 
field configurations along their trajectories to the detector
depending on the injection angle. 
This produces the spread in $\Delta t$ and $\Delta \alpha$
shown in Table \ref{tab:agasaC}. $\Delta t$ and 
$\Delta \alpha$ are expected to have a maximum uncertainty due
to the size of the detector of the order of 
$0.8~(r_{\rm src}/20~{\rm kpc})$ years and $0.5^\circ$ 
respectively. 
%%%%%%%%%%%%%%%%%%%%%%%%%%%%%%%%%%%%%%%%%%%%%%%%%%
\begin{table}[htb!]
\caption{
Energy, Galactic latitude ($b$) and longitude ($l$) and date
of arrival of the events belonging to clusters from the outer Galaxy as
detected by AGASA. Rows from 9 to 12 are the minimum and maximum values of
the time delay between the arrival of
a proton and a photon at the detector for sources
at Galactocentric 
distances $r_{\rm src}=12$, 15, 20 and 40 kpc. $\Delta \alpha$ is the
angular difference between the line of sight from the detector
to the source and the proton's arrival direction. It is also
computed for the same Galactocentric distances to the source 
as $\Delta t$. $r_{\oplus}$ is the distance from the 
potential source of the multiplet to the Earth, calculated
as well for the four Galactocentric distances.
\label{tab:agasaC}
}
\begin{center}
\footnotesize
\renewcommand{\arraystretch}{1.5}
\begin{tabular}{|c||c|c||c|c|c||c|c|} \hline
Cluster & \multicolumn{2}{c||}{~C1~} & \multicolumn{3}{c||}{~C2~} & \multicolumn{2}{c|}{~C4~}  \\ \cline{1-8}
%

% ENERGY
 E (eV)~& $2.13\times10^{20}$ & $5.07\times10^{19}$ & $7.76\times10^{19}$ & $5.50\times10^{19}$ & $5.35\times10^{19}$ & $5.47\times10^{19}$ & $4.89\times10^{19}$  \\ \hline

% GALACTIC LATITUDE
 b~& $-41.4^\circ$ & $-42.5^\circ$ & $55.1^\circ$ & $56.6^\circ$ & $56.2^\circ$ & $-11.2^\circ$ & $-10.8^\circ$  \\

% GALACTIC LONGITUDE
 l~& $130.5^\circ$ & $130.2^\circ$ & $145.5^\circ$ & $143.2^\circ$ & $147.5^\circ$ & $170.4^\circ$ & $171.1^\circ$  \\ \hline

% DATE
 yr/mo/da~& 93/12/03  & 95/10/29 & 95/01/26 & 92/08/01 & 98/04/04 & 86/01/05 & 95/11/15  \\ \hline

% DISTANCE FROM EARTH
$r_{\oplus}^{12}$ (kpc) & \multicolumn{2}{c||}{~5.2~} & \multicolumn{3}{c||}{~5.6~} & \multicolumn{2}{c|}{~3.5~}  \\
$r_{\oplus}^{15}$ (kpc) & \multicolumn{2}{c||}{~8.8~} & \multicolumn{3}{c||}{~9.2~} & \multicolumn{2}{c|}{~6.5~}  \\
$r_{\oplus}^{20}$ (kpc) & \multicolumn{2}{c||}{~14.4~} & \multicolumn{3}{c||}{~14.8~} & \multicolumn{2}{c|}{~11.5~}  \\
$r_{\oplus}^{40}$ (kpc) & \multicolumn{2}{c||}{~35.1~} & \multicolumn{3}{c||}{~35.6~} & \multicolumn{2}{c|}{~31.5~}  \\ \cline{1-8}

% DELAY
 $\Delta t_{12}$~(yr) & $0.6 - 0.7$ & $12.5 - 15.8$ & $5.7 - 8.0$ & $11.0 - 15.4$ & $11.7 - 17.0$ & $13.6 - 17.8$ & $16.7 - 23.3$ \\ 

 $\Delta t_{15}$~(yr) & $1.4 - 2.1$ & $24.4 - 33.4$ & $11.8 - 16.0$ & $22.1 - 31.8$ & $22.3 - 33.0$ & $30.1 - 51.6$ & $37.1 - 67.3$ \\

 $\Delta t_{20}$~(yr) & $1.6 - 2.7$ & $29.9 - 38.1$ & $14.2 - 20.8$ & $30.2 - 48.3$ & $31.4 - 48.3$ & $64.2 - 77.1$ & $72.7 - 93.0$ \\

 $\Delta t_{40}$~(yr) & $1.2 - 2.2$ & $32.4 - 47.0$ & $16.8 - 28.6$ & $34.3 - 55.7$ & $41.2 - 63.9$ & $79.7 - 108.2$ & $93.8 - 135.7$ \\ \hline

% DEFLECTION
 $\Delta \alpha_{12}$~(deg)& $0.4 - 1.6$ & $3.5 - 5.5$ & $2.5 - 3.7$ & $3.6 - 5.0$ & $3.9 - 5.3$ & $5.2 - 6.9$ & $5.2 - 7.2$ \\

$\Delta \alpha_{15}$~(deg)& $1.1 - 2.0$ & $5.6 - 7.1$ & $3.8 - 4.7$ & $5.1 - 6.6$ & $5.1 - 6.9$ & $6.6 - 9.5$ & $7.5 - 10.3$ \\

$\Delta \alpha_{20}$~(deg)& $1.6 - 2.7$ & $6.2 - 8.0$ & $4.0 - 5.1$ & $5.9 - 7.8$ & $6.1 - 7.8$ & $7.3 - 9.7$ & $7.9 - 10.8$ \\

$\Delta \alpha_{40}$~(deg)& $1.2 - 1.9$ & $6.3 - 8.8$ & $4.5 - 6.1$ & $6.6 - 8.5$ & $6.9 - 9.0$ & $5.5 - 7.9$ & $5.5 - 8.5$ \\ \hline
\end{tabular}
\end{center}
\end{table}
%%%%%%%%%%%%%%%%%%%%%%%%%%%%%%%%%%%%%%%%%%%%%%%%%%%%%%%%%%%%%%%%

As already discussed in Sec.~\ref{sec:single-point}, due to the spread in 
$\Delta t$, the lowest energy
event in the cluster does not necessarily arrive after 
the highest energy event and vice versa, as it would be expected from 
the larger angular deflection in the GMF of the lower energy event.
For this to happen the energy of the events has to be 
roughly within $\sim 30\%$ of each other as can be understood
by looking at Fig. \ref{fig:tdelay}.
For the sake of illustration of this point, take for instance the C4 doublet 
in Table \ref{tab:agasaC} and $r_{\rm src}=15$ kpc. Within the model
of the GMF we have here adopted, the low energy event in the doublet 
could arrive $\sim 40$ years after a photon emitted in the same
direction whereas for the high energy event this same delay could
be $\sim 50$ years. Then assuming they are emitted at the same
time, the high energy event would arrive 10 years after the low
energy one. This is of course not compatible with the difference
in arrival time observed by AGASA, however it is still possible and
illustrates that one should be careful when  
ruling out a simultaneous emission from a source on the basis
of the arrival times at the detector.

For a cluster to be explainable as coming from 
a bursting source within our GMF model, 
the predicted values of $\Delta t$ 
must be such that the time difference between
the arrival of the events in the cluster is of the order of that 
observed by AGASA. Their angular separation at the detector should 
also be smaller than AGASA's angular resolution of $2.5^\circ$. 
With these two criteria in mind, the doublet C1 cannot be produced 
by simultaneous emission from  
a source within the range of Galactocentric distances  
12 -- 40 kpc, since
the time between the arrival of both CRs in the doublet ($\sim 1.9$ yr) 
is always smaller than the time difference predicted by the model, even 
when the spread in $\Delta t$ is taken into account. This means that 
either the GMF strength is smaller than the one we are 
using by roughly an order of magnitude, or the 
source emits continuously and then the CRs are not emitted 
at the same time. It is also possible that the source is at a 
Galactocentric distance smaller than 12 kpc.
In the C4 doublet, the two CRs arrive
$\sim 9.8$ years apart in time and they can be 
produced by a bursting source for all source distances we have explored. The
angular difference between the event's tracks 
is smaller than $2.5^\circ$.  
The highest energy member of the C3 triplet arrives earlier than the 
event next in energy, but this arrival time inversion cannot be explained 
when accounting for the spread in $\Delta t$ unless $r_{\rm src}<15$ kpc. 
This points to  
a continuous source for the origin of the triplet, or to a nearby
short-lived source, assuming 
the GMF model is correct.

%%%%%%%%%%%%%%%%%%%%%%%%%%%%%%%%%%%%%%%%%%%%%%%%%%%%%%%%%%%%%%%%
\begin{table}[htb!]
\caption{
Energy, Galactic latitude ($b$) and longitude ($l$) and date
of arrival of the events detected by AGASA from the outer Galaxy that 
could belong to hypothetical clusters. 
Rows from 5 to 8 are the minimum and maximum values of
the time delay between the arrival of
a proton and a photon at the detector for sources
at Galactocentric 
distances $r_{\rm src}=12$, 15, 20 and 40 kpc. $\Delta \alpha$ is the
angular difference between the line of sight from the detector
to the source and the proton's arrival direction. It is also
computed for the same Galactocentric distances to the source
as $\Delta t$.
$r_{\oplus}$ is the distance from the
potential source of the multiplet to the Earth, calculated
as well for the four Galactocentric distances.
\label{tab:agasaD}  
}
\begin{center}
\footnotesize
\renewcommand{\arraystretch}{1.5}
\begin{tabular}{|c||c|c||c|c|c||c|c|} \hline
Cluster & \multicolumn{2}{c||}{~D1~} & \multicolumn{3}{c||}{~D2~} & \multicolumn{2}{c|}{~D3~}  \\ \cline{1-8}
%

% ENERGY
 E (eV)~& $6.81\times10^{19}$ & $4.95\times10^{19}$ & $7.16\times10^{19}$ & $6.11\times10^{19}$ & $4.29\times10^{19}$ & $9.10\times10^{19}$ & $4.78\times10^{19}$  \\ \hline

% GALACTIC LATITUDE
 b~& $-15.7^\circ$ & $-14.0^\circ$ & $-4.0^\circ$ & $-7.8^\circ$ & $-0.7^\circ$ & $25.6^\circ$ & $29.8^\circ$  \\

% GALACTIC LONGITUDE
 l~& $93.3^\circ$ & $98.8^\circ$ & $149.8^\circ$ & $152.4^\circ$ & $150.3^\circ$ & $108.8^\circ$ & $105.1^\circ$  \\ \hline

% DATE
 yr/mo/da~& 84/12/12  & 99/09/25 & 99/07/28 & 98/10/27 & 99/10/20 & 91/11/29 & 96/05/13  \\ \hline

% DISTANCE FROM EARTH
$r_{\oplus}^{12}$ (kpc) & \multicolumn{2}{c||}{~7.1~} & \multicolumn{3}{c||}{~3.8~} & \multicolumn{2}{c|}{~6.2~}  \\
$r_{\oplus}^{15}$ (kpc) & \multicolumn{2}{c||}{~11.0~} & \multicolumn{3}{c||}{~6.9~} & \multicolumn{2}{c|}{~9.9~}  \\
$r_{\oplus}^{20}$ (kpc) & \multicolumn{2}{c||}{~16.7~} & \multicolumn{3}{c||}{~12.0~} & \multicolumn{2}{c|}{~15.6~}  \\
$r_{\oplus}^{40}$ (kpc) & \multicolumn{2}{c||}{~37.6~} & \multicolumn{3}{c||}{~32.2~} & \multicolumn{2}{c|}{~36.4~}  \\ \cline{1-8}

% DELAY
 $\Delta t_{12}$~(yr) & $18.5 - 26.0$ & $37.3 - 52.9$ & $4.8 - 7.5$ & $6.8 - 10.6$ & $14.8 - 21.5$ & $5.3 - 8.0$ & $16.4 - 29.8$ \\ 

 $\Delta t_{15}$~(yr) & $15.6 - 28.5$ & $30.0 - 61.3$ & $13.7 - 23.2$ & $17.8 - 33.1$ & $49.0 - 86.2$ & $7.9 - 14.0$ & $29.7 - 54.2$ \\

 $\Delta t_{20}$~(yr) & $36.1 - 46.3$ & $62.7 - 93.8$ & $12.8 - 28.4$ & $16.1 - 41.3$ & $37.6 - 93.0$ & $7.9 - 15.1$ & $31.5 - 57.6$ \\

 $\Delta t_{40}$~(yr) & $77.4 - 254.1$ & $148.4 - 441.3$ & $25.1 - 39.9$ & $35.3 - 52.6$ & $63.0 - 105.0$ & $11.8 - 20.3$ & $37.9 - 79.7$ \\ \hline

% DEFLECTION
 $\Delta \alpha_{12}$~(deg)& $3.6 - 4.6$ & $4.9 - 6.8$ & $3.0 - 4.4$ & $3.7 - 5.0$ & $6.0 - 7.1$ & $1.6 - 2.9$ & $3.4 - 5.4$ \\

$\Delta \alpha_{15}$~(deg)& $4.1 - 4.9$ & $5.8 - 7.7$ & $5.1 - 6.9$ & $6.2 - 8.4$ & $9.5 - 12.7$ & $3.1 - 3.8$ & $5.8 - 6.9$ \\

$\Delta \alpha_{20}$~(deg)& $3.5 - 5.1$ & $4.8 - 6.5$ & $4.0 - 6.3$ & $5.0 - 8.0$ & $8.0 - 12.3$ & $2.6 - 3.7$ & $5.5 - 7.6$ \\

$\Delta \alpha_{40}$~(deg)& $0.8 - 4.7$ & $1.0 - 4.5$ & $2.7 - 5.4$ & $3.5 - 6.5$ & $4.9 - 9.8$ & $2.6 - 4.0$ & $3.8 - 8.9$ \\ \hline
\end{tabular}
\end{center}
\end{table}
%%%%%%%%%%%%%%%%%%%%%%%%%%%%%%%%%%%%%%%%%%%%%%%%%%%%%%%
Finally we show in Table \ref{tab:agasaD} the same information as
in Table \ref{tab:agasaC} for three hypothetical clusters
of events extracted from the AGASA's sample. They are constituted of 
events which 
are close to each other in the sky as can be seen in 
Fig.\ref{fig:AGASA-arr}. Their angular separation
is larger than $2.5^\circ$, being $\sim 5.6^\circ$ for the 
events in the D1 pair, $\sim 5.3^\circ$ for the  
events in D3 and from $\sim 3.3^\circ$ to 
$\sim 7.4^\circ$ for the three events in D2. 
However, the arrival directions follow the expected pattern of deflection
due to the GMF (see Fig.~\ref{fig:uniform-sky2}).

We explore the possibility of each of them being produced
by a distinct point source on the basis of their angular separation.  
Due to the spread in $\Delta \alpha$ the two 
members of the D1 hypothetical doublet could  
qualify as a doublet. 
The GMF model we use predicts that the lower energy protons
from a cluster should come from smaller values of $b$.  
The opposite is observed for 
the pair of events that constitute the D1 doublet, however 
the difference is still within the experimental uncertainty in the 
determination of arrival direction. 
%Their arrival time difference 
%($\sim 14.8$ years) is 
%not consistent with the predictions of the GMF model and  
%a simultaneous emission from the source is disfavored.
The range of angular separation between the members of the potential 
triplet D2 would in principle allow that they belong
to a triplet. However the sign of $\Delta b$ 
is the opposite of what is predicted in the GMF model and, 
unlike doublet D1, is much too large to be attributable to 
the experimental uncertainty. 
A possibility which is not yet excluded is that the low energy 
member of D2 is an antiproton which would explain why it is arriving
from a larger $b$ than the high energy members of the triplet. 
However if this is the case, the
sources have to be nearby otherwise the angular difference 
would be too large.  
The same comment applies to the 
constituents of the hypothetical doublet D3. The D1 doublet is
hence not yet ruled out as coming from a single 
source within the GMF model used in this paper. Clusters 
D2 and D3 are less favored but still possible in some 
scenarios. Future
accumulation of statistics by AGASA and other experiments might well
find actual clusters of events close to the directions of the three 
potential clusters we have just discussed.

%%%%%%%%%%%%%%%%%%%%%%%%%%%%%%%%%%%%%%%%%%%%%%%%%%%%%%%%%%%%%%

\section{Conclusions\label{sec:end}}

We have performed a detailed calculation of the UHECR flux expected in
several simple, limiting source scenarios.
We have implemented an approach different from that used
by other authors propagating the particles from the sources 
to the detector. 
This allows not only the investigation of the particle's deflection in
the GMF but also the detailed study of various source scenarios. We find the
self--correlation analyses in one and two dimensions
to be powerful tools for identification of
the sources of UHECR and their distribution.

The predictive power of our results 
is mainly limited by the uncertainty in the GMF model we use in terms
of structure, field strength and spatial extent. Nevertheless
we have obtained a number of interesting results which do not depend
very much on the parameters of the particular GMF model.

\begin{itemize}
\item
For an isotropic extragalactic flux one expects an isotropic flux at
Earth as long as cosmic rays are not trapped by the GMF.
The observed CRs would stem from sources being distributed
non-uniformly.
\item
The SGP plane is disfavoured as main source of UHECR as long as the
extragalactic magnetic field is weak.
\item
Point sources produce two-dimensional correlation ellipses, the slope of
which depends on the position of the source and not on its distance.
\item
The observed cosmic ray flux of point sources is energy-dependent
modulated. For example, cosmic rays with $10^{19}$ eV are suppressed
from sources in the Galactic plane.
\item
The size of the correlation in $\Delta l - \Delta b$ depends strongly on
the strength of the GMF and only weakly on the distance to
the sources.
\item
The energy-ordered correlation analysis 
reveals the general field direction/polarity in the vicinity of
the Solar system. It 
can be used to determine the
charge sign of the cosmic rays and also helps to estimate the importance
of statistical fluctuations in data. 
\item
The width of the time delay versus energy correlation allows for 
the inverted arrival time behaviour, i.e. 
the higher energy cosmic ray of a pair arrives earlier than
the lower energy one. This may happen if
the energies of the cosmic rays are within $\sim 30\%$ of
each other. 
\item
It seems conceivable that a scenario of multiple, uniformly
distributed sources can produce
correlation patterns as seen in AGASA data.
\item
The slope and the lobes of the 2D correlation ellipses
are not statistically significant for less than $\sim 200$ 
events from the outer Galaxy in the multiple sources scenario. 
\item
The energy-ordered 2D correlation distribution of the AGASA data does
not agree with the expectations for positively charged particles and the
BSS GMF model, however the small statistics of the data does not allow
us to draw any firm conclusions.
\item
The analysis of the AGASA multiplets within a model of point sources 
shows that only one doublet is excluded from being a pair of cosmic
rays simultaneously emitted from a source at a distance between 15 - 40
kpc. All other multiplets are within the uncertainties compatible with
coming from short-lived point sources.
\item
Further three possible multiplets were selected on the basis of the
expected self-correlation patterns.
\end{itemize}

In forthcoming work we plan to derive limits on UHECR models
such as $Z$-bursts and decay of super-heavy particles.
Supposing the self-correlation observed in AGASA data is not
a statistical fluctuation, we will extend our work to set
limits on the minimum number of nucleon-antinucleon pairs needed
per point source to produce doublets and triplets as
well as limits on the total number of point sources required
to sustain the observed UHECR flux.

%%%%%%%%%%%%%%%%%%%%%%%%%%%%%%%%%%%%%%%%%%%%%%%%%%%%%%%%%%%%%%%%

\vspace*{1cm}

\noindent
{\bf Acknowledgements}
The authors acknowledge fruitful discussions with P.L. Biermann, 
T.K. Gaisser, P.P. Kronberg and M. Teshima. This research is supported in
part by NASA Grant NAG5-7009.
RE\&TS are also supported by the US Department of Energy contract
DE-FG02 91ER 40626. 
TS also acknowledges the hospitality of PCC, Coll\`ege de France and
a grant from the Minist\`ere de la Recherche of France.
The simulations presented here were performed on DEC Alpha and 
Beowulf clusters funded by NSF grant ATM-9977692. 

%%%%%%%%%%%%%%%%%%%%%%%%%%%%%%%%%%%%%%%%%%%%%%%%%%%%%%%%%%%%%%%%

%\bibliographystyle{zpc}
%\bibliography{astro-1}

\end{document}